\documentclass[11pt,hidelinks,colorlinks=true]{article}

\usepackage{color,soul}

\pdfoutput=1
\usepackage{latexsym}
\usepackage{amsmath}
\usepackage{amssymb}
\usepackage{amsfonts,amsthm}
\usepackage[top=1in, bottom=1in, left=1in, right=1in]{geometry}
\usepackage[lined, algoruled, linesnumbered]{algorithm2e}

\newcommand{\ignore}[1]{}

\usepackage{stackrel}
\usepackage{breqn}
\usepackage{enumitem}
\usepackage{amsthm}

\usepackage{hyperref,xcolor}
\definecolor{winered}{rgb}{0.5,0,0}
\hypersetup
{
    pdfborder={0 0 0},
    colorlinks=true,
    linkcolor={winered},
    urlcolor={winered},
    filecolor={winered},
    citecolor={winered},
    linktoc=all,
}

\newcommand{\strong}[3]{$#1\stackrel[]{#3} \leftrightarrow #2$}
\newcommand{\tstrong}[3]{$#1\stackrel[]{#3} \leftrightarrow_{t} #2$}

\newcommand{\strongEC}[3]{$#1\stackrel[]{#3} \leftrightarrow_{\mathit{2e}} #2$}
\newcommand{\tstrongEC}[3]{$#1\stackrel[]{#3} \leftrightarrow_{\mathit{2et}} #2$}

\newtheorem{theorem}{Theorem}[section]
\newtheorem{lemma}[theorem]{Lemma}

\newtheorem{corollary}[theorem]{Corollary}

\newtheorem{remark}[theorem]{Remark}

\newtheorem{property}[theorem]{Property}
\newtheorem{proposition}[theorem]{Proposition}

\usepackage{authblk}

\begin{document}
\title{\bf On 2-strong connectivity orientations of mixed graphs and related problems\thanks{Research supported by the Hellenic Foundation for Research and Innovation (H.F.R.I.) under the ``First Call for H.F.R.I. Research Projects to support Faculty members and Researchers and the procurement of high-cost research equipment grant'', Project FANTA (eFficient Algorithms for NeTwork Analysis), number HFRI-FM17-431.}}

\author{Loukas Georgiadis}
\author{Dionysios Kefallinos}
\author{Evangelos Kosinas}
\affil{Department of Computer Science \& Engineering, University of Ioannina, Greece.\\ \small{E-mail: \{loukas,dkefallinos,ekosinas\}@cse.uoi.gr}}

\maketitle

\begin{abstract}
A mixed graph $G$ is a graph that consists of both undirected and directed edges.
An orientation of $G$ is formed by orienting all the undirected edges of $G$, i.e., converting each undirected edge $\{u,v\}$ into a directed edge that is either $(u,v)$ or $(v,u)$.
The problem of finding an orientation of a mixed graph that makes it strongly connected is well understood and can be solved in linear time.
Here we introduce the following orientation problem in mixed graphs. Given a mixed graph $G$, we wish to compute its maximal sets of vertices $C_1,C_2,\ldots,C_k$ with the property that
by removing any edge $e$ from $G$ (directed or undirected), there is an orientation $R_i$ of $G\setminus{e}$ such that all vertices in $C_i$ are strongly connected in $R_i$.
We study properties of those sets and show how to solve this problem in linear time by reducing it to the computation of the $2$-edge twinless strongly connected components of a directed graph.

A directed graph $G=(V,E)$ is twinless strongly connected if it contains a strongly connected spanning subgraph without any pair of antiparallel (or \emph{twin}) edges. The twinless strongly connected components (TSCCs) of a directed graph $G$ are its maximal twinless strongly connected subgraphs.
A \emph{$2$-edge twinless strongly connected component (2eTSCC) of $G$} is a maximal subset of vertices $C$ such that any two vertices $u, v \in C$ are in the same twinless strongly connected component of $G \setminus e$, for any edge $e$.
These concepts are motivated by several diverse applications, such as the design of road and telecommunication networks, and the structural stability of buildings.
Our algorithm is based on two notions: (i) a collection of auxiliary graphs $\mathcal{H}$ that preserve the 2eTSCCs
of $G$ and, for any $H \in \mathcal{H}$, the strongly connected components of $H$ after the deletion of any edge have a very simple structure, and (ii) a reduction to the problem of computing the connected components of an undirected graph after the deletion of certain vertex-edge cuts.
\end{abstract}

\section{Introduction}
\label{sec:intro}

\begin{figure*}[t!]
\begin{center}
\centerline{\includegraphics[trim={0 0 0 11cm}, clip=true, width=1\textwidth]{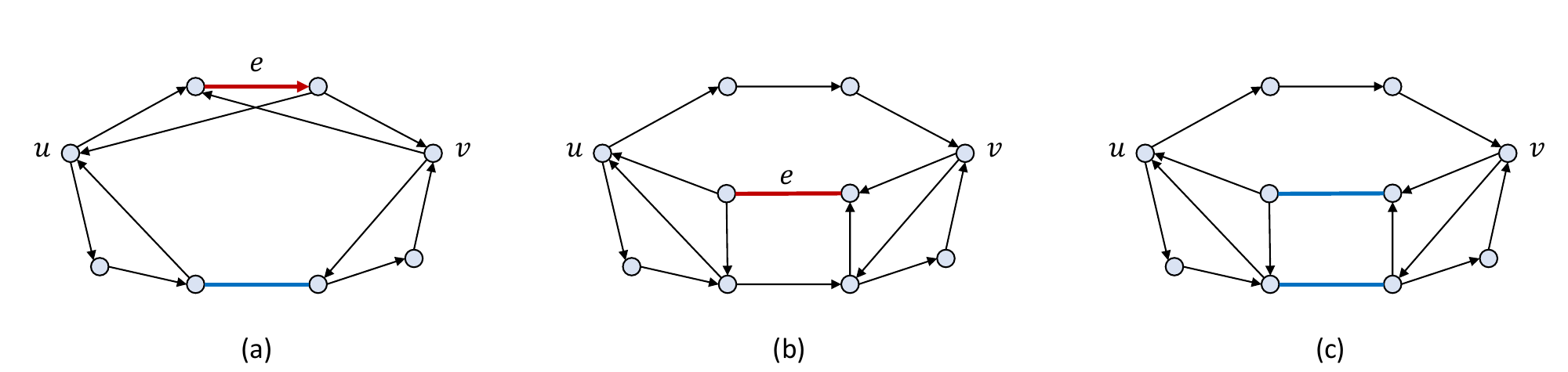}}
\caption{Examples illustrating the notion of edge-resilient strongly orientable blocks of a mixed graph $G$; undirected edges are shown in blue color and the specified edge $e$ is shown in red color.
(a)-(b) Vertices $u$ and $v$ are not in the same edge-resilient strongly orientable block of $G$. After the deletion of edge $e$ (which is directed in (a) and undirected in (b)), there is no orientation of $G \setminus{e}$ such that $u$ and $v$ are strongly connected. (c) Here $u$ and $v$ are in the same edge-resilient strongly orientable block of $G$, since for any edge $e$, there is an orientation of $G\setminus{e}$ such that $u$ and $v$ are strongly connected.\label{figure:MixedGraphsExamples}}
\end{center}
\vspace{-.5cm}
\end{figure*}

In this paper, we investigate some connectivity problems in mixed graphs and in directed graphs (digraphs).
A mixed graph $G$ contains both undirected edges and directed edges.
We denote an edge with endpoints $u$ and $v$ by $\{u,v\}$ if it is undirected, and by $(u,v)$ if it is directed from $u$ to $v$.
An \emph{orientation} $R$ of $G$ is formed by orienting all the undirected edges of $G$, i.e., converting each undirected edge $\{u,v\}$ into a directed edge that is either $(u,v)$ or $(v,u)$.
Several (undirected or mixed)
graph orientation problems have been studied in the literature, depending on the properties that we wish
an orientation $R$ of $G$ to have.
See, e.g.,~\cite{digraphs:jensen-gutin,CombinatorialOptimization:Frank,nash-williams1960,schrijver-book}.
An orientation $R$ of $G$ such that $R$ is strongly connected is called a \emph{strong orientation of $G$}.
More generally, an orientation $R$ of $G$ such that $R$ is $k$-edge strongly connected is called a \emph{$k$-edge strong orientation of $G$}.
Motivated by recent work in $2$-edge strong connectivity in digraphs~\cite{2ECC:GILP:TALG,StrongConnectivityFailures:SICOMP,2CC:HenzingerKL:ICALP15}, we introduce the following 
strong connectivity orientation problem in mixed graphs. Given a mixed graph $G$, we wish to compute its maximal sets of vertices $C_1,C_2,\ldots,C_k$ with the property that for every $i\in\{1,\dots,k\}$, and every edge $e$ of $G$ (directed or undirected), there is an orientation $R$ of $G\setminus{e}$ such that all vertices of $C_i$ are strongly connected in $R$.
We refer to these maximal vertex sets as the \emph{edge-resilient strongly orientable blocks} of $G$.
See Figure~\ref{figure:MixedGraphsExamples}.
Note that when $G$ contains only directed edges, then this definition coincides with the usual notion of $2$-edge strong connectivity, i.e., each $C_i$ is a $2$-edge strongly connected
component of $G$.
We show how to solve this problem in linear time, by providing a linear-time algorithm for computing the $2$-edge twinless strongly connected components~\cite{JABERI2021102969}, that we define next.  Moreover, as a consequence of our algorithm, it follows that $\{C_1,\dots,C_k\}$ is a partition of $V$.

We recall some concepts in directed graphs. A digraph $G=(V,E)$ is \emph{strongly connected} if there is a directed path from each vertex to every other vertex.
The \emph{strongly connected components} (SCCs) of $G$ are its maximal strongly connected subgraphs. Two vertices $u,v \in V$  are \emph{strongly connected} if they belong to the same strongly connected component of $G$.
We refer to a pair of antiparallel edges, $(x,y)$ and $(y,x)$, of $G$ as \emph{twin edges}.
A digraph $G=(V,E)$ is \emph{twinless strongly connected} if it contains a strongly connected spanning subgraph $(V,E')$ without any pair of twin edges.
The \emph{twinless strongly connected components} (TSCCs) of $G$ are its maximal twinless strongly connected subgraphs. Two vertices $u,v \in V$  are \emph{twinless strongly connected} if they belong to the same twinless strongly connected component of $G$.
Equivalently, $u$ and $v$ are twinless strongly connected if $G$ contains a path from $u$ to $v$ and a path from $v$ to $u$ so that the union of these two paths does not contain any pair of twin edges.
Raghavan~\cite{Raghavan2006} provided a characterization of twinless strongly connected digraphs, and, based on this characterization, presented a linear-time algorithm for computing the TSCCs of a digraph.

An edge (resp., a vertex) of a digraph $G$ is a \emph{strong bridge} (resp., a \emph{strong articulation point}) if its removal increases the number of strongly connected components.
A strongly connected digraph $G$ is \emph{$2$-edge strongly connected} if it has no strong bridges, and it is
\emph{$2$-vertex strongly connected} if it has at least three vertices and no strong articulation points.
Two vertices $u, v\in V$ are said to be \emph{$2$-edge strongly connected} (resp., \emph{$2$-vertex strongly connected}) if there are two edge-disjoint  (resp., two internally vertex-disjoint) directed paths from $u$ to $v$ and two edge-disjoint  (resp., two internally vertex-disjoint) directed paths from $v$ to $u$ (note that a path from $u$ to $v$ and a path from $v$ to $u$ need not be edge- or vertex-disjoint).
Equivalently, by Menger's theorem~\cite{Menger} we have that $u$ and $v$ are $2$-edge strongly connected if they remain in the same SCC after the deletion of any edge.
A \emph{$2$-edge strongly connected component} (resp., \emph{$2$-vertex strongly connected component}) of a digraph $G=(V,E)$ is defined as a maximal subset $C \subseteq V$ such that every two vertices $u, v \in C$ are $2$-edge strongly connected (resp., $2$-vertex strongly connected). Also, note that the subgraph induced by $C$ is not necessarily $2$-edge strongly connected (resp., $2$-vertex strongly connected).

The above notions extend naturally to the case of twinless strong connectivity.
An edge $e \in E$ is a \emph{twinless strong bridge} of $G$ if the deletion of $e$ increases the number of TSCCs of $G$. Similarly,
a vertex $v \in V$ is a \emph{twinless strong articulation point} of $G$ if the deletion of $v$ increases the number of TSCCs of $G$.
A linear-time algorithm for detecting all twinless strong bridges can be derived by combining the linear-time algorithm of Italiano et al.~\cite{Italiano2012} for computing all the strong bridges of a digraph, and a linear-time algorithm for computing all the edges that belong to a cut-pair in a $2$-edge-connected undirected graph~\cite{TwinlessSAP}.
Georgiadis and Kosinas~\cite{TwinlessSAP} showed that the computation of twinless strong articulation points reduces to the following problem in undirected graphs, which is also of independent interest:
Given a $2$-vertex-connected (biconnected) undirected graph $H$, find all vertices $v$ that belong to a vertex-edge cut-pair, i.e., for which there exists an edge $e$ such that
$H \setminus \{v,e\}$ is not connected.
Then, \cite{TwinlessSAP} presented a linear-time algorithm that not only finds all such vertices $v$, but also computes the number of vertex-edge cut-pairs of $v$ (i.e., the number of edges $e$ such that $H \setminus \{v,e\}$ is not connected).
Alternatively, it is possible to compute the vertices that form a vertex-edge cut-pair by exploiting the structure of the triconnected components of $H$, represented by an SPQR tree~\cite{SPQR:triconnected,SPQR:planarity} of $H$.

\begin{figure*}[t!]
\begin{center}
\centerline{\includegraphics[trim={0 0 0 11cm}, clip=true, width=1\textwidth]{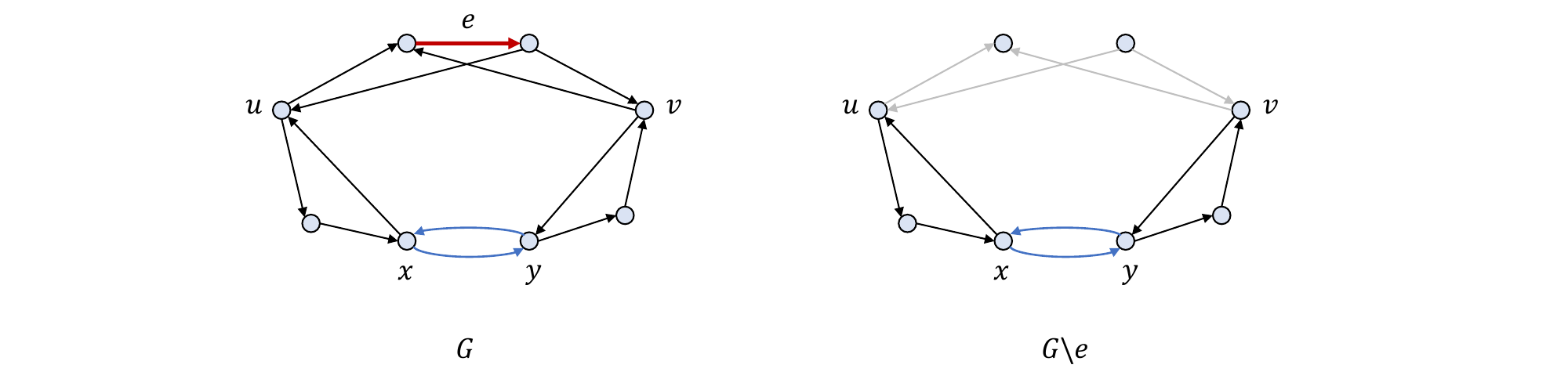}}
\caption{Vertices $u$ and $v$ of the strongly connected digraph $G$ are $2$-edge strongly connected but not $2$-edge twinless strongly connected. The deletion of the edge $e$ leaves $u$ and $v$ in a strongly connected subgraph that must contain both twin edges $(x,y)$ and $(y,x)$.\label{figure:example0}}
\end{center}
\vspace{-.5cm}
\end{figure*}

A \emph{$2$-edge twinless strongly connected component (2eTSCC) of $G$} is a maximal subset of vertices $C$ such that any two vertices $u, v \in C$ are in the same TSCC of $G \setminus e$, for any edge $e$.
Two vertices $u$ and $v$ are \emph{$2$-edge twinless strongly connected} if they belong to the same 2eTSCC. See Figure~\ref{figure:example0}.
Jaberi~\cite{JABERI2021102969} studied some properties of the $2$-edge twinless strongly connected components and presented an $O(mn)$-time algorithm for a digraph with $m$ edges and $n$ vertices.
We provide a linear-time algorithm that is based on two notions: (i) a collection of auxiliary graphs $\mathcal{H}$ that preserve the $2$-edge twinless strongly connected components of $G$ and, for any $H \in \mathcal{H}$, the SCCs of $H$ after the deletion of any edge have a very simple structure, and (ii) a reduction to the problem of computing the connected components of an undirected graph after the deletion of certain vertex-edge cuts.

The notions of twinless strong connectivity and mixed graph orientations are indeed related and are motivated by several diverse applications, such as the design of road and telecommunication networks, the structural stability of buildings~\cite{TrailOrientations,DirectingRoadNetworks,StrongOrientations,Raghavan2006}, and the analysis of biological networks~\cite{ELBERFELD201396}.
Given a mixed graph $G$, it is natural to ask whether it has an orientation so that the resulting directed graph is strongly connected~\cite{StrongOrientations,MixedStrongOrientations} or, more generally, $k$-edge strongly connected~\cite{FRANK198297,CostScalingSubmodularFlow,ArcConnectivityOrientations}.
Raghavan~\cite{Raghavan2006} noted that testing whether a digraph $G$ is twinless strongly connected is equivalent to testing whether a mixed graph has a strong orientation.
We can also observe that the computation of the twinless strongly connected components is equivalent to the following generalization of strong orientations of a mixed graph: Given a mixed graph $G$, we wish to compute its maximal sets of vertices $C_1,C_2,\ldots,C_k$ with the property that for every $i\in\{1,\dots,k\}$ there is an orientation $R$ of $G$ such that all vertices of $C_i$ are strongly connected in $R$. We call those sets the \emph{strongly orientable blocks} of $G$. 
Similarly, we show that the computation of the edge-resilient strongly orientable blocks reduces to the computation of the $2$-edge twinless strongly connected components.

The computation of edge-resilient strongly orientable blocks is related to $2$-edge strong orientations of mixed graphs in the following sense.
A mixed graph $G$ has a $2$-edge strong orientation only if it consists of a single edge-resilient strongly orientable block.
While finding a strong orientation of a mixed graph is well understood and can be solved in linear time~\cite{StrongOrientations,MixedStrongOrientations}, computing a $k$-edge strong orientation for $k>1$ seems much harder.
Frank~\cite{FRANK198297} gave a polynomial-time algorithm for this problem based on the concept of submodular flows.
Faster algorithms were later presented by Gabow~\cite{CostScalingSubmodularFlow}, and by Iwata and Kobayashi~\cite{ArcConnectivityOrientations}.
More efficient algorithms exist for computing a $k$-edge strong orientation of an undirected graph~\cite{FastEdgeOrientation,GabowSplitting,NISplitting}. In particular, the algorithm of Bhalgat and Hariharan~\cite{FastEdgeOrientation} runs in $\tilde{O}(nk^4 +m)$ time\footnote{The notation $\widetilde{O}(\cdot)$ hides poly-logarithmic factors.}, for an undirected graph with $n$ vertices and $m$ edges.
We refer to Section~\ref{sec:concluding} for a discussion of related concepts and open problems.

\section{Preliminaries}
\label{sec:preliminaries}

Let $G$ be a (directed or undirected) graph. In general, we allow $G$ to have multiple edges, unless otherwise specified. We denote by $V(G)$ and $E(G)$, respectively, the vertex set and the edge set of $G$.
For a set of edges (resp., vertices) $S$, we let $G \setminus S$ denote the graph that results from $G$ after deleting the edges in $S$ (resp., the vertices in $S$ and their incident edges). We extend this notation for mixed sets $S$, that may contain both vertices and edges of $G$, in the obvious way. Also, if $S$ has only one element $x$, we abbreviate $G \setminus S$ by $G \setminus x$.
Let $C \subseteq V(G)$.
The induced subgraph of $C$, denoted by $G[C]$, is the subgraph of $G$ with vertex set $C$ and edge set $\{e\in E\mid \mbox{both endpoints of } e \mbox{ are in } C\}$.

Let $G$ be a directed graph (digraph).
A \emph{path} $P$ in $G$ is an alternating sequence $v_1,e_1,v_2,\dots,e_{k-1},v_k$ of vertices and edges of $G$ (i.e., $v_i\in V(G)$ for every $i\in\{1,\dots,k\}$, and $e_i\in E(G)$ for every $i\in\{1,\dots,k-1\}$), such that $k\geq 1$, $e_i=(v_i,v_{i+1})$ for every $i\in\{1,\dots,k-1\}$, and all edges $e_1,\dots,e_{k-1}$ are distinct\footnote{In some textbooks, what we define here as paths is referred to as ``trails'', and the term ``path'' refers to a trail with no repeated vertices. Here we use this definition because it is convenient for our purposes.}.
We say that $P$ is a path from $v_1$ to $v_k$, and $v_1,v_k$ are called the \emph{endpoints} of $P$. We may write $v\in P$ or $e\in P$, if $P$ uses a vertex $v$ or an edge $e$, respectively. A subpath is a subsequence of a path that is also a path (i.e., it starts and ends with a vertex), and a circuit is a path whose endpoints coincide. If $k=1$ (i.e., $P$ consists of a single vertex), then we call $P$ a \emph{trivial} path.
Given a path $P$ from $x$ to $y$ and a path $Q$ from $y$ to $z$ in $G$, we let $P+Q$ denote the path from $x$ to $z$ that results from the concatenation of $P$ and $Q$ in $G$.

\begin{remark}
\normalfont{Note that since we work on digraphs that may have multiple edges, in general, it is not forbidden for a path to go from a vertex $x$ to a vertex $y$ using an edge $(x,y)$ multiple times, provided that the times that $(x,y)$ is used does not exceed its multiplicity in the graph. Therefore, if an edge $(x,y)$ has multiplicity one (e.g. if it is a strong bridge), then it is correct to argue that a path can use $(x,y)$ only once.}
\end{remark}

In a digraph $G$, we say that a vertex $x$ \emph{reaches} $y$ if there is a path in $G$ from $x$ to $y$.
If $x$ reaches $y$ and $y$ reaches $x$, we say that $x$ and $y$ are strongly connected, and we denote this fact as \strong{x}{y}{G}.
Also, the notation \tstrong{x}{y}{G} means that $x$ and $y$ are twinless strongly connected in $G$.
We omit the reference graph $G$ from the \strong{}{}{G} notation when it is clear from the context. Thus we may simply write \strong{x}{y}{} and \tstrong{x}{y}{}.
Similarly, we let \strongEC{x}{y}{G} and \tstrongEC{x}{y}{G} denote, respectively, that the vertices $x$ and $y$ are $2$-edge strongly connected and $2$-edge twinless strongly connected in $G$.
Let $G=(V,E)$ be a strongly connected digraph.
The \emph{reverse digraph} of $G$, denoted by $G^R=(V, E^R)$, is the digraph that results from $G$ by reversing the direction of all edges.
We say that an edge $e$ of a strongly connected digraph $G$ \emph{separates} two vertices $x$ and $y$ if $x$ and $y$ belong to different strongly connected components of $G\setminus{e}$. (We consider the strongly connected components of a digraph as subgraphs.)

For any digraph $G$, the associated undirected graph $G^u$ is the \emph{simple} undirected graph with vertices $V(G^u)=V(G)$ and edges $E(G^u) = \{ \{u,v\} \mid (u,v) \in E(G) \vee (v,u) \in E(G) \}$.
Let $H$ be an undirected graph. An edge $e \in E(H)$ is a \emph{bridge} if its removal increases the number of connected components of $H$. A connected graph $H$ is $2$-edge-connected if it contains no bridges.
Raghavan~\cite{Raghavan2006} proved the following characterization of twinless strongly connected digraphs.

\begin{theorem}(\cite{Raghavan2006})
\label{theorem:TwinlessCharacterization}
Let $G$ be a strongly connected digraph. Then $G$ is twinless strongly connected if and only if its underlying undirected graph $G^u$ is $2$-edge-connected.
\end{theorem}

We introduce the concept of \textit{marked vertex-edge blocks} of an undirected graph, which will be needed in our algorithm for computing the $2$-edge twinless strongly connected components.
We note that Heinrich et al.~\cite{2halfconnectivity} introduced the related concept of the \emph{$2.5$-connected components} of a biconnected graph.
Let $G$ be an undirected graph where some vertices of $G$ are marked. Let $V'$ be the set of the marked vertices of $G$. Then, a marked vertex-edge block of $G$ is a maximal subset $B$ of $V(G)\setminus{V'}$ with the property that all vertices of $B$ remain connected in $G\setminus\{v,e\}$, for every marked vertex $v$ and any edge $e$. In Section~\ref{sec:spqr} we provide a linear-time algorithm for computing the marked-vertex edge blocks of a biconnected undirected graph $G$, by exploiting properties of the SPQR-tree of the triconnected components of $G$~\cite{SPQR:triconnected,SPQR:planarity}.

\section{Reduction to the computation of the $2$-edge twinless strongly connected components}
\label{sec:reduction}
\begin{figure*}[t!]
\begin{center}
\centerline{\includegraphics[trim={0 0 0 15cm}, clip=true, width=\textwidth]{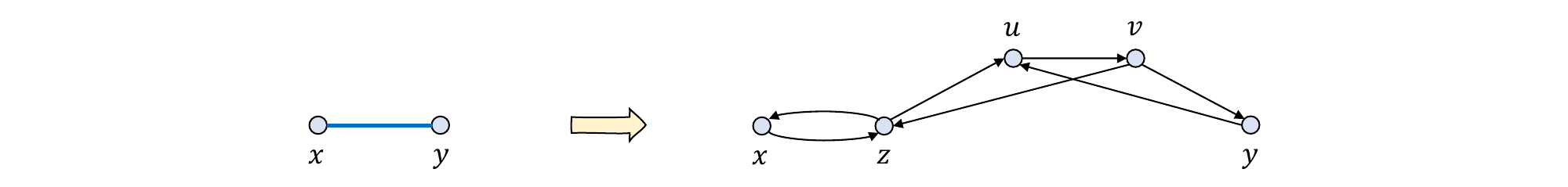}}
\vspace{-0.3cm}
\caption{Replacing an undirected edge $\{x,y\}$ with a gadget.}
\label{figure:gadget}
\end{center}
\vspace{-.5cm}
\end{figure*}

Let $G$ be a mixed graph.
By \emph{splitting} a directed edge $(x,y)$ of a graph $G$, we mean that we remove $(x,y)$ from $G$, and we introduce a new auxiliary vertex $z$ and two edges $(x,z),(z,y)$.
By \emph{replacing with a gadget} an undirected edge $\{x,y\}$ of a graph $G$, we mean that we remove $\{x,y\}$ from $G$, and we introduce three new auxiliary vertices $z,u,v$ and the edges $(x,z),(z,x),(z,u),(u,v),(v,y),(y,u),(v,z)$. See Figure~\ref{figure:gadget}.
Note that the definition of gadget replacement is not symmetric for $x,y$, but this does not affect the correctness of our approach.
(We can assume an arbitrary order of the vertices of $G$, so when we perform the gadget replacements of the undirected edges, the role of each vertex in the gadgets is predetermined.)
We refer to a non-auxiliary vertex as an \emph{ordinary vertex}.
Also, we call $(u,v)$ the \emph{critical edge} of the gadget. The idea of using this gadget is twofold. Firstly, by removing the critical edge, we simulate the operation of removing $\{x,y\}$ from the original graph.
Secondly, if we remove an edge that does not belong to the gadget, then the only paths from $x$ to $y$ and from $y$ to $x$ inside the gadget must use the pair of twin edges $(x,z)$ and $(z,x)$. These properties are useful in order to establish a correspondence between orientations and twinless strong connectivity for our applications.

Now we can reduce the computation of the edge-resilient strongly orientable blocks of a mixed graph to the computation of the 2eTSCC of a digraph.
As a warm-up, we show how to compute the strongly orientable blocks of a mixed graph via a reduction to the computation of the TSCCs of a digraph.
This is achieved by Algorithm~\ref{algorithm:blocks1}, whose correctness follows easily from known results.
As a consequence of this method for computing $C_1,\dots,C_k$, we can see that $\{C_1,\dots,C_k\}$ is a partition of $V$. Furthermore, $C_1,\dots,C_k$ satisfy the stronger property, that there is an orientation $R$ of $G$ such that, for every $i\in\{1,\dots,k\}$, $R[C_i]$ is strongly connected.
\vspace{.5cm}

\begin{algorithm}[H]
\caption{\textsf{A linear-time algorithm for computing the strongly orientable blocks of a mixed graph $G$}}
\label{algorithm:blocks1}
\LinesNumbered
\DontPrintSemicolon
split every directed edge of $G$\;
\label{line:blocks1-line1}
replace every undirected edge $\{x,y\}$ of $G$ with a pair of twin edges $(x,y),(y,x)$\;
\label{line:blocks1-line2}
compute the TSCCs $T_1,\dots,T_k$ of the resulting graph\;
\textbf{return} the sets of ordinary vertices of $T_1,\dots,T_k$\;
\end{algorithm}
\vspace{.5cm}

\begin{proposition}
Algorithm~\ref{algorithm:blocks1} is correct.
\end{proposition}
\begin{proof}
Let $G$ be the input graph, let $G'$ be the graph derived from $G$ after we have performed steps \ref{line:blocks1-line1} and \ref{line:blocks1-line2}, and let $C$ be one of the sets returned by the algorithm. First we will show that there is an orientation $R$ of $G$ such that all vertices of $C$ are strongly connected in $R$. Then we will show that $C\cup\{s\}$ does not have this property for any $s\in V(G)\setminus{C}$.

Since $C$ is a twinless strongly connected component of $G'$, by \cite{Raghavan2006} we know that there is a subgraph $G_0$ of $G'$ that contains no pair of twin edges and is such that $C$ is strongly connected in $G_0$. Let $\{x,y\}$ be an undirected edge of $G$, and let $(x,y),(y,x)$ be the pair of twin edges of $G'$ that replaced $\{x,y\}$ in Step~\ref{line:blocks1-line2}. Assume w.l.o.g. that $G_0$ contains $(x,y)$. Then we orient $\{x,y\}$ in $G$ as $(x,y)$. We do this for every undirected edge of $G$, and let $R$ be the resulting orientation. Now it is not difficult to see that all vertices of $C$ are strongly connected in $R$, and so $C$ is contained within a strongly orientable block of $G$.

To establish the maximality of $C$, let $s$ be a vertex in $V(G)\setminus{C}$. This means that $s$ is not twinless strongly connected with the vertices of $C$ in $G'$. Then we have that either $(1)$ $s$ and (the vertices of) $C$ are not strongly connected, or $(2)$ there is a pair of twin edges $(x,y),(y,x)$ of $G'$, such that for every path $P$ from $s$ to (a vertex of) $C$ in $G'$ and every path $Q$ from (a vertex of) $C$ to $s$ in $G'$, we have $(x,y)\in P$ and $(y,x)\in Q$. In case $(1)$ we have that $s$ is not strongly connected with $C$ in $G$, even if we allow every undirected edge to be replaced by a pair of twin edges. In case $(2)$, let $\{x,y\}$ be the undirected edge of $G$ that was replaced in $G'$ with the pair of twin edges $(x,y),(y,x)$. (Notice that, due to Step~\ref{line:blocks1-line1}, all pairs of twin edges in $G'$ are due to Step~\ref{line:blocks1-line2}.)
Then we have that, even if we replace every undirected edge of $G$ - except $\{x,y\}$ - with a pair of twin edges, then we still have to replace $\{x,y\}$ with the pair of twin edges $(x,y),(y,x)$ in order to have $s$ strongly connected with $C$. In any case, we have that there is no orientation $R$ of $G$ such that $s$ is strongly connected with $C$ in $R$.

Notice that this argument also shows that there is no orientation of $G$ such that a vertex $t\in C$ becomes strongly connected with a vertex $s\in V(G)\setminus{C}$. Thus, the sets returned by Algorithm~\ref{algorithm:blocks1} are all the strongly orientable blocks of $G$.
\end{proof}

Now Algorithm \ref{algorithm:blocks2} shows how we can compute all the edge-resilient strongly orientable blocks
$C_1,\dots,C_k$ of a mixed graph. 
As a consequence of this method for computing $C_1,\dots,C_k$, we can see that the edge-resilient strongly orientable blocks partition the vertex set $V$.
\vspace{.5cm}

\begin{algorithm}[H]
\caption{\textsf{A linear-time algorithm for computing the edge-resilient strongly orientable blocks of a mixed graph $G$}}
\label{algorithm:blocks2}
\LinesNumbered
\DontPrintSemicolon
split every directed edge of $G$\;
\label{line:blocks2-line1}
replace every undirected edge of $G$ with a gadget\;
\label{line:blocks2-line2}
compute the 2eTSCCs $T_1,\dots,T_k$ of the resulting graph\;
\textbf{return} the sets of ordinary vertices of $T_1,\dots,T_k$\;
\end{algorithm}
\vspace{.5cm}

\begin{proposition}
Algorithm~\ref{algorithm:blocks2} is correct.
\end{proposition}
\begin{proof}
Let $G$ be the input graph, let $G'$ be the graph derived from $G$ after we have performed steps \ref{line:blocks2-line1} and \ref{line:blocks2-line2}, and let $C$ be one of the sets returned by the algorithm. First, we will show that, for every edge $e$, there is an orientation $R$ of $G\setminus{e}$ such that all vertices of $C$ are strongly connected in $R$. Then we will show that $C\cup\{s\}$ does not have this property for any $s\in V(G)\setminus{C}$.

Now let $e$ be an edge of $G$. Suppose first that $e$ is a directed edge of the form $(x,y)$. Let $(x,z),(z,y)$ be the two edges of $G'$ into $e$ was split. Then we have that all vertices of $C$ are twinless strongly connected in $G'\setminus{(x,z)}$. By \cite{Raghavan2006}, this implies that there is a subgraph $G_0$ of $G'\setminus{(x,z)}$ that contains no pair of twin edges and is such that all vertices of $C$ are strongly connected in it. Let $\{x',y'\}$ be an undirected edge of $G$, and assume w.l.o.g. that $(x',z'),(z',x')$ is the pair of twin edges in the gadget of $G'$ that replaced $\{x',y'\}$. Assume w.l.o.g. that $G_0$ contains $(x',z')$. Then we orient $\{x',y'\}$ in $G$ as $(x',y')$. We do this for every undirected edge of $G$, and let $R$ be the resulting orientation of $G\setminus{(x,y)}$. Then it is not difficult to see that all vertices of $C$ are strongly connected in $R$. Now suppose that $e$ is an undirected edge of the form $\{x,y\}$. Let $(u,v)$ be the critical edge of the gadget of $G'$ that replaced the undirected edge $\{x,y\}$ of $G$. Then we have that all vertices of $C$ are twinless strongly connected in $G'\setminus{(u,v)}$. Now let $G_0$ and $R$ be defined similarly as above. Then it is not difficult to see that all vertices of $C$ are strongly connected in $R$.

To establish the maximality of $C$, let $s$ be a vertex in $V(G)\setminus{C}$. This means that $s$ is not $2$-edge twinless strongly connected with the vertices of $C$ in $G'$, and so there is an edge $e$ of $G'$ such that $s$ is not in the same twinless strongly connected component of $G'\setminus{e}$ that contains $C$.
Assume first that $e$ is either $(x,z)$ or $(z,y)$, where $(x,z),(z,y)$ is the pair of edges of $G'$ into which a directed edge $(x,y)$ of $G$ was split.
Then we have that either $(1)$ $s$ and (the vertices of) $C$ are not strongly connected, or $(2)$ there is a pair of twin edges $(x',z'),(z',x')$ of $G'$, such that for every path $P$ from $s$ to (a vertex of) $C$ in $G'\setminus{e}$ and every path $Q$ from (a vertex of) $C$ to $s$ in $G'\setminus{e}$, we have $(x',z')\in P$ and $(z',x')\in Q$. In case $(1)$ we have that $s$ is not strongly connected with $C$ in $G\setminus{(x,y)}$, even if we allow every undirected edge to be replaced by a pair of twin edges. In case $(2)$, let $\{x',y'\}$ be the undirected edge of $G$ that was replaced with the gadget of $G'$ that contains the pair of edges $(x',z'),(z',x')$. Then we have that, even if we replace every undirected edge of $G$ - except $\{x',y'\}$ - with a pair of twin edges, then we still have to replace $\{x',y'\}$ with the pair of twin edges $(x',y'),(y',x')$ in order to have $s$ strongly connected with $C$ in $G\setminus{(x,y)}$. In any case, we have that there is no orientation $R$ of $G\setminus{(x,y)}$ such that $s$ is strongly connected with $C$ in $R$. Now, if $e$ is an edge of a gadget that replaced an undirected edge $\{x,y\}$ of $G$, then with a similar argument we can show that there is no orientation $R$ of $G\setminus\{x,y\}$ such that $s$ is strongly connected with $C$ in $R$.

Notice that this argument also shows that if $t$ is a vertex in $C$ and $s$ is a vertex in $V(G)\setminus C$, then there is an edge $e$ (directed or undirected) such that there is no orientation of $G\setminus e$ that makes $s$ and $t$ strongly connected. Thus, the sets returned by Algorithm~\ref{algorithm:blocks2} are all the edge-resilient strongly orientable blocks of $G$.
\end{proof}

Our goal in the following sections is to provide a linear-time implementation of Algorithm~\ref{algorithm:blocks2}.
To achieve this, it suffices to provide a linear-tine algorithm for computing the $2$-edge twinless strongly connected components of a digraph.
%

\section{Connectivity-preserving auxiliary graphs}
\label{sec:auxiliary}

In this section, we describe how to construct a set of auxiliary graphs that preserve the $2$-edge twinless strongly connected components of a twinless strongly connected digraph, and moreover have the property that their strongly connected components after the deletion of any edge have a very simple structure.
We base our construction on the auxiliary graphs defined in \cite{2ECC:GILP:TALG} for computing the $2$-edge strongly connected components of a digraph, and perform additional operations in order to achieve the desired properties.
We note that a similar construction was given in \cite{3ECC:SODA23} to derive auxiliary graphs (referred to as 2-connectivity-light graphs) that enable the fast computation of the $3$-edge strongly connected components of a digraph.
Still, we cannot apply directly the construction of \cite{3ECC:SODA23}, since we also need to maintain twinless strong connectivity.

\subsection{Flow graphs and dominator trees}
\label{sec:dominators}

A \emph{flow graph} is a directed graph with a distinguished \emph{start vertex} $s$ such that every vertex is reachable from $s$.
For a digraph $G$, we use the notation $G_s$ in order to emphasize the fact that we consider $G$ as a flow graph with source $s$.
Let $G=(V,E)$ be a strongly connected graph.
We will let $s$ be a fixed but arbitrary start vertex of $G$.
Since $G$ is strongly connected, all vertices are reachable from $s$ and reach $s$, so we can refer to the flow graphs $G_s$ and $G_s^R$.

\begin{figure*}[t!]
\begin{center}
\centerline{\includegraphics[trim={0 0 0 10.5cm}, clip=true, width=\textwidth]{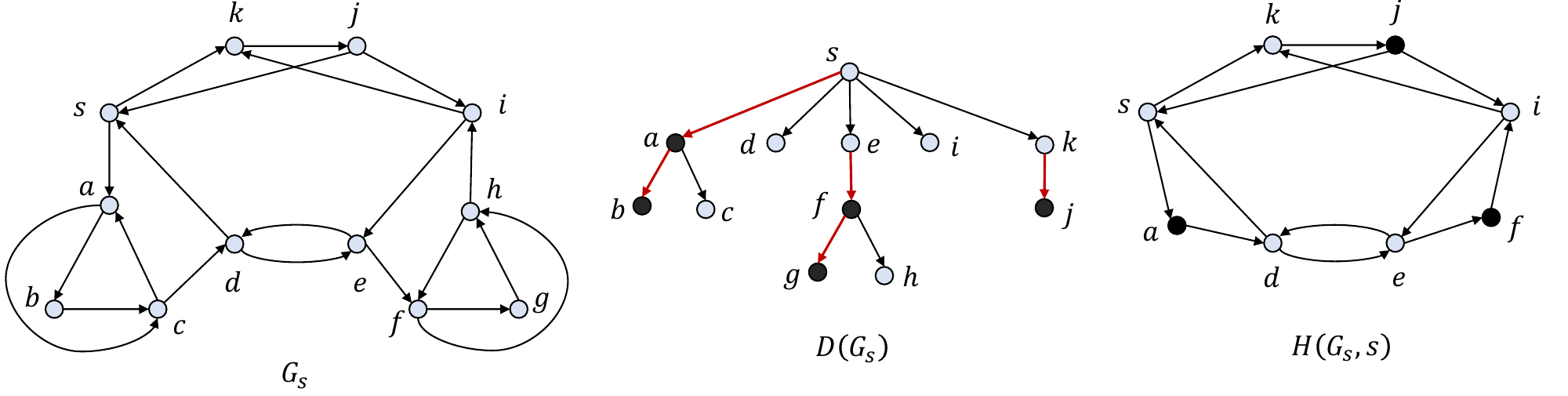}}
\vspace{-0.3cm}
\caption{A flow graph $G_s$ with start vertex $s$, its dominator tree $D(G_s)$, and the auxiliary graph $H(G_s,s)$. The bridges of $G_s$ are colored red in $D(G_s)$. Marked vertices and auxiliary vertices are colored black in $D(G_s)$ and $H(G_s,s)$, respectively.}
\label{figure:domtree}
\end{center}
\vspace{-.5cm}
\end{figure*}

Let $G_s$ be a flow graph with start vertex $s$.
A vertex $u$ is a \emph{dominator} of a vertex $v$ ($u$ \emph{dominates} $v$) if every path from $s$ to $v$ in $G_s$ contains $u$; $u$ is a \emph{proper dominator} of $v$ if $u$ dominates $v$ and $u \not= v$.
The dominator relation is reflexive and transitive. Its transitive reduction is a rooted tree, the \emph{dominator tree} $D(G_s)$: $u$ dominates $v$ if and only if $u$ is an ancestor of $v$ in $D(G_s)$.
See Figure~\ref{figure:domtree}.
For every vertex $x\neq s$ of $G_s$, $d(x)$ is the immediate dominator of $x$ in $G_s$ (i.e., the parent of $x$ in $D(G_s)$). For every vertex $r$ of $G_s$, we let $D(r)$ denote the subtree of $D(G_s)$ rooted at $r$.
Lengauer and Tarjan~\cite{domin:lt} presented an algorithm for computing dominators in  $O(m \alpha(m,n))$ time for a flow graph with $n$ vertices and $m$ edges, where $\alpha$ is a functional inverse of Ackermann's function~\cite{dsu:tarjan}.
Subsequently, several linear-time algorithms
were discovered~\cite{domin:ahlt,dominators:bgkrtw,dominators:Fraczak2013,Gabow:Poset:TALG}.

An edge $(u,v)$ is a \emph{bridge} of a flow graph $G_s$ if all paths from $s$ to $v$ include $(u,v)$.\footnote{Throughout the paper, to avoid confusion we use consistently the term  \emph{bridge} to refer to a bridge of a flow graph and the term \emph{strong bridge} to refer to a strong bridge in the original graph.}
The following properties were proved in~\cite{Italiano2012}.
\begin{property}
	\label{property:strong-bridge} \emph{(\cite{Italiano2012})}
	Let $s$ be an arbitrary start vertex of $G$. An edge $e=(u,v)$ is a strong bridge of $G$ if and only if it is a bridge of $G_s$, in which case $u=d(v)$, or a bridge of $G_s^R$, in which case $v=d^R(u)$, or both.
\end{property}

Let $G_s$ be a strongly connected digraph. For every bridge $(x,y)$ of $G_s$, we say that $y$ is a \emph{marked} vertex. (Notice that $s$ cannot be marked.)
Property~\ref{property:strong-bridge} implies that the bridges of $G_s$ induce a decomposition of $D(G_s)$ into rooted subtrees. More precisely, for every bridge $(x,y)$ of $G_s$, we remove the edge $(x,y)$ from $D(G_s)$. (By Property~\ref{property:strong-bridge}, this is indeed an edge of $D(G_s)$.) Thus we have partitioned $D(G_s)$ into subtrees. Every tree $T$ in this decomposition inherits the parent relation from $D(G_s)$, and thus it is rooted at a vertex $r$. We denote $T$ as $T(r)$ to emphasize the fact that the root of $T$ is $r$. Observe that the root $r$ of a tree $T(r)$ is either a marked vertex or $s$. Conversely, for every vertex $r$ that is either marked or $s$, there is a tree $T(r)$.

In our constructions, we will use the next lemma, which follows from well-known properties of the dominator tree. (See, e.g., ~\cite{StrongConnectivityFailures:SICOMP,2ECC:GILP:TALG}.)
\ignore{
\begin{lemma}
\label{lemma:internal_paths}
Let $G_s$ be a strongly connected digraph, and let $r$ be a marked vertex of $G_s$ or $s$.
Then, for any vertex $x$ in $D(r)$, there is a path from $r$ to $x$ in $G_s$ that contains only vertices in $D(r)$. Furthermore, if $r \not= s$, then for any vertex $x$ in $D(s)\setminus D(r)$, there is a path from $x$ to $d(r)$ in $G_s$ that contains only vertices in $D(s)\setminus D(r)$.
\end{lemma}
}


\begin{lemma}
\label{lemma:internal_paths}
Let $G_s$ be a strongly connected digraph, and let $r$ be a marked vertex of $G_s$. Then we have the following.
\begin{enumerate}
\item{For every vertex $x$ in $D(r)$, there is a path from $r$ to $x$ in $G_s$ that uses only vertices from $D(r)$.}
\item{For every vertex $x$ in $D(s)\setminus D(r)$, there is a path from $s$ to $x$ in $G_s$ that uses only vertices from $D(s)\setminus D(r)$.}
\item{For every vertex $x$ in $D(s)\setminus D(r)$, there is a path from $x$ to $d(r)$ in $G_s$ that uses only vertices from $D(s)\setminus D(r)$.}
\item{If $(x,y)$ is an edge of $G_s$ such that $x\in D(s)\setminus D(r)$ and $y\in D(r)$, then $(x,y)=(d(r),r)$.}
\end{enumerate}
\end{lemma}

\subsection{Construction of auxiliary graphs}
\label{sec:auxiliary-construction}

Now let $G_s$ be a strongly connected digraph, and let $r$ be either a marked vertex of $G_s$ or $s$. We define the \emph{auxiliary} graph $H(G_s,r)$ as follows. In $G_s$ we shrink every $D(z)$, where $z$ is a marked vertex such that $d(z)\in T(r)$ into $z$. Also, if $r\neq s$, we shrink $D(s)\setminus{D(r)}$ into $d(r)$. During those shrinkings, we maintain all edges, except for self-loops.
Also, in \cite{2ECC:GILP:TALG} multiple edges are converted into single edges. Here, multiple edges are converted into double edges, in order to avoid introducing new strong bridges in the auxiliary graphs.
The resulting graph is $H(G_s,r)$. We consider $H(G_s,r)$ as a flow graph with start vertex $r$. Notice that it consists of the subgraph of $G_s$ induced by $T(r)$, plus some extra vertices and edges. To be specific, the vertex set of $H(G_s,r)$ consists of the vertices of $T(r)$, plus all marked vertices $z$ of $G_s$ such that $d(z)\in T(r)$, plus $d(r)$ if $r\neq s$. The vertices of $T(r)$ are called \emph{ordinary} in $H(G_s,r)$. The vertices of $H(G_s,r)\setminus{T(r)}$ are called \emph{auxiliary} in $H(G_s,r)$. In particular, if $r\neq s$, $d(r)$ is called the \emph{critical} vertex of $H(G_s,r)$, and $(d(r),r)$ is called the \emph{critical} edge of $H(G_s,r)$. (Thus, $H(G_s,s)$ is the only auxiliary graph of $G_s$ that has no critical vertex and no critical edge.)

\ignore{
The above construction guarantees that each path in $G_s$ whose endpoints lie in some auxiliary graph $H(G_s,r)$ has a corresponding path in $H(G_s,r)$ with the same endpoints and vice versa.  In particular, this implies that each $H(G_s,r)$ is strongly connected. Moreover, we have the following results:
}

Now we will give a more precise description of the edges of $H(G_s,r)$. First, let $z_1,\dots,z_k$ be the collection of all marked vertices of $G_s$ such that $d(z_i)\in T(r)$, for $i\in\{1,\dots,k\}$. Then we can partition the vertex set of $G_s$ into $T(r)$, $D(z_1),\dots,D(z_k)$, and $D(s)\setminus{D(r)}$, if $r\neq s$. Now, due to Lemma~\ref{lemma:internal_paths}, an edge $(x,y)$ of $G_s$ satisfies one of:
\begin{enumerate}[label={(\roman*)}]
\item{$x\in T(r)$ and $y\in T(r)$}
\item{$x\in T(r)$ and $y\in D(s)\setminus{D(r)}$ (assuming that $r\neq s$)}
\item{$x\in D(z)$ and $y\in D(z)$, for some $z\in\{z_1,\dots,z_k\}$}
\item{$x\in D(z)$ and $y\in T(r)$, for some $z\in\{z_1,\dots,z_k\}$}
\item{$x\in D(z)$ and $y\in D(s)\setminus{D(r)}$, for some $z\in\{z_1,\dots,z_k\}$ (assuming that $r\neq s$)}
\item{$(x,y)=(d(z),z)$, for some $z\in\{z_1,\dots,z_k\}$}
\item{$x\in D(s)\setminus{D(r)}$ and $y\in D(s)\setminus{D(r)}$ (assuming that $r\neq s$)}
\item{$(x,y)=(d(r),r)$ (assuming that $r\neq s$)}
\end{enumerate}

Each of those cases provides the \textit{corresponding} edge of $H(G_s,r)$ as follows. $(i)\mapsto (x,y)$, $(ii)\mapsto (x,d(r))$, $(iii)\mapsto\mathit{none}$. $(iv)\mapsto (z,y)$. $(v)\mapsto (z,d(r))$. $(vi)\mapsto (d(z),z)$. $(vii)\mapsto\mathit{none}$. $(viii)\mapsto (d(r),r)$. For every edge $\tilde{e}$ of $H(G_s,r)$ we say that an edge $e$ of $G_s$ \textit{corresponds} to $\tilde{e}$ if $\tilde{e}$ is the edge corresponding to $e$. (Notice that the correspondences $(i)$, $(vi)$ and $(viii)$ between edges of $G_s$ and edges of $H(G_s,r)$ are bijective.) Using this correspondence, for every path $P$ in $G_s$ whose endpoints are vertices in $H(G_s,r)$, there is a \textit{corresponding} path $P_H$ in $H(G_s,r)$ that has the same endpoints. The following Lemma summarizes the properties of the corresponding path and establishes a kind of inverse to it.

\begin{lemma}
\label{lemma:cor_paths}
Let $G_s$ be a strongly connected digraph, and let $r$ be either a marked vertex of $G_s$ or $s$. Let $P$ be a path in $G_s$ whose endpoints lie in $H(G_s,r)$. Then there is a corresponding path $P_H$ in $H(G_s,r)$ with the same endpoints. In particular, every occurrence of an edge $(x,y)\in P$, such that either $(1)$ $x$ and $y$ are ordinary vertices in $H(G_s,r)$, or $(2)$ or $(x,y)=(d(z),z)$, where $z$ is a marked vertex of $G_s$ with $d(z)\in T(r)$, or $(3)$ $(x,y)=(d(r),r)$ (if $r\neq s$), is maintained in $P_H$. Conversely, for every path $\tilde{P}$ in $H(G_s,r)$, there is a path $P$ in $G_s$ such that $P_H=\tilde{P}$.
\end{lemma}
\begin{proof}
To construct $P_H$ we simply apply the above correspondence, which maps edges of $G_s$ to edges of $H(G_s,r)$, to every edge of $P$. (Notice that some edges of $P$ may be nullified.) It is a simple matter to verify that $P_H$ is indeed a path, and it also satisfies the claimed property. Now, given a path $\tilde{P}$, we will construct a path $P$ in $G_s$ such that $P_H=\tilde{P}$. This is done essentially by maintaining some of the edges of $\tilde{P}$ and by expanding some of its edges to paths of $G_s$. In particular, every edge $(x,y)$ of $\tilde{P}$ satisfying $(1)$, $(2)$, or $(3)$, is maintained in $P$. Now, suppose that $\tilde{P}$ contains $(d(z),z)$ such that $z$ is a marked vertex of $G_s$ with $d(z)\in T(r)$. Then, the successor of $(d(z),z)$ in $\tilde{P}$, if it exists, is either an edge of the form $(z,y)$, where $y\in T(r)$, or $(z,d(r))$ (assuming that $r\neq s$). In the first case, let $(x,y)$ be an edge of $G_s$ that corresponds to $(z,y)$ in $H(G_s,r)$. Then we have that $x$ is a descendant of $z$ in $D(G_s)$, and so, by Lemma~\ref{lemma:internal_paths}, there is a path in $G_s$ from $z$ to $x$ that uses only vertices from $D(z)$. Thus we replace $(z,y)$ in $\tilde{P}$ with precisely this path. In the second case, let $(x,y)$ be an edge of $G_s$ that corresponds to $(z,d(r))$ in $H(G_s,r)$. Then we have $x\in D(z)$ and $y\in D(s)\setminus{D(r)}$. Now, by Lemma~\ref{lemma:internal_paths}, there is a path $P_1$ from $z$ to $x$ that uses only vertices from $D(z)$. Furthermore, by the same Lemma, there is a path $P_2$ from $y$ to $d(r)$ that uses only vertices from $D(s)\setminus{D(r)}$. Thus we replace $(z,d(r))$ in $\tilde{P}$ with $P_1+(x,y)+P_2$. Now suppose that $r\neq s$ and $\tilde{P}$ contains $(d(r),r)$. Then, the immediate predecessor of $(d(r),r)$ in $\tilde{P}$, if it exists, is either an edge of the form $(x,d(r))$, where $x\in T(r)$, or $(z,d(r))$, where $z$ in a marked vertex of $G_s$ with $d(z)\in T(r)$. Now we can treat those cases as before, using Lemma~\ref{lemma:internal_paths} in order to expand the corresponding edges of $(x,d(r))$ or $(z,d(r))$ in $G_s$ into paths of $G_s$.
\end{proof}

An immediate consequence of Lemma~\ref{lemma:cor_paths} is the following.

\begin{corollary}
\label{corollary:sc_of_H}
Let $G_s$ be a strongly connected digraph, and let $r$ be either a marked vertex of $G_s$ or $s$. Then $H(G_s,r)$ is strongly connected.
\end{corollary}

Furthermore, \cite{2ECC:GILP:TALG} provided the following result.

\begin{theorem}
\label{theorem:aux_linear_time_main} \emph{(\cite{2ECC:GILP:TALG})}
Let $G_s$ be a strongly connected digraph, and let $r_1,\dots,r_k$ be the marked vertices of $G_s$.
\begin{enumerate}[label={(\roman*)}]
\item{For any two vertices $x$ and $y$ of $G_s$, \strongEC{x}{y}{G_s} if and only if there is a vertex $r$ (a marked vertex of $G_s$ or $s$), such that $x$ and $y$ are both ordinary vertices of $H(G_s,r)$ and \strongEC{x}{y}{H(G_s,r)}.}
\item{The collection $H(G_s,s)$, $H(G_s,r_1),\dots,H(G_s,r_k)$ of all the auxiliary graphs of $G_s$ can be computed in linear time.}
\end{enumerate}
\end{theorem}

We provide the analogous result for $2$-edge twinless strong connectivity.

\begin{proposition}
\label{proposition:H-2et_main}
Let $x,y$ be two vertices of a strongly connected digraph $G_s$. Then \tstrongEC{x}{y}{G_s} if and only if there is a vertex $r$ (a marked vertex of $G_s$ or $s$), such that $x$ and $y$ are both ordinary vertices of $H(G_s,r)$ and \tstrongEC{x}{y}{H(G_s,r)}.
\end{proposition}
\begin{proof}
($\Rightarrow$) Suppose that $x\leftrightarrow_{2et}y$ in $G_s$. Then $x$ and $y$ are not separated by any bridge of $G_s$, and so they both belong to the same graph $H(G_s,r)$, for some $r$ (a marked vertex of $G_s$ or $s$), as ordinary vertices. Now let $e=(u,v)$ be an edge of $H(G_s,r)$. We will prove that $x\leftrightarrow_{t}y$ in $H(G_s,r)\setminus{e}$. We distinguish five different cases for $e$.
\begin{enumerate}[label={(\arabic*)}]
\item{$u$ and $v$ are both ordinary vertices of $H(G_s,r)$.}
\item{$v$ is an auxiliary vertex of $H(G_s,r)$, but not $d(r)$ (if $r\neq s$).}
\item{$u$ is an auxiliary vertex of $H(G_s,r)$, but not $d(r)$ (if $r\neq s$).}
\item{$(u,v)=(d(r),r)$ (if $r\neq s$).}
\item{$v=d(r)$ (if $r\neq s$).}
\end{enumerate}
We will now show how to handle cases $(1)$ to $(5)$.

$(1)$ Let $u$ and $v$ be ordinary vertices of $H(G_s,r)$. Then $e=(u,v)$ is an edge of $G_s$. Since $x\leftrightarrow_{2et}y$ in $G_s$, there exist two paths $P$ and $Q$ in $G_s\setminus{e}$, such that $P$ starts from $x$ and ends in $y$, $Q$ starts from $y$ and ends in $x$, and for every edge $(z,w)\in P$ we have that $(w,z)\notin Q$. Now consider the induced paths $P_H$ and $Q_H$ of $P$ and $Q$, respectively, in $H(G_s,r)$. These have the property that $P_H$ starts from $x$ and ends in $y$, and $Q_H$ starts from $y$ and ends in $x$. Now we will show that there are subpaths $P'_H$ and $Q'_H$ of $P_H$ and $Q_H$, respectively, that have the same endpoints as $P_H$ and $Q_H$, respectively, and the property that for every $(z,w)\in P'_H$ we have that $(w,z)\notin Q'_H$. We will construct $P'_H$ and $Q'_H$ by eliminating some segments of $P_H$ and $Q_H$, respectively, that form loops. So let $(z,w)$ be an edge of $P_H$ with the property that $(w,z)\in Q_H$. Then we have that $z$ and $w$ cannot both be ordinary vertices of $H(G_s,r)$ (for otherwise we would have that $(z,w)\in P$ and $(w,z)\in Q$). Thus, let us take first the case that $(z,w)=(d(h),h)$, where $h$ is an auxiliary vertex of $H(G_s,r)$. Then we have that $(h,d(h))\in Q_H$. But, in order for $Q_H$ to use the edge $(h,d(h))$, it must first pass from the bridge $(d(h),h)$. Thus we can simply discard the segment $(d(h),h),(h,d(h))$ from $Q_H$, as this contributes nothing for $Q_H$ in order to reach $y$ from $x$. Now let us take the case that $z$ is an auxiliary vertex of $H(G_s,r)$, but $(z,w)\neq(d(r),r)$ (if $r\neq s$). If $w=d(z)$ we work as before. But this supposition exhausts this case since $(w,z)$ is an edge of $H(G_s,r)$, and so by Lemma~\ref{lemma:internal_paths} we have that $w$ must necessarily be $d(z)$. Now let us take the case that $(z,w)=(d(r),r)$ (assuming that $r\neq s$). Then we have that $(r,d(r))\in Q_H$. But then, in order for $Q_H$ to end at an ordinary vertex, it must immediately afterwards use the bridge $(d(r),r)$. Thus we can eliminate the segment $(r,d(r)),(d(r),r)$ from $Q_H$. Finally, the case $(z,w)=(r,d(r))$ is treated similarly as $(z,w)=(d(r),r)$. This shows that $x\leftrightarrow_{t}y$ in $H(G_s,r)\setminus{e}$.

In cases $(2)$ and $(4)$, the challenge in proving $x\leftrightarrow_{t}y$ in $H(G_s,r)\setminus{e}$ is the same as that in $(1)$, since $e$ happens to be an edge of $G_s$. Thus, cases $(1)$, $(2)$ and $(4)$ have essentially the same proof. So we are left to consider cases $(3)$ and $(5)$.

$(3)$ Let's assume that $u$ is an auxiliary vertex of $H(G_s,r)$ with $u\neq d(r)$ (if $r\neq s$). Then, since $x\leftrightarrow_{2et}y$ in $G_s$, there exist two paths $P$ and $Q$ in $G_s\setminus{(d(u),u)}$, such that $P$ starts from $x$ and ends in $y$, $Q$ starts from $y$ and ends in $x$, and for every edge $(z,w)\in P$ we have that $(w,z)\notin Q$. Now consider the induced paths $P_H$ and $Q_H$ of $P$ and $Q$, respectively, in $H(G_s,r)$. By Lemma~\ref{lemma:cor_paths}, these paths do not pass from the bridge $(d(u),u)$. Therefore, they also cannot use the edge $(u,v)$. Now we can reason as in the proof for case $(1)$, to show that there are subpaths $P'_H$ and $Q'_H$ of $P_H$ and $G_H$, respectively, with the same endpoints, respectively, and the property that for every $(z,w)\in P'_H$ we have that $(w,z)\notin Q'_H$. This shows that $x\leftrightarrow_{t}y$ in $H(G_s,r)\setminus{(x,y)}$.

$(5)$ Let's assume that $r\neq s$ and $v=d(r)$. Then, since $x\leftrightarrow_{2et}y$ in $G_s$, there exist two paths $P$ and $Q$ in $G_s\setminus{(d(r),r)}$, such that $P$ starts from $x$ and ends in $y$, $Q$ starts from $y$ and ends in $x$, and for every edge $(z,w)\in P$ we have that $(w,z)\notin Q$. Now consider the induced paths $P_H$ and $Q_H$ of $P$ and $Q$, respectively, in $H(G_s,r)$. By Lemma~\ref{lemma:cor_paths}, these paths do not pass from the bridge $(d(r),r)$. Therefore, they also cannot use the edge $(v,u)$. Now we can reason as in the proof for case $(1)$, to show that there are subpaths $P'_H$ and $Q'_H$ of $P_H$ and $G_H$, respectively, with the same endpoints, respectively, and the property that for every $(z,w)\in P'_H$ we have that $(w,z)\notin Q'_H$. This shows that $x\leftrightarrow_{t}y$ in $H(G_s,r)\setminus{(x,y)}$.

Now, since cases $(1)$ to $(5)$ exhaust all possibilities for $e$, we infer that $x\leftrightarrow_{t}y$ in $H(G_s,r)\setminus{e}$. Due to the generality of $e$, we conclude that $x\leftrightarrow_{2et}y$ in $H(G_s,r)$.
\\
($\Leftarrow$) Suppose that $x\not\leftrightarrow_{2et}y$ in $G_s$. If there is no $r$ such that $x$ and $y$ are both in $H(G_s,r)$ as ordinary vertices, then we are done. So let us assume that there is an $r$ such that $x$ and $y$ are both in $H(G_s,r)$ as ordinary vertices. 
By \cite{2ECC:GILP:TALG}, we have that if $x\not\leftrightarrow_{2e}y$ in $G_s$, then also $x\not\leftrightarrow_{2e}y$ in $H(G_s,r)$, and therefore $x\not\leftrightarrow_{2et}y$ in $H(G_s,r)$.
Now let's assume that $x\leftrightarrow_{2e}y$ in $G_s$. Then $x\not\leftrightarrow_{2et}y$ in $G_s$ implies that there is an edge $e$ in $G_s$ such that, for every two paths $P$ and $Q$ in $G_s\setminus{e}$ such that $P$ starts from $x$ and ends in $y$, and $Q$ starts from $y$ and ends in $x$, we have that there exists an edge $(u,v)\in P$ such that $(v,u)\in Q$. We will strengthen this by showing that there exists an edge $(u,v)\in P$ such that $(v,u)\in Q$ and $u,v$ are ordinary vertices of $H(G_s,r)$. We will achieve this by showing that, if either $u$ or $v$ is not an ordinary vertex of $H(G_s,r)$, then we can replace either the part of $P$ that contains $(u,v)$ with one that doesn't contain it, or the part of $Q$ that contains $(v,u)$ with one that doesn't contain it. (So that, finally, since the resulting paths must have the property that they share a pair of antiparallel edges, this pair cannot but join ordinary vertices of $H(G_s,r)$.) We will distinguish three cases, depending on the location of $u$ in $D(G_s)$. $(1)$ Let's assume that $u$ is a descendant in $D(G_s)$ of an auxiliary vertex $z$ of $H(G_s,r)$, with $z\neq d(r)$ (if $r\neq s$). Let's assume first that $(u,v)=(z,d(z))$, where $z$ is an auxiliary vertex in $H(G_s,r)$. Since $P$ starts from $x$ (an ordinary vertex in $H(G_s,r)$), we have that $P$ must cross the bridge $(d(z),z)$ in order to reach $z$. But, since $P$ eventually returns again to $d(z)$ through $(u,v)=(z,d(z))$, it is unnecessary to traverse this part from $d(z)$ to $d(z)$ (that uses $(u,v)=(z,d(z))$), and so we can skip it. Now let's assume that $v\neq d(z)$. Then, by Lemma~\ref{lemma:internal_paths}, we have that $v$ must also be a descendant in $D(G_s)$ of $z$. Now, in order for $P$ to use the edge $(u,v)$, it must first pass from the bridge $(d(z),z)$ (since it starts from $x$, an ordinary vertex of $H(G_s,r)$). Furthermore, in order for $Q$ to use the edge $(v,u)$, it must also pass from the bridge $(d(z),z)$. But this means that $v$ is reachable from $z$ without using $(u,v)$, and so we can replace the part of $P$ from $z$ to $(u,v)$, with that of $Q$ from $z$ to $v$. $(2)$ Now let's assume that $u$ is not a descendant of $r$ in $D(G_s)$ (of course, this presupposes that $r\neq s$). Suppose, first, that $(u,v)=(d(r),r)$. Then, since $Q$ uses the edge $(v,u)=(r,d(r))$, but then has to return to $H(G_s,r)$, it has no other option but to pass eventually from the bridge $(d(r),r)$, after visiting vertices which are not descendants of $r$ in $D(G_s)$. Thus we can simply skip this part of $Q$ that starts from $r$ and returns again to $r$ (and uses $(v,u)=(r,d(r))$). Now let's assume that $(u,v)\neq (d(r),r)$. Then, by Lemma~\ref{lemma:internal_paths} we have that $v$ is also not a descendant of $r$ in $D(G_s)$. Since $P$ reaches $v$, in order to end at $y$ (inside $H(G_s,r)$) it has to pass from the bridge $(d(r),r)$. The same is true for $Q$, since it ends at $x$ (inside $H(G_s,r)$). But then $Q$ can avoid using $(v,u)$, and we replace the part of $Q$ that starts from $v$ and ends in $d(r)$ with that of $P$ that starts from $v$ and ends in $d(r)$. $(3)$ Finally, we are left to assume that $u$ is an ordinary vertex of $H(G_s,r)$. If $v$ is also an ordinary vertex of $H(G_s,r)$, we are done. Otherwise, we can revert back to the previous two cases, interchanging the roles of $u$ and $v$.

Thus we have shown that there is an edge $e$ in $G_s$ such that, for every two paths $P$ and $Q$ in $G_s\setminus{e}$ such that $P$ starts from $x$ and ends in $y$, and $Q$ starts from $y$ and ends in $x$, we have that there exists an edge $(u,v)\in P$ such that $(v,u)\in Q$, and $u,v$ are both ordinary vertices of $H(G_s,r)$ $(*)$.
Now we distinguish three cases, depending on the location of the endpoints of $e$ relative to $D(G_s)$. $(1)$ Let us assume that both endpoints of $e$ are ordinary vertices in $H(G_s,r)$. Now let's assume for the sake of contradiction that there is a path $P'$ in $H(G_s,r)\setminus{e}$ from $x$ to $y$, and a path $Q'$ in $H(G_s,r)\setminus{e}$ from $y$ to $x$, such that for every edge $(z,w)\in P'$ we have that $(w,z)\notin Q'$ $(**)$. By Lemma~\ref{lemma:cor_paths}, there is a path $P$ in $G_s$ from $x$ to $y$ such that $P_H=P'$, and a path $Q$ in $G_s$ from $y$ to $x$ such that $Q_H=Q'$. Then, by Lemma~\ref{lemma:cor_paths}, both $P$ and $Q$ avoid $e$. But also by the same Lemma, we have that an edge $(u,v)\in H(G_s,r)$ with both endpoints ordinary in $H(G_s,r)$ is used by $P$ (resp. by $Q$) if and only if it is used by $P'$ (resp. $Q'$). But then the property $(**)$ of $P'$ and $Q'$ violates the property $(*)$ of $P$ and $Q$. This shows that $x\not\leftrightarrow_{t}y$ in $H(G_s,r)\setminus{e}$ in this case. $(2)$ Now let's assume that both endpoints of $e$ are descendants of $r$ in $D(G_s)$, but also that one endpoint of $e$ is a descendant in $D(G_s)$ of an auxiliary vertex $z$ in $H(G_s,r)$, where $z\neq d(r)$ (if $r\neq s$). Then, by removing $(d(z),z)$ from $G_s$, we have that no ordinary vertex from $H(G_s,r)$ can reach the edge $e$. Therefore, $(*)$ implies that for every path $P$ from $x$ to $y$ in $G_s\setminus{(d(z),z)}$ and every path $Q$ from $y$ to $x$ in $G_s\setminus{(d(z),z)}$, we have that there exists an edge $(u,v)\in P$ such that $(v,u)\in Q$, and $u,v$ are both ordinary vertices of $H(G_s,r)$. Now let's assume for the sake of contradiction that there is a path $P'$ in $H(G_s,r)\setminus\{(d(z),z)\}$ from $x$ to $y$, and a path $Q'$ in $H(G_s,r)\setminus\{(d(z),z)\}$ from $y$ to $x$, such that for every edge $(z,w)\in P'$ we have that $(w,z)\notin Q'$ $(**)$ Then, by Lemma~\ref{lemma:cor_paths}, both $P$ and $Q$ avoid $(d(z),z)$. But also by the same Lemma, we have that an edge $(u,v)\in H(G_s,r)$ with both endpoints ordinary in $H(G_s,r)$ is used by $P$ (resp. by $Q$) if and only if it is used by $P'$ (resp. $Q'$). But then the property $(**)$ of $P'$ and $Q'$ violates the property $(*)$ of $P$ and $Q$. This shows that $x\not\leftrightarrow_{t}y$ in $H(G_s,r)\setminus{e}$ in this case. $(3)$ There remains to consider the case that one endpoint of $e$ is not a descendant of $r$ in $D(G_s)$ (assuming, of course, that $r\neq s$). Then, by removing $(d(r),r)$ from $G_s$, we have that no path that starts from an ordinary vertex of $H(G_s,r)$ and uses the edge $e$ can return to $H(G_s,r)$. Therefore, $(*)$ implies that for every path $P$ from $x$ to $y$ in $G_s\setminus{(d(r),r)}$ and every path $Q$ from $y$ to $x$ in $G_s\setminus{(d(r),r)}$, we have that there exists an edge $(u,v)\in P$ such that $(v,u)\in Q$, and $u,v$ are both ordinary vertices of $H(G_s,r)$. Now let's assume for the sake of contradiction that there is a path $P'$ in $H(G_s,r)\setminus\{(d(r),r)\}$ from $x$ to $y$, and a path $Q'$ in $H(G_s,r)\setminus\{(d(r),r)\}$ from $y$ to $x$, such that for every edge $(z,w)\in P'$ we have that $(w,z)\notin Q'$ $(**)$ Then, by Lemma~\ref{lemma:cor_paths}, both $P$ and $Q$ avoid $(d(r),r)$. But also by the same Lemma, we have that an edge $(u,v)\in H(G_s,r)$ with both endpoints ordinary in $H(G_s,r)$ is used by $P$ (resp. by $Q$) if and only if it is used by $P'$ (resp. $Q'$). But then the property $(**)$ of $P'$ and $Q'$ violates the property $(*)$ of $P$ and $Q$. This shows that $x\not\leftrightarrow_{t}y$ in $H(G_s,r)\setminus{e}$ in this case. Any of those cases implies that $x\not\leftrightarrow_{2e}y$ in $H(G_s,r)$, and therefore $x\not\leftrightarrow_{2et}y$ in $H(G_s,r)$.
\end{proof}

\begin{figure*}[t!]
\begin{center}
\centerline{\includegraphics[trim={0 0 0 8cm}, clip=true, width=\textwidth]{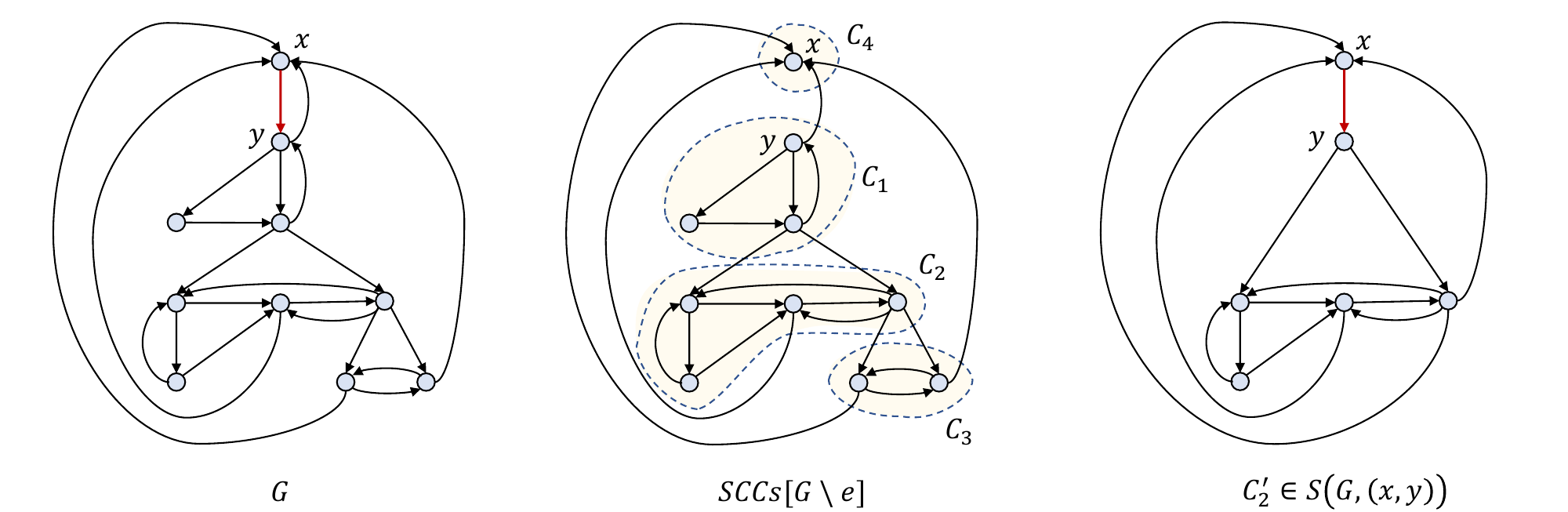}}
\vspace{-0.3cm}
\caption{A strongly connected digraph $G$ with a strong bridge $e=(x,y)$ shown red. The deletion of $e$ splits $G$ into four strongly connected components $C_1$, $C_2$, $C_3$ and $C_4$ (numbered in topological order). $C'_2$ is the digraph in $S(G,e)$ that corresponds to $C_2$ after attaching $x$ and $y$ to it, and all the edges due to the $S$-operation.}
\label{figure:s-operation}
\end{center}
\vspace{-.5cm}
\end{figure*}

Now let $G$ be a strongly connected digraph and let $(x,y)$ be a strong bridge of $G$. We will define the \emph{$S$-operation} on $G$ and $(x,y)$, which produces a set of digraphs as follows. Let $C_1,\dots,C_k$ be the strongly connected components of $G\setminus{(x,y)}$. Now let $C\in\{C_1,\dots,C_k\}$. We will construct a graph $C'$ as follows. First, notice that either $x\notin C$ and $y\in C$, or $y\notin C$ and $x\in C$, or $\{x,y\}\cap C=\emptyset$. Then we set $V(C')=V(C)\cup\{x\}$, or $V(C')=V(C)\cup\{y\}$, or $V(C')=V(C)\cup\{x,y\}$, respectively.
Every edge of $G$ with both endpoints in $C$ is included in $C'$. Furthermore, for every edge $(u,v)$ of $G$ such that $u\in C$ and $v\notin C$, we add the edge $(u,x)$ to $C'$. Also, for every edge $(u,v)$ of $G$ such that $u\notin C$ and $v\in C$, we add the edge $(y,v)$ to $C'$. Finally, we also add the edge $(x,y)$ to $C'$. Now we define $S(G,(x,y)):=\{C'_1,\dots,C'_k\}$.
See Figure~\ref{figure:s-operation}.

Intuitively, we can describe the idea of the $S$-operation on $G$ and $(x,y)$ as follows. Let $\mathcal{D}$ be the directed acyclic graph (DAG) of the strongly connected components of $G\setminus{(x,y)}$. Then we have that $\mathcal{D}$ has a single source that contains $y$, and a single sink that contains $x$. Now, for every node $C$ of $\mathcal{D}$, we merge all nodes of $\mathcal{D}\setminus\{C\}$ that are reachable from $C$ into $x$, and all nodes of $\mathcal{D}\setminus\{C\}$ that reach $C$ into $y$. (We ignore all the other nodes of $\mathcal{D}$.) We also reconnect this graph by adding the edge $(x,y)$. We perform this operation for every node of $\mathcal{D}$, and thus we get the collection of graphs $S(G,(x,y))$.

The purpose of this operation is to maintain the connectivity information within every strongly connected component of $G\setminus{(x,y)}$. In particular, we have the following correspondence between paths of $G$ and paths in the graphs of $S(G,(x,y))$. Let $u,v$ be two vertices of the same strongly connected component $C$ of $G\setminus{(x,y)}$ and let $P$ be a path in $G$ from $u$ to $v$. Then we construct the \textit{compressed} path $\tilde{P}$ of $P$ as follows. First, every edge of $P$ with both endpoints in $C$ is maintained in $\tilde{P}$. If $P$ lies entirely within $C$, we are done. Otherwise, let $(z_1,w_1),\dots,(z_t,w_t)$ be a maximal segment of $P$ consisting of edges that have at least one of their endpoints outside of $C$. (Thus we have $z_1\in C$ and $w_t\in C$.) Notice that this segment is unique because if a path leaves $C$ it has to use the edge $(x,y)$ in order to return to $C$ (and $P$ can use $(x,y)$ only once). 
Now, if $(z_1,w_1)=(x,y)$, we replace this segment in $\tilde{P}$ with $(x,y),(y,w_t)$. Similarly, if $(z_t,w_t)=(x,y)$, we replace this segment with $(z_1,x),(x,y)$. Otherwise, we replace this segment with $(z_1,x),(x,y),(y,w_t)$. Then, it should be clear that $\tilde{P}$ is a path in $C'$ from $u$ to $v$.  The construction of the compressed path, together with a kind of converse to it, is summarized in the following.

\begin{lemma}
\label{lemma:compressed_path}
Let $G$ be a strongly connected digraph and let $(x,y)$ be a strong bridge of $G$. Let $C$ be a strongly connected component of $G\setminus{(x,y)}$, and let $C'$ be the corresponding graph in $S(G,(x,y))$. Let also $u$ and $v$ be two vertices in $C$. Then, for every path $P$ in $G$ from $u$ to $v$, there is a (compressed) path $\tilde{P}$ in $C'$ from $u$ to $v$ with the following three properties. $(1)$ Every edge $e\in P$ that lies in $C$ is maintained in $P$. Also, every occurrence of $(x,y)$ in $P$ is maintained in $\tilde{P}$. $(2)$ If $x\notin C$ and $(z,w)$ is an edge of $P$ such that $z\in C$ and $w\notin C$, then $\tilde{P}$ contains $(z,x)$. $(3)$ If $y\notin C$ and $(z,w)$ is an edge of $P$ such that $z\notin C$ and $w\in C$, then $\tilde{P}$ contains $(y,w)$. Conversely, for every path $P'$ in $C'$ from $u$ to $v$, there is a path $P$ in $G$ from $u$ to $v$ such that $\tilde{P}=P'$.
\end{lemma}
\begin{proof}
The first part was discussed above. For the converse, we construct the path $P$ by replacing some edges of $P'$ with paths of $G$ (if needed) as follows. First, every edge of $P'$ that lies in $C$ is maintained in $P$. Furthermore, every occurrence of $(x,y)$ in $P'$ is also maintained in $P$. Now, if $x\notin C$ and there is an edge $(z,x)\in P'$, then we trace a path from $z$ to $x$ in the DAG $\mathcal{D}$ of the strongly connected components of $G\setminus{(x,y)}$ (possibly using internal paths from the intermediate components in $\mathcal{D}$), and we let it replace $(z,x)$ in $P$. Similarly, if $y\notin C$ and there is an edge $(y,z)\in P'$, then we trace a path from $y$ to $z$ in $\mathcal{D}$ (possibly using internal paths from the intermediate components in $\mathcal{D}$), and we let it replace $(y,z)$ in $P$. It should be clear that $P$ so constructed is indeed a path in $G$ from $u$ to $v$ such that $\tilde{P}=P'$.
\end{proof}

Lemma~\ref{lemma:compressed_path} implies the following.

\begin{corollary}
Let $G$ be a strongly connected digraph and let $e$ be a strong bridge of $G$. Then every graph of $S(G,e)$ is strongly connected.
\end{corollary}

The following two lemmas are useful strengthenings of Lemma~\ref{lemma:compressed_path}.

\begin{lemma}
\label{lemma:compressed_path2}
Let $G$ be a strongly connected digraph and let $(x,y)$ be a strong bridge of $G$. Let $C$ be a strongly connected component of $G\setminus (x,y)$, and let $C'$ be the corresponding digraph in $S(G,(x,y))$. Let $P'$ be a path in $C'$ with both endpoints in $C$. Then there is a path $P$ in $G$ with $\tilde{P}=P'$ such that for every pair of twin edges $(z,w)\in P$ and $(w,z)\in P$ we have that both $(z,w)$ and $(w,z)$ are in $C'$.
\end{lemma}
\begin{proof}
By Lemma~\ref{lemma:compressed_path} we have that there is a path $P$ in $G$ such that $\tilde{P}=P'$. In particular, the endpoints of $P$ lie in $C$. Now let $(z,w)$ be an edge such that $(z,w)\in P$, $(w,z)\in P$, and $(z,w)\notin C'$. (If no such edge exists, then we are done.) We will show that we can eliminate all those instances from $P$, and maintain the property that $\tilde{P}=P'$. First, let us suppose, for the sake of contradiction, that $(x,y)\in\{(z,w),(w,z)\}$. Since $(z,w)\notin C'$, we have $(z,w)\neq (x,y)$. Therefore, $(w,z)=(x,y)$ and $(z,w)=(y,x)$. Since $(z,w)=(y,x)\notin C'$, we have that neither $x$ nor $y$ is in $C$. Thus, since $P$ starts from $C$, in order to reach $y$ it must pass from $(x,y)$. Then, since it uses $(y,x)$, in order to reach $C$ it must again use $(x,y)$. But this contradicts the fact that $P$ may use an edge only once. This shows that $(x,y)\notin\{(z,w),(w,z)\}$.

Now, since $(x,y)\notin\{(z,w),(w,z)\}$, the existence of the edges $(z,w)$ and $(w,z)$ in $G$ implies that $z$ and $w$ belong to the same strongly connected component $D$ of $G\setminus (x,y)$. Notice that $D$ must be different from $C$, because otherwise, we would have $(z,w)\in C$. Thus, due to the fact that $P$ may use $(x,y)$ only once, we have that $D$ is accessed by $P$ only once (because we need to pass from $(x,y)$ in order to either reach $D$ from $C$, or reach $C$ from $D$). Now, if $P$ first uses $(z,w)$ and afterwards $(w,z)$, we have a circuit that starts with $(z,w)$ and ends with $(w,z)$ that can be eliminated from $P$. Similarly, if $P$ first uses $(w,z)$ and then $(z,w)$, we have a circuit that starts with $(w,z)$ and ends with $(z,w)$ that can be eliminated from $P$. In both cases, $\tilde{P}=P'$ is maintained.

Thus, after all those eliminations, we have that, if $(z,w)$ and $(w,z)$ are edges of $P$, then both of them must be in $C'$.
\end{proof}

\begin{lemma}
\label{lemma:compressed_path3}
Let $G$ be a strongly connected digraph and let $(x,y)$ be a strong bridge of $G$. Let $C$ be a strongly connected component of $G\setminus (x,y)$, and let $C'$ be the corresponding digraph in $S(G,(x,y))$. Let $P'$ and $Q'$ be two paths in $C'$ whose endpoints lie in $C$. Then there are paths $P$ and $Q$ in $G$ with $\tilde{P}=P'$ and $\tilde{Q}=Q'$, such that for every pair of edges $(z,w)\in P$ and $(w,z)\in Q$ we have that both $(z,w)$ and $(w,z)$ are in $C'$.
\end{lemma}
\begin{proof}
By Lemma~\ref{lemma:compressed_path2} we have that there is a path $P$ in $G$ with $\tilde{P}=P'$ such that for every pair of twin edges $(z,w)\in P$ and $(w,z)\in P$ we have that both $(z,w)$ and $(w,z)$ are in $C'$. Similarly, by Lemma~\ref{lemma:compressed_path2} we have that there is a path $Q$ in $G$ with $\tilde{Q}=Q'$ such that for every pair of twin edges $(z,w)\in Q$ and $(w,z)\in Q$ we have that both $(z,w)$ and $(w,z)$ are in $C'$.

Now let us suppose that there is an edge $(z,w)\in P$, such that $(w,z)\in Q$, and at least one of $(z,w)$ and $(w,z)$ is not in $C'$. (If no such edge exists, then we are done.) Let us suppose, for the sake of contradiction, that $(x,y)\in\{(z,w),(w,z)\}$. We may assume w.l.o.g. that $(z,w)=(x,y)$. Now, since at least one of $(z,w)$ and $(w,z)$ is not in $C'$, we have that $(w,z)=(y,x)\notin C'$. This implies that $y\notin C$. Thus, since $Q$ starts from $C$ and reaches $y$, it must use the edge $(x,y)$. But then, since $(x,y)\in Q$ and $(y,x)\in Q$ and $(y,x)\notin C'$, we have a contradiction to the property of $Q$ provided by Lemma~\ref{lemma:compressed_path2}. This shows that $(x,y)\notin\{(z,w),(w,z)\}$.

Since $(x,y)\notin\{(z,w),(w,z)\}$ and $(z,w),(w,z)$ are edges of the graph, we have that $z$ and $w$ belong to the same strongly connected component $D$ of $G\setminus (x,y)$, and therefore both $(z,w)$ and $(w,z)$ lie entirely within $D$. Notice that $D\neq C$, because otherwise we would have that both $(z,w)$ and $(w,z)$ are edges of $C$. Then we have that either $(1)$ $C$ reaches $D$ in $G\setminus (x,y)$, or $(2)$ $D$ reaches $C$ in $G\setminus (x,y)$, or $(3)$ neither $C$ reaches $D$ in $G\setminus (x,y)$, nor $D$ reaches $C$ in $G\setminus (x,y)$. Notice that case $(3)$ is impossible due to the existence of $P$ (and $Q)$: because in this case, since $P$ starts from $C$, in order to reach $D$ it must use $(x,y)$, and then, in order to return to $C$, it must again use $(x,y)$. Thus, it is sufficient to consider cases $(1)$ and $(2)$. (Notice that $(1)$ implies that $x\notin C$, and $(2)$ implies that $y\notin C$.)

Notice that since both $P$ and $Q$ go outside of $C$ at some point, in order to return back to $C$ they have the following structure: by the time they exit $C$, they use a path to reach $x$ (possibly the trivial path, if $x\in C$), then they use $(x,y)$, and finally they use another path to return to $C$ (possibly the trivial path, if $y\in C$). Thus, the parts of $P$ and $Q$ from $C$ to $x$ and from $y$ to $C$ can be substituted by any other path in $G$ (that has the same endpoints), and the property $\tilde{P}=P'$ and $\tilde{Q}=Q'$ is maintained. In the following, we will exploit this property.

Let $u$ and $v$ be the first and the last vertex, respectively, of the subpath of $P$ inside $D$, and let $u'$ and $v'$ be the first and the last vertex, respectively, of the subpath of $Q$ inside $D$. Suppose first that $(1)$ is true. Let $P[v,x]$ be the subpath of $P$ from the last occurrence of $v$ to the first (and only) occurrence of $x$, and let $Q[v',x]$ be the subpath of $Q$ from the last occurrence of $v'$ to the first (and only) occurrence of $x$. Since $D$ is strongly connected, there is a path $S$ from $u$ to $v$ inside $D$, and we may assume that all vertices of $S$ are distinct. Then, we replace the part of $P$ inside $D$ with $S$. (Observe that $\tilde{P}=P'$ is maintained.) Since $D$ is strongly connected, there is also at least one path $S'$ from $u'$ to a vertex $p$ in $S$, that uses each vertex only once, and intersects with $S$ only at $p$. Then we consider the path $S''$ from $u'$ to $v$ in $D$ that is formed by first using $S'$ from to $u'$ to $p$, and then the subpath of $S$ from $p$ to $v$. We let $S''$ replace the part of $Q$ inside $D$, and we substitute $Q[v',x]$ with $P[v,x]$. (Observe that $Q$ is still a path, and satisfies $\tilde{Q}=Q'$.) Now, we have that there is no pair of twin edges $(z',w')\in P$ and $(w',z')\in Q$ such that both $(z',w')$ and $(w',z')$ belong to $D$. Furthermore, since we replaced the part of $Q$ from $D$ to $x$ with the subpath of $P$ from $D$ to $x$, we have that there is no pair of twin edges $(z',w')\in P$ and $(w',z')\in Q$ such that $(z',w')$ and $(w',z')$ are on this path, due to Lemma~\ref{lemma:compressed_path2}.

Now let us suppose that $(2)$ is true. (The argument here is analogous to the previous one.) Let $P[y,u]$ be the subpath of $P$ from the first (and only) occurrence of $y$ to the first occurrence of $u$, and let $Q[y,u']$ be the subpath of $Q$ from the first (and only) occurrence of $y$ to the first occurrence of $u'$. Since $D$ is strongly connected, there is a path $S$ from $u$ to $v$ inside $D$, and we may assume that all vertices of $S$ are distinct. Then, we replace the part of $P$ inside $D$ with $S$. (Observe that $\tilde{P}=P'$ is maintained.) Since $D$ is strongly connected, there is also at least one path $S'$ in $D$ from a vertex $p\in S$ to $v'$, that uses each vertex only once and intersects with $S$ only in $p$. Then we consider the path $S''$ from $u$ to $v'$ in $D$ that is formed by first using the subpath of $S$ from to $u$ to $p$, and then the path $S'$ from $p$ to $v'$. We let $S''$ replace the part of $Q$ inside $D$, and we substitute $Q[y,u']$ with $P[y,u]$. (Observe that $Q$ is still a path, and satisfies $\tilde{Q}=Q'$.) Now, we have that there is no pair of twin edges $(z',w')\in P$ and $(w',z')\in Q$ such that both $(z',w')$ and $(w',z')$ belong to $D$. Furthermore, since we replaced the part of $Q$ from $y$ to $D$ with the subpath of $P$ from $y$ to $D$, we have that there is no pair of twin edges $(z',w')\in P$ and $(w',z')\in Q$ such that $(z',w')$ and $(w',z')$ are on this path, due to Lemma~\ref{lemma:compressed_path2}.

Thus, we have basically shown that we can eliminate all instances where there are edges $(z,w)\in P$ and $(w,z)\in Q$ such that at least one of $(z,w)$ and $(w,z)$ is not in $C'$, while still maintaining $\tilde{P}=P'$ and $\tilde{Q}=Q'$, by substituting parts of $Q$ with parts of $P$. This shows that for every pair of edges $(z,w)\in P$ and $(w,z)\in Q$ we have that both $(z,w)$ and $(w,z)$ are in $C'$.
\end{proof}

Now we can show that the $S$-operation maintains the relation of $2$-edge twinless strong connectivity.


\begin{proposition}
\label{proposition:S-2et_main}
Let $G$ be a strongly connected digraph and let $(x,y)$ be a strong bridge of $G$. Then, for any two vertices $u,v \in G$, we have
\tstrongEC{u}{v}{G} if and only if $u$ and $v$ belong to the same graph $C$ of $S(G,(x,y))$ and \tstrongEC{u}{v}{C}.
\end{proposition}
\begin{proof}
($\Rightarrow$) Let $u\leftrightarrow_{2et}v$ in $G$. This implies that $u$ and $v$ remain strongly connected after the removal of any edge of $G$. In particular, $u$ and $v$ must belong to the same strongly connected component $C$ of $G\setminus{(x,y)}$, and thus $u$ and $v$ belong to the same graph $C'$ of $S(G,(x,y))$. Now let $e$ be an edge of $C'$. Suppose first that both endpoints of $e$ lie in $C$. Then, $u\leftrightarrow_{2et}v$ in $G$ implies that there are paths $P$ and $Q$ in $G\setminus{e}$ such that $P$ goes from $u$ to $v$, $Q$ goes from $v$ to $u$, and there is no edge $(z,w)$ such that $(z,w)\in P$ and $(w,z)\in Q$.  Now consider the compressed paths $\tilde{P}$ and $\tilde{Q}$, of $P$ and $Q$, respectively, in $C'$. Then we have that $\tilde{P}$ is a path from $u$ to $v$, $\tilde{Q}$ is a path from $v$ to $u$, and neither $\tilde{P}$ nor $\tilde{Q}$ contains $e$. Furthermore, for every $(z,w)\in\tilde{P}$ that lies in $C$, we have that $(w,z)\notin\tilde{Q}$.
Now take an edge $(z,w)\in\tilde{P}$ such that at least one of $z$ and $w$ lies outside of $C$.
There are three cases to consider: either $x\in C$ and $y\notin C$, or $x\notin C$ and $y\in C$, or $C\cap\{x,y\}=\emptyset$. Let us consider first the case that $x\in C$ and $y\notin C$. Then we must have that either $(z,w)=(x,y)$, or $z=y$ and $w\in C$. Suppose that $(z,w)=(x,y)$ and $(y,x)\in\tilde{Q}$. But since $y\notin C$, in order for $\tilde{Q}$ to reach $y$ it must necessarily pass from the edge $(x,y)$. Thus the segment $(x,y),(y,x)$ from $\tilde{Q}$ can be eliminated. Now, if $z=y$ and $w\in C$, then we can either have $w=x$ or $w\neq x$. If $w=x$, then we can reason for $\tilde{P}$ as we did before for $\tilde{Q}$, in order to see that the segment $(w,z),(z,w)$ can be eliminated from $\tilde{P}$. Otherwise, we simply have that $(w,y)$ cannot be an edge of $C'$, and therefore $(w,z)$ is not contained in $\tilde{Q}$. Thus we have demonstrated that there are subpaths of $\tilde{P}$ and $\tilde{Q}$, that have the same endpoints, respectively, and do not share any pair of antiparallel edges. We can use the same argument for the other two cases (i.e., for the cases $x\notin C$ and $y\in C$, and $C\cap\{x,y\}=\emptyset$).

Now suppose that at least one of the endpoints of $e$ lies outside of $C$. Then, since $u\leftrightarrow_{2et}v$ in $G$, we have that there are paths $P$ and $Q$ in $G\setminus{(x,y)}$ such that $P$ goes from $u$ to $v$, $Q$ goes from $v$ to $u$, and there is no edge $(z,w)$ such that $(z,w)\in P$ and $(w,z)\in Q$. Since $P$ and $Q$ avoid the edge $(x,y)$, it is impossible that they use edges with at least one endpoint outside of $C$ (for otherwise they cannot return to $C$ using edges of the DAG of the strongly connected components of $G\setminus{(x,y)}$). Thus $P$ and $Q$ lie within $C$, and so they avoid $e$. This concludes the proof that $u\leftrightarrow_{2et}v$ in $C'$.\\
($\Leftarrow$) Suppose that $u\not\leftrightarrow_{2et}v$ in $G$. If $u$ and $v$ are in two different strongly connected components of $G\setminus{(x,y)}$, then we are done. Otherwise, suppose that $u$ and $v$ belong to the same strongly connected component $C$ of $G\setminus{(x,y)}$, and let $C'$ be the corresponding graph in $S(G,(x,y))$. Suppose first that $u\not\leftrightarrow_{2e} v$ in $G$. Then we may assume, without loss of generality, that there is an edge $e$ such that $u$ cannot reach $v$ in $G\setminus{e}$. Obviously, we cannot have $e=(x,y)$. Furthermore, it cannot be the case that at least one of the endpoints of $e$ lies outside of $C$ (precisely because $(x,y)$ cannot separate $u$ and $v$). Thus, $e$ lies within $C$. Now let us assume, for the sake of contradiction, that $u$ can reach $v$ in $C'\setminus{e}$. So let $P'$ be a path in $C'\setminus{e}$ from $u$ to $v$. Then, by Lemma~\ref{lemma:compressed_path}, there is a path $P$ in $G$ from $u$ to $v$ such that $\tilde{P}=P'$. Since, $e\in C$ but $e\notin P'$, by the same lemma we have that $P$ avoids $e$, a contradiction. Thus, $v$ is not reachable from $u$ in $C'\setminus{e}$, and so $u\not\leftrightarrow_{2et}v$ in $C'$.

So let us finally assume that $u\leftrightarrow_{2e} v$ in $G$. Since $u\not\leftrightarrow_{2et}v$ in $G$, this means that there is an edge $e\in G$ such that for every path $P$ in $G\setminus{e}$ from $u$ to $v$, and every path $Q$ in $G\setminus{e}$ from $v$ to $u$, there is a pair of twin edges $(z,w)\in P$ and $(w,z)\in Q$ $(*)$.
Suppose first that $e$ does not lie in $C$. Now let $P'$ be a path in $C'\setminus{(x,y)}$ from $u$ to $v$, and let $Q'$ be a path in $C'\setminus{(x,y)}$ from $v$ to $u$. Then, by Lemma~\ref{lemma:compressed_path} there are paths $P$ and $Q$ in $G$ such that $\tilde{P}=P'$ and $\tilde{Q}=Q'$. Thus, $P$ is a path from $u$ to $v$, and $Q$ is a path from $v$ to $u$. Since $(x,y)\notin P'$ and $(x,y)\notin Q'$, Lemma~\ref{lemma:compressed_path} implies that $(x,y)\notin P$ and $(x,y)\notin Q$. Thus, since $P$ and $Q$ have both of their endpoints in $C$ and avoid $(x,y)$, they cannot but lie entirely within $C$. Thus, $P=\tilde{P}=P'$ and $Q=\tilde{Q}=Q'$, and both $P$ and $Q$ avoid $e$. Then, by $(*)$, there is a pair of twin edges $(z,w)\in P'$ and $(w,z)\in Q'$. This shows that $u\not\leftrightarrow_{t}v$ in $C'\setminus{(x,y)}$.

Now suppose that $e$ lies in $C$. Let $P'$ be a path in $C'\setminus{e}$ from $u$ to $v$, and let $Q'$ be a path in $C'\setminus{e}$ from $v$ to $u$. Then, by Lemma~\ref{lemma:compressed_path3} there are paths $P$ and $Q$ in $G$ with $\tilde{P}=P'$ and $\tilde{Q}=Q'$, such that for every pair of twin edges $(z,w)\in P$ and $(w,z)\in Q$ we have that both $(z,w)$ and $(w,z)$ are in $C'$. Since $\tilde{P}=P'$ and $e\notin P'$, Lemma~\ref{lemma:compressed_path} implies that $P$ is a path from $u$ to $v$ that avoids $e$. Similarly, since $\tilde{Q}=Q'$ and $e\notin Q'$, Lemma~\ref{lemma:compressed_path} implies that $Q$ is a path from $v$ to $u$ that avoids $e$. Thus, by $(*)$ we have that there is a pair of twin edges $(z,w)\in P$ and $(w,z)\in Q$. Then, we have that both $(z,w)$ and $(w,z)$ are in $C'$. Thus, since $\tilde{P}=P'$ and $\tilde{Q}=Q'$, by Lemma~\ref{lemma:compressed_path} we have that $(z,w)\in P'$ and $(w,z)\in Q'$. This shows that $u\not\leftrightarrow_{t}v$ in $C'\setminus{e}$.
In any case, we conclude that $u\not\leftrightarrow_{2et}v$ in $C'$.
\end{proof}

We can combine Propositions \ref{proposition:H-2et_main} and \ref{proposition:S-2et_main} in order to derive some auxiliary graphs that maintain the relation of $2$-edge twinless strong connectivity of the original graph. Then we can exploit the properties of those graphs in order to provide a linear-time algorithm for computing the $2$-edge twinless strongly connected components.
First, we introduce some notation. Let $G_s$ be a strongly connected digraph, and let $r$ be either a marked vertex of $G_s$ or $s$. Then we denote $H(G_s,r)$ as $H_r$. Furthermore, if $r'$ is either a marked vertex of $H_r^R$ 
or $r$, we denote $H(H_r^R,r')$ as $H_{rr'}$. 
In the following, when we refer to the immediate dominator $d(x)$ of a vertex $x$, or to the subtree $T(x)$ of a dominator tree decomposition, we specify the corresponding underlying graph (e.g., we say that $d(x)$ is the immediate dominator of $x$ in $H_{rr'}$).

Now let $x$ be a vertex in $H_{rr'}$. Then $x$ is also a vertex in $H_r$. We distinguish four cases, depending on whether $x$ is ordinary or auxiliary in $H_{rr'}$ and $H_r$. If $x$ is ordinary in $H_{rr'}$ and ordinary in $H_r$, it is called an $\mathit{oo}$-vertex. If $x$ is ordinary in $H_{rr'}$ and auxiliary in $H_r$, it is called an $\mathit{oa}$-vertex. If $x$ is auxiliary in $H_{rr'}$ and ordinary in $H_r$, it is called $\mathit{ao}$-vertex. And if $x$ is auxiliary in both $H_{rr'}$ and $H_r$, it is called an $\mathit{aa}$-vertex.

Now we have the following.

\begin{corollary}
\label{corollary:2et_char}
Let $G_s$ be a strongly connected digraph, and let $x,y$ be two vertices of $G_s$. Then $x\leftrightarrow_{2et}y$ in $G_s$ if and only if $x$ and $y$ are both $\mathit{oo}$-vertices in $A$ and $x\leftrightarrow_{2et}y$ in $A$, where $A$ is either $(1)$ $H_{ss}$, or $(2)$ a graph in $S(H_{sr},(d(r),r))$, or $(3)$ $H_{rr}$, or $(4)$ a graph in $S(H_{rr'},(d(r'),r'))$ (where $r$ and $r'$ are marked vertices, $d(r)$ in $(2)$ denotes the immediate dominator of $r$ in $H_s^R$, and $d(r')$ in $(4)$ denotes the immediate dominator of $r'$ in $H_r^R$).
\end{corollary}
\begin{proof}
This is an immediate consequence of Propositions \ref{proposition:H-2et_main} and \ref{proposition:S-2et_main}.
\end{proof}

We note that the graphs $H_{rr'}$ were used in \cite{2ECC:GILP:TALG} to compute the $2$-edge strongly connected components in linear time. Specifically, it was shown in \cite{2ECC:GILP:TALG} that the $2$-edge strongly connected components of $G_s$ are given by the $\mathit{oo}$-vertices of: $(1)$ the graph $H_{ss}$, $(2)$ the strongly connected components of all graphs $H_{sr}\setminus{(d(r),r)}$, where $r$ is a marked vertex of $G_s$, and $(d(r),r)$ is the critical edge of $H_{sr}$, $(3)$ the graphs $H_{rr}$, for all marked vertices $r$ of $G_s$, and $(4)$ the strongly connected components of all graphs $H_{rr'}\setminus{(d(r'),r')}$, where $r$ is a marked vertex of $G_s$, $r'$ is a marked vertex of $H_r^R$, and $(d(r'),r')$ is the critical edge  of $H_{rr'}$. Since we can compute all those auxiliary graphs in linear time (Theorem~\ref{theorem:aux_linear_time_main}), this implies that we can compute all $2$-edge strongly connected components in linear time.

The following series of lemmas provide useful information concerning the reachability of vertices in the auxiliary graphs when we remove an edge. All of them are basically a consequence of Lemma~\ref{lemma:cor_paths}, which establishes a two-way correspondence between paths of $G_s$ and paths of $H(G_s,r)$.

\begin{lemma}
\label{lemma:r-reachability}
Let $G_s$ be a strongly connected digraph and let $r$ be either a marked vertex of $G_s$ or $s$. Let $(x,y)$ be an edge of $H_r$ such that $y$ is an ordinary vertex of $H_r$. Then $r$ can reach every vertex of $H_r$ in $H_r\setminus{(x,y)}$.
\end{lemma}
\begin{proof}
Suppose first that $x$ is an ordinary vertex of $H_r$. Then $(x,y)$ is also an edge of $G_s$, which is not a bridge of $G_s$. Now let $u$ be a vertex of $H_r$, but not $d(r)$ (if $r\neq s$). Then there is a path $P$ from $s$ to $u$ that avoids $(x,y)$. If $r\neq s$, then, by Lemma~\ref{lemma:internal_paths}, $P$ must necessarily pass from the bridge $(d(r),r)$, and every time it moves out of $D(r)$ it must later on pass again through $(d(r),r)$, in order to reach eventually $u$. Thus $P$ contains a subpath $Q$ from $r$ to $u$ that visits only vertices of $D(r)$. Now consider the corresponding path $Q_H$ in $H_r$. By Lemma~\ref{lemma:cor_paths}, this is precisely a path from $r$ to $u$ in $H_r$ that avoids $(x,y)$. Now let $r\neq s$. Since $H_r$ is strongly connected (by Corollary~\ref{corollary:sc_of_H}), we have that $r$ can reach $d(r)$. Since $y$ is ordinary, $(x,y)$ is not an incoming edge to $d(r)$ in $H_r$. And since $r$ can reach every vertex of $H_r\setminus\{d(r)\}$ in $H_r\setminus{(x,y)}$, we conclude that $d(r)$ is also reachable from $r$ in $H_r\setminus{(x,y)}$.

Now suppose that $x$ is an auxiliary vertex of $H_r$. If $x=d(r)$ (assuming that $r\neq s$), then we must have that $y=r$. Since $H_r$ is strongly connected (by Corollary~\ref{corollary:sc_of_H}), we have that $r$ can reach every vertex of $H_r$. Thus, even by removing the incoming edge $(d(r),r)$ of $r$, we still have that $r$ can reach every vertex of $H_r\setminus{(d(r),r)}$. So let us assume that $x\neq d(r)$ (if $r\neq s$). Then we have that $x$ is a marked vertex of $G_s$ such that $d(x)\in T(r)$. Now let $u$ be a vertex of $H_r$ but not $d(r)$ (if $r\neq s$). If $u\neq x$, then $u$ is not a descendant of $x$ in $D(G_r)$. Therefore, as a consequence of Lemma~\ref{lemma:internal_paths}, there is a path $P$ in $G_s$ from $s$ to $u$ that avoids $(d(x),x)$. Arguing as above, we have that there is a subpath $Q$ of $P$ from $r$ to $u$. Now consider the corresponding path $Q_H$ of $Q$ in $H_r$. Then, by Lemma~\ref{lemma:cor_paths}, $Q_H$ avoids $(d(x),x)$, and therefore $(x,y)$. Now, if $u=x$, then there is definitely a path from $r$ to $u$ in $H_r$ (by Corollary~\ref{corollary:sc_of_H}). But then we can also avoid using $(x,y)$ in order to reach $u$ from $r$ in $H_r$. Finally, we can argue as above, in order to conclude that $d(r)$ is also reachable from $r$ in $H_r\setminus{(x,y)}$. This completes the proof.
\end{proof}

\begin{lemma}
\label{lemma:bridge-r-reachability}
Let $G_s$ be a strongly connected digraph and let $r$ be either a marked vertex of $G_s$ or $s$. Let $(d(z),z)$ be an edge of $H_r$ such that $z$ is a marked vertex of $G_s$ with $d(z)\in T(r)$. Then $r$ can reach every vertex of $H_r$ in $H_r\setminus{(x,y)}$, except $z$, which is unreachable from $r$ in $H_r\setminus{(x,y)}$, and possibly $d(r)$ (if $r\neq s$).
\end{lemma}
\begin{proof}
We can argue as in the proof of Lemma~\ref{lemma:r-reachability} in order to show that, for every vertex $u$ in $H_r$ that is not $z$ or $d(r)$ (if $r\neq s$), there is a path from $r$ to $u$ in $H_r$ that avoids $(d(z),z)$. Now let's assume for the sake of contradiction that there is a path $\tilde{P}$ in $H_r$ from $r$ to $z$ that avoids $(d(z),z)$. Then, by Lemma~\ref{lemma:cor_paths} there is path $P$ in $G_s$ from $r$ to $z$ that corresponds to $\tilde{P}$ and it also avoids $(d(z),z)$. Now, since $d(z)$ is not a dominator of $r$ in $G_s$, there is a path $Q$ from $s$ to $r$ in $G_s$ that avoids $d(z)$. But now the concatenation of $Q$ with $P$ is a path in $G_s$ from $s$ to $z$ that avoids $(d(z),z)$, contradicting the fact that this edge is a bridge in $G_s$.
\end{proof}

\begin{lemma}
\label{lemma:r-reachability-cor_edge}
Let $G_s$ be a strongly connected digraph and let $x$ be a marked vertex of $G_s$ such that $d(x)\in T(s)$ in $G_s$. Let $(x_0,y)$ be an edge of $G_s$ such that $x_0$ is a descendant of $x$ in $D(G_s)$ and $y$ is not a descendant of $x$ in $D(G_s)$. Let $u$ be a vertex of $H_s$. If $u$ reaches $s$ in $G_s\setminus{(x_0,y)}$, then $u$ reaches $s$ in $H_s\setminus{(x,y)}$.
\end{lemma}
\begin{proof}
Let $u$ be a vertex of $H_s$ and let $P$ be a path from $u$ to $s$ in $G_s\setminus{(x_0,y)}$. Consider the corresponding path $P_H$ of $P$ in $H_s$. $P_H$ has the same endpoints as $P$. Thus, if $P_H$ avoids the edge $(x,y)$ of $H_s$, then we are done. Otherwise, this means that $P$ uses an edge $(x_1,y)$ of $G_s$, distinct from $(x_0,y)$ (although we may have $x_1=x_0$, since we allow multiple edges), such that $x_1$ is a descendant of $x$ in $D(G_s)$. But then we must have that $(x,y)$ is a double edge of $H_s$, and therefore $H_s\setminus{(x,y)}$ is strongly connected (since $H_s$ is strongly connected, by Corollary~\ref{corollary:sc_of_H}). Thus $u$ can reach $s$ in $H_s\setminus{(x,y)}$.
\end{proof}

\begin{lemma}
\label{lemma:r-reachability2}
Let $G_s$ be a strongly connected digraph and let $r$ be either a marked vertex of $G_s$ or $s$. Let $z$ by a marked vertex of $G_s$ such that $d(z)\in T(r)$. Let $\{(z,y_1),\dots,(z,y_k)\}$ be the set of all edges that stem from $z$ in $H_r$. Then $r$ can reach every vertex of $H_r$ in $H_r\setminus\{(z,y_1),\dots,(z,y_k)\}$, except possibly $d(r)$ (if $r\neq s$).
\end{lemma}
\begin{proof}
We can argue as in the proof of Lemma~\ref{lemma:r-reachability} in order to show that, for every vertex $u$ in $H_r$ that is not $z$ or $d(r)$ (if $r\neq s$), there is a path from $r$ to $u$ in $H_r$ that avoids $(d(z),z)$, and therefore all of the edges $\{(z,y_1),\dots,(z,y_k)\}$. Furthermore, there is certainly a path in $H_r$ from $r$ to $z$ (by Corollary~\ref{corollary:sc_of_H}), and so we can definitely reach $z$ from $r$ in $H_r\setminus\{(z,y_1),\dots,(z,y_k)\}$.
\end{proof}

\begin{lemma}
\label{lemma:uv-reachability}
Let $G_s$ be a strongly connected digraph and let $r$ be either a marked vertex of $G_s$ or $s$. Let $u$ and $v$ be two vertices of $H_r$, and let $(x,y)$ be an edge of $G_s$ such that either $(1)$ both $x$ and $y$ are ordinary vertices of $H_r$, or $(2)$ $y$ is a marked vertex of $G_s$ and $x=d(y)$, or $(3)$ $(x,y)=(d(r),r)$ (assuming that $r\neq s$). Then $u$ can reach $v$ in $G_s\setminus{(x,y)}$ if and only if $u$ can reach $v$ in $H_r\setminus{(x,y)}$.
\end{lemma}
\begin{proof}
$(\Rightarrow)$ Let $P$ be a path in $G_s\setminus{(x,y)}$ that starts from $u$ and ends in $v$. Now consider the corresponding path $P_H$ of $P$ in $H_r$. Then, since either one of $(1)$, $(2)$, or $(3)$ holds for $(x,y)$, by Lemma~\ref{lemma:cor_paths} we have that $(x,y)$ is not contained in $P_H$. Thus $u$ can reach $v$ in $H_r\setminus{(x,y)}$.\\
$(\Leftarrow)$ Let us assume that $u$ cannot reach $v$ in $G_s\setminus{(x,y)}$. Now let's assume for the sake of contradiction that there is path $\tilde{P}$ in $H_r\setminus{(x,y)}$ that starts from $u$ and ends in $v$. By Lemma~\ref{lemma:cor_paths}, there is a path $P$ in $G_s$ such that $P_H=\tilde{P}$. Since either one of $(1)$, $(2)$, or $(3)$ holds for $(x,y)$, by the same Lemma we have that $(x,y)$ is not contained in $P$. But this means that $u$ can reach $v$ in $G_s\setminus{(x,y)}$ - a contradiction.
\end{proof}

\begin{corollary}
\label{corollary:2et_char_main}
Let $G_s$ be a strongly connected digraph, and let $x,y$ be two vertices of $G_s$. Then \tstrongEC{x}{y}{G_s} if and only if $x$ and $y$ are both ordinary vertices in $H$ and
\tstrongEC{x}{y}{H}, where $H$ is either $(1)$ $H_{ss}$, or $(2)$ $H_{rr}$, or $(3)$ a graph in $S(H_{sr},(d(r),r))$, or $(4)$ a graph in $S(H_{rr'},(d(r'),r'))$ (where $r$, $r'$, $d(r)$ and $d(r')$ are as in the statement of Corollary~\ref{corollary:2et_char}).
\end{corollary}
\begin{proof}
This is an immediate consequence of Propositions \ref{proposition:H-2et_main} and \ref{proposition:S-2et_main}.
\end{proof}

Now we can describe the structure of the strongly connected components of the graphs that appear in Corollary~\ref{corollary:2et_char_main} when we remove a strong bridge from them.

\begin{lemma}
\label{lemma:Hss}
Let $G_s$ be a strongly connected digraph. Let $(x,y)$ be a strong bridge of $H_{ss}$. Then the strongly connected components of $H_{ss}\setminus{(x,y)}$ are given by one of the following:
\begin{enumerate}[label={(\roman*)}]
\item{$\{x\}$ and $H_{ss}\setminus\{x\}$, where $x$ is auxiliary in $H_s$ and $y=d(x)$ in $G_s$ (see case $(1.3)$ below)}
\item{$\{x\}$ and $H_{ss}\setminus\{x\}$, where $x$ is auxiliary in $H_{ss}$ and $(x,y)$ is the only outgoing edge of $x$ in $H_{ss}$ (see case $(3.3)$ below)}
\item{$\{y\}$ and $H_{ss}\setminus\{y\}$, where $y$ is auxiliary in $H_{ss}$ and $x=d(y)$ in $H_s^R$ (see cases $(2.1)$ and $(2.2)$ below)}
\item{$\{x\}$, $\{y\}$ and $H_{ss}\setminus\{x,y\}$, where $x$ is auxiliary in $H_s$, $y$ is auxiliary in $H_{ss}$, $x=d(y)$ in $H_s^R$ and $y=d(x)$ in $G_s$ (see case $(2.3)$ below)}
\end{enumerate}

\end{lemma}

\begin{proof}
Let $(x,y)$ be an edge of $H_{ss}$.
We start by distinguishing four different cases, depending on whether $x$ and $y$ are ordinary or auxiliary vertices of $H_{ss}$.

$(1)$ Let $x$ and $y$ be both ordinary vertices of $H_{ss}$.
Here we distinguish four different cases, depending on whether $x$ and $y$ are ordinary or auxiliary in $H_s$. Notice that, in any case, $(y,x)$ is an edge of $H_s$. $(1.1)$ Let $x$ and $y$ be both $\mathit{oo}$-vertices. Then, by Lemma~\ref{lemma:r-reachability}, $s$ can reach all vertices of $H_s$ in $H_s\setminus{(y,x)}$. Thus, with the removal of $(x,y)$ from $H_s^R$, every vertex of $H_s^R$ can reach $s$. Thus, Lemma~\ref{lemma:uv-reachability} implies that, even by removing $(x,y)$ from $H_{ss}$, every vertex from $H_{ss}$ can reach $s$. Now, since $x$ and $y$ are both ordinary vertices of $H_{ss}$, by Lemma~\ref{lemma:r-reachability} we have that $s$ can reach every vertex in $H_{ss}\setminus{(x,y)}$. We conclude that $H_{ss}\setminus{(x,y)}$ is strongly connected. $(1.2)$ Let $x$ be $\mathit{oo}$ and $y$ be $\mathit{oa}$. Then we use precisely the same argument as in $(1.1)$ to conclude that $H_{ss}\setminus{(x,y)}$ is strongly connected. $(1.3)$ Let $x$ be $\mathit{oa}$ and $y$ be $\mathit{oo}$. Then, since there is no critical vertex in $H_s$, $x$ must be a marked vertex of $G_s$ such that $d(x)\in T(s)$. Since $(y,x)$ is an edge of $H_s$, we must have that $y=d(x)$ in $G_s$. Thus, by Lemma~\ref{lemma:bridge-r-reachability}, $s$ can reach all vertices of $H_s$ in $H_s\setminus{(y,x)}$, except $x$. This implies that $s$ is reachable from all vertices of $H_s^R$ in $H_s^R\setminus{(x,y)}$, except from $x$. By Lemma~\ref{lemma:uv-reachability}, this implies that $s$ is reachable from all vertices of $H_{ss}$ in $H_{ss}\setminus{(x,y)}$, except from $x$. Furthermore, Lemma~\ref{lemma:uv-reachability} also implies that all vertices of $H_{ss}$ are reachable from $s$ in $H_{ss}\setminus{(x,y)}$. Thus, the strongly connected components of $H_{ss}\setminus{(x,y)}$ are $\{x\}$ and $H_{ss}\setminus\{x\}$.  $(1.4)$ Let $x$ and $y$ be both $\mathit{oa}$. This implies that $(y,x)$ is an edge of $H_s$ both of whose endpoints are auxiliary. But this is impossible to occur since there is no critical vertex in $H_s$.

$(2)$ Let $x$ be an ordinary vertex of $H_{ss}$ and let $y$ be an auxiliary vertex of $H_{ss}$. Since $H_{ss}$ does not contain a critical vertex, this means that $y$ is a marked vertex of $H_s^R$ and $x=d(y)$ in $H_s^R$. Thus, $(y,x)$ is also an edge of $H_s$. Now we distinguish four different cases, depending on whether $x$ and $y$ are ordinary or auxiliary in $H_s$. $(2.1)$ Let $x$ be $\mathit{oo}$ and $y$ be $\mathit{ao}$. Then, by Lemma~\ref{lemma:r-reachability}, $s$ can reach all vertices of $H_s$ in $H_s\setminus{(y,x)}$. This implies that $s$ is reachable from all vertices of $H_s^R$ in $H_s^R\setminus{(x,y)}$. By Lemma~\ref{lemma:uv-reachability}, this implies that $s$ is reachable from all vertices of $H_{ss}$ in $H_{ss}\setminus{(x,y)}$. Now, since $x=d(y)$ in $H_s^R$, by Lemma~\ref{lemma:bridge-r-reachability} we have that $s$ can reach all vertices of $H_{ss}$ in $H_{ss}\setminus{(x,y)}$, except $y$. Thus, the strongly connected components of $H_{ss}\setminus{(x,y)}$ are $\{y\}$ and $H_{ss}\setminus\{y\}$. $(2.2)$ Let $x$ be $\mathit{oo}$ and $y$ be $\mathit{aa}$. Then we use precisely the same argument as in $(2.1)$ in order to show that the strongly connected components of $H_{ss}\setminus{(x,y)}$ are $\{y\}$ and $H_{ss}\setminus\{y\}$. $(2.3)$ Let $x$ be $\mathit{oa}$ and $y$ be $\mathit{ao}$. This means that $y$ is an ordinary vertex of $H_s$ and $x$ is an auxiliary vertex of $H_s$. Thus, $x$ is a marked vertex of $G_s$, and $y=d(x)$ in $G_s$. By Lemma~\ref{lemma:bridge-r-reachability}, we have that $s$ can reach all vertices of $H_s$ in $H_s\setminus{(y,x)}$, except $x$. This implies that $s$ is reachable from all vertices of $H_s^R$ in $H_s^R\setminus{(x,y)}$, except from $x$. By Lemma~\ref{lemma:uv-reachability}, this implies that $s$ is reachable from all vertices of $H_{ss}$ in $H_{ss}\setminus{(x,y)}$, except from $x$. Now, since $x=d(y)$ in $H_s^R$, Lemma~\ref{lemma:bridge-r-reachability} implies that $s$ can reach all vertices of $H_{ss}$ in $H_{ss}\setminus{(x,y)}$, except $y$. We conclude that the strongly connected components of $H_{ss}\setminus{(x,y)}$ are $\{x\}$, $\{y\}$, and $H_{ss}\setminus\{x,y\}$. $(2.4)$. Let $x$ be $\mathit{oa}$ nd $y$ be $\mathit{aa}$. This implies that $(y,x)$ is an edge of $H_s$ both of whose endpoints are auxiliary. But this is impossible to occur since there is no critical vertex in $H_s$.

$(3)$ Let $x$ be an auxiliary vertex of $H_{ss}$ and let $y$ be an ordinary vertex of $H_{ss}$. Since $H_{ss}$ has no critical vertex, this means that $x$ is a marked vertex of $H_s^R$ with $d(x)\in T(s)$ in $H_s^R$.
Now we distinguish four different cases, depending on whether $x$ and $y$ are ordinary or auxiliary in $H_s$. In any case, by Lemma~\ref{lemma:r-reachability} we have that $s$ can reach every vertex of $H_{ss}$ in $H_{ss}\setminus{(x,y)}$.
$(3.1)$ Let $x$ be $\mathit{ao}$ and $y$ be $\mathit{oo}$. Let $(x_0,y)$ be an edge of $H_s^R$ that corresponds to $(x,y)$ in $H_{ss}$. Then $(y,x_0)$ is an edge of $H_s$.
If $x_0$ is an ordinary vertex of $H_s$, then by Lemma~\ref{lemma:r-reachability} we have that $s$ reaches all vertices of $H_s$ in $H_s\setminus{(y,x_0)}$. This implies that every vertex of $H_s^R$ can reach $s$ in $H_s^R\setminus{(x_0,y)}$. Then Lemma~\ref{lemma:r-reachability-cor_edge} implies that $x$ can reach $s$ in $H_{ss}\setminus{(x,y)}$. Thus, every vertex of $H_{ss}$ can reach $s$ in $H_{ss}\setminus{(x,y)}$, and so we have that $H_{ss}\setminus{(x,y)}$ is strongly connected.
Similarly, if $x_0$ is an auxiliary vertex of $H_s$, then, by Lemma~\ref{lemma:bridge-r-reachability}, we have that $s$ reaches all vertices of $H_s$ in $H_s\setminus{(y,x_0)}$, except $x_0$, which is unreachable from $s$ in $H_s\setminus{(y,x_0)}$. This implies that every vertex of $H_s^R$ can reach $s$ in $H_s^R\setminus{(x_0,y)}$, except $x_0$, which cannot reach $s$ in $H_s^R\setminus{(x_0,y)}$. Since $x$ is an ordinary vertex of $H_s$, we have that $x\neq x_0$, and so $x$ can reach $s$ in $H_s^R\setminus{(x_0,y)}$. Now Lemma~\ref{lemma:r-reachability-cor_edge} implies that $x$ can reach $s$ in $H_{ss}\setminus{(x,y)}$. Thus, every vertex of $H_{ss}$ can reach $s$ in $H_{ss}\setminus{(x,y)}$, and so we have that $H_{ss}\setminus{(x,y)}$ is strongly connected.
$(3.2)$ Let $x$ be $\mathit{ao}$ and $y$ be $\mathit{oa}$. Now let $\{(x_1,y),\dots,(x_k,y)\}$ be the set of all edges of $H_s^R$ that correspond to $(x,y)$. Then $\{(y,x_1),\dots,(y,x_k)\}$ is a set of edges of $H_s$ that stem from $y$. Thus, by Lemma~\ref{lemma:r-reachability2} we have that $s$ can reach all vertices of $H_s$ in $H_s\setminus\{(y,x_1),\dots,(y,x_k)\}$.
This implies that $s$ can be reached by all vertices of $H_s^R$ in $H_s^R\setminus\{(x_1,y),\dots,(x_k,y)\}$. In particular, there is a path $P$ from $x$ to $s$ in $H_s^R\setminus\{(x_1,y),\dots,(x_k,y)\}$. Then, by Lemma~\ref{lemma:cor_paths}, the corresponding path $P_H$ of $P$ in $H_{ss}$ goes from $x$ to $s$ and avoids the edge $(x,y)$ of $H_{ss}$. This means that $s$ is reachable from $x$ in $H_{ss}\setminus{(x,y)}$, and therefore every vertex of $H_{ss}$ reaches $s$ in $H_{ss}\setminus{(x,y)}$. Thus we have established that $H_{ss}\setminus{(x,y)}$ is strongly connected.
$(3.3)$ Let $x$ be $\mathit{aa}$ and $y$ be $\mathit{oo}$. Let $(x_0,y)$ be an edge of $H_s^R$ that corresponds to $(x,y)$ in $H_{ss}$. Then $(y,x_0)$ is an edge of $H_s$. Now, if $x_0$ is ordinary in $H_s$ or it is auxiliary in $H_s$ but $x_0\neq x$, then we can argue as in $(3.1)$ in order to establish that $H_{ss}\setminus{(x,y)}$ is strongly connected. Otherwise, suppose that $x_0=x$. Then, since $x$ is auxiliary in $H_s$ and $y$ is ordinary in $H_s$ (and $H_s$ contains no critical vertex), we have that $y=d(x)$ in $G_s$. Now, by Lemma~\ref{lemma:bridge-r-reachability} we have that $s$ reaches all vertices of $H_s$ in $H_s\setminus{(y,x)}$, except $x$, which is unreachable from $s$ in $H_s\setminus{(y,x)}$. This implies that every vertex of $H_s^R$ can reach $s$ in $H_s^R\setminus{(x,y)}$, except $x$, which cannot reach $s$ in $H_s^R\setminus{(x,y)}$. Now Lemma~\ref{lemma:r-reachability-cor_edge} implies that every vertex of $H_{ss}$ can reach $s$ in $H_{ss}\setminus{(x,y)}$, except possibly $x$. Thus, we have that either $H_{ss}\setminus{(x,y)}$ is strongly connected, or its strongly connected components are $\{x\}$ and $H_{ss}\setminus\{x\}$.
%
Furthermore, since $(y,x)$ is the only incoming edge to $x$ in $H_s$, we have that $(x,y)$ is the only outgoing edge from $x$ in $H_s^R$. Therefore, since $y$ is ordinary in $H_{ss}$, we have that $x$ has no descendants in $D(H_s^R)$. This means that $(x,y)$ is the only outgoing edge of $x$ in $H_{ss}$.
$(3.4)$ We do not have to consider the case that $x$ is $\mathit{aa}$ and $y$ is $\mathit{oa}$, because this implies that $(y,x)$ is an edge between two auxiliary vertices in $H_s$, which is impossible to occur. 

$(4)$ It is impossible that $x$ and $y$ are both auxiliary vertices of $H_{ss}$, since $H_{ss}$ does not contain a critical vertex.
\end{proof}

\begin{lemma}
\label{lemma:Hrr}
Let $G_s$ be a strongly connected digraph, and let $r$ be a marked vertex of $G_s$. Let $(x,y)$ be a strong bridge of $H_{rr}$. Then the strongly connected components of $H_{rr}\setminus{(x,y)}$ are given by one of the following:
\begin{enumerate}[label={(\roman*)}]
\item{$\{x\}$ and $H_{rr}\setminus\{x\}$, where $x$ is auxiliary in $H_r$ and $y=d(x)$ in $G_s$ (see case $(1.3)$ below)}
\item{$\{x\}$, $\{d(r)\}$ and $H_{rr}\setminus\{x,d(r)\}$, where $x$ is $\mathit{oa}$ in $H_{rr}$, $y$ is $\mathit{oo}$ in $H_{rr}$, $y=d(x)$ in $G_s$, and $(d(r),x)$ is the only kind of outgoing edge of $d(r)$ in $H_{rr}$ (see case $(1.3)$ below)}
\item{$\{x\}$ and $H_{rr}\setminus\{x\}$, where $x$ is auxiliary in $H_{rr}$ and $(x,y)$ is the only outgoing edge of $x$ in $H_{rr}$ (see cases $(3.3)$ and $(3.4)$ below)}
\item{$\{y\}$ and $H_{rr}\setminus\{y\}$, where $y$ is auxiliary in $H_{rr}$ and $x=d(y)$ in $H_r^R$ (see the beginning of $(2)$, and cases $(2.1)$ and $(2.2)$ below)}
\item{$\{x\}$, $\{y\}$ and $H_{rr}\setminus\{x,y\}$, where $x$ is $\mathit{oa}$ in $H_{rr}$, $y$ is $\mathit{ao}$ in $H_{rr}$, $x=d(y)$ in $H_r^R$ and $y=d(x)$ in $G_s$ (see case $(2.3)$ below)}
\item{$\{x\}$, $\{y\}$, $\{d(r)\}$ and $H_{rr}\setminus\{x,y,d(r)\}$, where $x$ is $\mathit{oa}$ in $H_{rr}$, $y$ is $\mathit{ao}$ in $H_{rr}$, $x=d(y)$ in $H_r^R$, $y=d(x)$ in $G_s$, and $(d(r),x)$ is the only kind of outgoing edge of $d(r)$ in $H_{rr}$ (see case $(2.3)$ below)}
\end{enumerate}
\end{lemma}
\begin{proof}
First notice that $d(r)$ is an auxiliary vertex of $H_r$ and has only one outgoing edge $(d(r),r)$. Thus, vertex $d(r)$ in $H_r^R$ has only one incoming edge $(r,d(r))$, and so this must be a bridge of $H_r^R$. This means that $d(r)$ is an $\mathit{aa}$-vertex of $H_{rr}$. Now let $(x,y)$ be an edge of $H_{rr}$. We distinguish four different cases, depending on whether $x$ and $y$ are ordinary or auxiliary vertices of $H_{rr}$.

$(1)$ Let $x$ and $y$ be both ordinary vertices of $H_{rr}$. Here we distinguish four different cases, depending on whether $x$ and $y$ are ordinary or auxiliary in $H_r$. Notice that, in any case, $(y,x)$ is an edge of $H_r$.
$(1.1)$ Let $x$ and $y$ be both $\mathit{oo}$ in $H_{rr}$. By Lemma~\ref{lemma:r-reachability} we have that $r$ can reach all vertices of $H_r$ in $H_r\setminus{(y,x)}$. Thus we infer that $r$ is reachable from all vertices of $H_r^R$ in $H_r^R\setminus{(x,y)}$, and so Lemma~\ref{lemma:uv-reachability} implies that $r$ is reachable from all vertices of $H_{rr}$ in $H_{rr}\setminus{(x,y)}$. Now, since, $x$ and $y$ are both ordinary vertices of $H_{rr}$, by Lemma~\ref{lemma:r-reachability} we have that $r$ reaches all vertices of $H_{rr}\setminus{(x,y)}$. We conclude that $H_{rr}\setminus{(x,y)}$ is strongly connected.
$(1.2)$ Let $x$ be $\mathit{oo}$ and $y$ be $\mathit{oa}$. Then we can argue as in $(1.1)$ in order to conclude that $H_{rr}\setminus{(x,y)}$ is strongly connected.
$(1.3)$ Let $x$ be $\mathit{oa}$ and $y$ be $\mathit{oo}$. By Lemma~\ref{lemma:bridge-r-reachability} we have that $r$ can reach all vertices of $H_r$ in $H_r\setminus{(y,x)}$, except $x$, which is unreachable from $r$ in $H_r\setminus{(y,x)}$, and possibly $d(r)$. In any case, since $d(r)$ has only one outgoing edge $(d(r),r)$ in $H_r$, we also have that $d(r)$ cannot reach $x$ in $H_r\setminus{(y,x)}$. This implies that $r$ is reachable from all vertices of $H_r^R$ in $H_r^R\setminus{(x,y)}$, except from $x$, and possibly $d(r)$; and that, in any case, $x$ cannot reach $d(r)$ in $H_r^R\setminus{(x,y)}$. Now Lemma~\ref{lemma:uv-reachability} implies that $r$ is reachable from all vertices of $H_{rr}$ in $H_{rr}\setminus{(x,y)}$, except from $x$, and possibly $d(r)$, and that, in any case, $x$ cannot reach $d(r)$ in $H_{rr}$. We conclude that there are two possibilities: the strongly connected components of $H_{rr}\setminus{(x,y)}$ are either $\{x\}$ and $H_{rr}\setminus\{x\}$, or $\{x\}$, $\{d(r)\}$ and $H_{rr}\setminus\{x,d(r)\}$.
$(1.4)$ Let $x$ and $y$ be both $\mathit{oa}$. This implies that $(y,x)$ is an edge of $H_s$ both of whose endpoints are auxiliary. But this is impossible to occur since none of those vertices is $d(r)$ (which is $\mathit{aa}$ in $H_{rr}$).

$(2)$ Let $x$ be ordinary in $H_{rr}$ and let $y$ be auxiliary in $H_{rr}$. If $y=d(r)$, then $x$ must necessarily be $r$, since $d(r)$ has only one incoming edge $(r,d(r))$ in $H_r^R$. So let us consider first the case that $(x,y)=(r,d(r))$, which is the only case in which $d(r)$ can appear as an endpoint of $(x,y)$ if $y$ is auxiliary in $H_{rr}$. By Corollary~\ref{corollary:sc_of_H}, we have that $H_r^R$ is strongly connected. Thus, even by removing the edge $(r,d(r))$, we still have that $r$ is reachable from every vertex of $H_r^R$. By Lemma~\ref{lemma:uv-reachability}, this implies that every vertex of $H_{rr}$ is reachable from $r$ in $H_{rr}\setminus{(r,d(r))}$. Now, by Lemma~\ref{lemma:bridge-r-reachability} we have that $r$ reaches every vertex of $H_{rr}\setminus{(r,d(r))}$, except $d(r)$, which is unreachable from $r$ in $H_{rr}\setminus{(r,d(r))}$. Thus we conclude that the strongly connected components of $H_{rr}\setminus{(r,d(r))}$ are $\{d(r)\}$ and $H_{rr}\setminus\{d(r)\}$. Now, due to the preceding argument, in what follows we can assume that none of $x$ and $y$ are $d(r)$. This means that $y$ is a marked vertex of $H_r^R$ and $x=d(y)$ in $H_r^R$. Thus, $(y,x)$ is an edge of $H_r$. Now we distinguish four possibilities, depending on whether $x$ and $y$ are ordinary or auxiliary in $H_r$.
$(2.1)$ Let $x$ be $\mathit{oo}$ and $y$ be $\mathit{ao}$. Since $x$ and $y$ are ordinary vertices of $H_r$, and $(y,x)$ is an edge of $H_r$, we can argue as in $(1.1)$ in order to show that $r$ is reachable from every vertex of $H_{rr}$ in $H_{rr}\setminus{(x,y)}$. Now, by Lemma~\ref{lemma:bridge-r-reachability} we have that $r$ reaches every vertex of $H_{rr}$ in $H_{rr}\setminus{(x,y)}$, except $y$, which is unreachable from $r$ in $H_{rr}\setminus{(x,y)}$. We conclude that the strongly connected components of $H_{rr}\setminus{(x,y)}$ are $\{y\}$ and $H_{rr}\setminus\{y\}$.
$(2.2)$ Let $x$ be $\mathit{oo}$ and $y$ be $\mathit{aa}$. Then we can argue as in $(2.1)$ in order to show that the strongly connected components of $H_{rr}\setminus{(x,y)}$ are $\{y\}$ and $H_{rr}\setminus\{y\}$.
$(2.3)$ Let $x$ be $\mathit{oa}$ and $y$ be $\mathit{ao}$. This means that $y=d(x)$ in $H_r$, and so, by Lemma~\ref{lemma:bridge-r-reachability}, we have that $r$ can reach every vertex of $H_r$ in $H_r\setminus{(y,x)}$, except $x$, which is unreachable from $r$ in $H_r\setminus{(y,x)}$, and possibly $d(r)$. In any case, we have that $d(r)$ cannot reach $x$ in $H_r\setminus{(y,x)}$, since there is only one outgoing edge $(d(r),r)$ of $d(r)$ in $H_r$. Thus we have that $r$ is reachable from every vertex of $H_r^R$ in $H_r^R\setminus{(x,y)}$, except from $x$, which cannot reach $r$ in $H_r^R$, and possibly $d(r)$. In any case, we have that $d(r)$ is not reachable from $x$ in $H_r^R\setminus{(x,y)}$. By Lemma~\ref{lemma:uv-reachability}, this implies that $r$ is reachable from every vertex of $H_{rr}$ in $H_{rr}\setminus{(x,y)}$, except from $x$, which cannot reach $r$ in $H_r^R$, and possibly $d(r)$. In any case, we have that $d(r)$ is not reachable from $x$ in $H_{rr}\setminus{(x,y)}$. Now, by Lemma~\ref{lemma:bridge-r-reachability} we have that $r$ can reach every vertex of $H_{rr}\setminus{(x,y)}$, except $y$, which is unreachable from $r$ in $H_{rr}\setminus{(x,y)}$. Furthermore, no vertex of $H_{rr}$ can reach $y$ in $H_{rr}\setminus{(x,y)}$, since $(x,y)$ is the only incoming edge of $y$ in $H_{rr}$. We conclude that the strongly connected components of $H_{rr}\setminus{(x,y)}$ are either $\{x\}$, $\{y\}$ and $H_{rr}\setminus\{x,y\}$, or $\{x\}$, $\{y\}$, $\{d(r)\}$ and $H_{rr}\setminus\{x,y,d(r)\}$.
$(2.4)$ Let $x$ be $\mathit{oa}$ and let $y$ be $\mathit{aa}$. This means that both $x$ and $y$ are auxiliary vertices of $H_r$, and so one of them must be $d(r)$. But we have forbidden this case to occur since it was treated in the beginning of $(2)$.

$(3)$ Let $x$ be an auxiliary vertex of $H_{rr}$ and let $y$ be an ordinary vertex of $H_{rr}$. Since $H_{rr}$ has no critical vertex, this means that $x$ is a marked vertex of $H_r^R$ with $d(x)\in T(r)$ in $H_r^R$. Now we
distinguish four different cases, depending on whether $x$ and $y$ are ordinary or auxiliary in $H_r$. In any case, by Lemma~\ref{lemma:r-reachability} we have that $r$ can reach every vertex of $H_{rr}$ in $H_{rr}\setminus{(x,y)}$.
$(3.1)$ Let $x$ be $\mathit{ao}$ and $y$ be $\mathit{oo}$. Let $(x_0,y)$ be an edge of $H_r^R$ that corresponds to $(x,y)$ in $H_{rr}$. Then $(y,x_0)$ is an edge of $H_r$. If $x_0$ is an ordinary vertex of $H_r$, then by Lemma~\ref{lemma:r-reachability} we have that $r$ reaches all vertices of $H_r$ in $H_r\setminus{(y,x_0)}$. This implies that every vertex of $H_r^R$ can reach $r$ in $H_r^R\setminus{(x_0,y)}$. Then Lemma~\ref{lemma:r-reachability-cor_edge} implies that $x$ can reach $r$ in $H_{rr}\setminus{(x,y)}$. Thus, every vertex of $H_{rr}$ can reach $r$ in $H_{rr}\setminus{(x,y)}$, and so we have that $H_{rr}\setminus{(x,y)}$ is strongly connected. Similarly, if $x_0$ is an auxiliary vertex of $H_r$, then, by Lemma~\ref{lemma:bridge-r-reachability}, we have that $r$ reaches all vertices of $H_r$ in $H_r\setminus{(y,x_0)}$, except $x_0$, which is unreachable from $r$ in $H_r\setminus{(y,x_0)}$, and possibly $d(r)$.
This implies that every vertex of $H_r^R$ can reach $r$ in $H_r^R\setminus{(x_0,y)}$, except $x_0$, which cannot reach $r$ in $H_r^R\setminus{(x_0,y)}$, and possibly $d(r)$. Since $x$ is an ordinary vertex of $H_r$, we have that $x\neq x_0$, and so $x$ can reach $r$ in $H_r^R\setminus{(x_0,y)}$. Now Lemma~\ref{lemma:r-reachability-cor_edge} implies that $x$ can reach $r$ in $H_{rr}\setminus{(x,y)}$. Thus, since $H_{rr}$ is strongly connected, we have that every vertex of $H_{rr}$ can reach $r$ in $H_{rr}\setminus{(x,y)}$, and so we have that $H_{rr}\setminus{(x,y)}$ is strongly connected.
%
$(3.2)$ Let $x$ be $\mathit{ao}$ and $y$ be $\mathit{oa}$. Let $\{(x_1,y),\dots,(x_k,y)\}$ be the collection of all edges of $H_r^R$ that correspond to $(x,y)$ in $H_{rr}$. Then $\{(y,x_1),\dots,(y,x_k)\}$ is a set of outgoing edges from $y$ in $H_r$. Thus, Lemma~\ref{lemma:r-reachability2} implies that $r$ can reach all vertices of $H_r$ in $H_r\setminus\{(y,x_1),\dots,(y,x_k)\}$, except possibly $d(r)$. This implies that $r$ can be reached from all vertices of $H_r^R$ in $H_r^R\setminus\{(x_1,y),\dots,(x_k,y)\}$, except possibly from $d(r)$. In particular, there is a path $P$ from $x$ to $r$ in $H_r^R$ that avoids all edges in $\{(x_1,y),\dots,(x_k,y)\}$. Then, by Lemma~\ref{lemma:cor_paths} we have that the corresponding path $P_H$ in $H_{rr}$ avoids $(x,y)$. This means that $r$ is reachable from $x$ in $H_{rr}\setminus{(x,y)}$. Since $H_{rr}$ is strongly connected (by Corollary~\ref{corollary:sc_of_H}), we know that all vertices of $H_{rr}$ can reach $r$. Thus they can reach $r$ by avoiding the edge $(x,y)$. We conclude that $H_{rr}\setminus{(x,y)}$ is strongly connected.
$(3.3)$ Let $x$ be $\mathit{aa}$ and $y$ be $\mathit{oo}$.
%
First, let us assume that $x=d(r)$ (in $G_s$). Since $H_{rr}$ is strongly connected and $(r,d(r))$ is the only incoming edge to $d(r)$ in $H_{rr}$, we have that every vertex from $H_{rr}$ can reach $r$ in $H_{rr}\setminus{(x,y)}$, except possibly $d(r)$. Thus, we have that $H_{rr}\setminus{(x,y)}$ is strongly connected if and only if $d(r)$ has at least two outgoing edges in $H_{rr}$. Otherwise, $(x,y)$ is the only outgoing edge from $d(r)$ in $H_{rr}$, and then the strongly connected components of $H_{rr}\setminus{(x,y)}$ are $\{x\}$ and $H_{rr}\setminus\{x\}$.
Now let us assume that $x\neq d(r)$. Notice that $x$ is a marked vertex of $H_r^R$.
Let $(x_0,y)$ be an edge of $H_r^R$ that corresponds to $(x,y)$ in $H_{rr}$. Then $(y,x_0)$ is an edge of $H_r$. Now, if $x_0$ is ordinary in $H_r$ or it is auxiliary in $H_r$ but $x_0\neq x$, then we can argue as in $(3.1)$ in order to establish that $H_{rr}\setminus{(x,y)}$ is strongly connected. Otherwise, suppose that $x_0=x$. Then, since $x$ is auxiliary in $H_r$ and $y$ is ordinary in $H_r$ (and $x\neq d(r)$), we have that $y=d(x)$ in $G_s$. Now, by Lemma~\ref{lemma:bridge-r-reachability} we have that $r$ reaches all vertices of $H_r$ in $H_r\setminus{(y,x)}$, except $x$, which is unreachable from $r$ in $H_r\setminus{(y,x)}$, and possibly $d(r)$. This implies that every vertex of $H_r^R$ can reach $r$ in $H_r^R\setminus{(x,y)}$, except $x$, which cannot reach $r$ in $H_r^R\setminus{(x,y)}$, and possibly $d(r)$. Now Lemma~\ref{lemma:r-reachability-cor_edge} implies that every vertex of $H_{rr}$ can reach $r$ in $H_{rr}\setminus{(x,y)}$, except possibly $x$ and $d(r)$. 
But since $H_{rr}$ is strongly connected (by Corollary~\ref{corollary:sc_of_H}) and there is no edge $(d(r),x)$ in $H_{rr}$ (since $x$ and $d(r)$ are both auxiliary, but not critical, vertices of $H_{rr}$), we also have that $d(r)$ reaches $r$ in $H_{rr}\setminus{(x,y)}$.
Thus we conclude that either $H_{rr}\setminus{(x,y)}$ is strongly connected, or its strongly connected components are $\{x\}$ and $H_{rr}\setminus\{x\}$.
$(3.4)$ Let $x$ be $\mathit{aa}$ and $y$ be $\mathit{oa}$. If $x\neq d(r)$, then we can argue as in $(3.2)$ in order to conclude that $H_{rr}\setminus{(x,y)}$ is strongly connected. Otherwise, let $x=d(r)$.
Then, since $H_{rr}$ is strongly connected (by Corollary~\ref{corollary:sc_of_H}), we have that $r$ is reachable from every vertex of $H_{rr}$. Since $d(r)$ has only one incoming edge $(r,d(r))$ in $H_{rr}$, this implies that every vertex of $H_{rr}\setminus\{d(r)\}$ can reach $r$ by avoiding $d(r)$. Thus, if $d(r)$ has more than one outgoing edge in $H_{rr}$, we have that $H_{rr}\setminus{(x,y)}$ is strongly connected. Otherwise, the strongly connected components of $H_{rr}\setminus{(x,y)}$ are $\{d(r)\}$ and $H_{rr}\setminus\{d(r)\}$.

$(4)$ It is impossible that both $x$ and $y$ are auxiliary vertices of $H_{rr}$, since $H_{rr}$ does not contain a critical vertex.
\end{proof}

We can simplify the result of Lemma~\ref{lemma:Hrr} by essentially eliminating cases $(ii)$ and $(vi)$. This is done by observing that, in those cases, the only outgoing edges from $d(r)$ have the form $(d(r),x)$, where $x$ is an $\mathit{oa}$ vertex in $H_{rr}$. To achieve the simplification, we remove $d(r)$ from $H_{rr}$, and we add a new edge $(r,x)$. This is in order to replace the path $(r,d(r)),(d(r),x)$ of $H_{rr}$ with $(r,x)$. We call the resulting graph $\widetilde{H}_{rr}$. The following two lemmas demonstrate that $\widetilde{H}_{rr}$ maintains all the information that we need for our purposes.

\begin{lemma}
\label{lemma:tilde_hrr_2etsc}
For any two $\mathit{oo}$ vertices $u,v$ of $\widetilde{H}_{rr}$ we have $u\stackrel[]{\widetilde{H}_{\mathit{rr}}}\leftrightarrow_{2et}v$ if and only if $u\stackrel[]{H_{\mathit{rr}}}\leftrightarrow_{2et}v$.
\end{lemma}
\begin{proof}
First, observe that $x$ has only one outgoing edge in $H_{rr}$. This is because $x$ is an auxiliary vertex of $H_r$, and thus (since $x\neq d(r)$) it has only one incoming edge in $H_r$. This means that $x$ has only one outgoing edge in $H_r^R$, and thus it has only one outgoing edge in $H_{rr}$, since it is ordinary in $H_{rr}$.

Now let $u,v$ be two $\mathit{oo}$ vertices of $\widetilde{H}_{rr}$. Suppose first that $u\stackrel[]{H_{\mathit{rr}}}\leftrightarrow_{2et}v$. Let $e$ be an edge of $\widetilde{H}_{rr}$ different from $(r,x)$. Then $e$ is also an edge of $H_{rr}$, and $u,v$ are twinless strongly connected in $H_{rr}\setminus{e}$. This means that there is a path $P$ in $H_{rr}\setminus{e}$ from $u$ to $v$, and a path $Q$ in $H_{rr}\setminus{e}$ from $v$ to $u$, such that there is no edge $(z,w)\in P$ with $(w,z)\in Q$. Now, if $P$ and $Q$ do not pass from $d(r)$, then they are paths in $\widetilde{H}_{rr}$, and this implies that $u,v$ are twinless strongly connected in $\widetilde{H}_{rr}\setminus{e}$. Otherwise, at least one of $P,Q$  has $(r,d(r)),(d(r),x)$ as a subpath. Then we replace every occurrence of this subpath with $(r,x)$. If the outgoing edge of $x$ in $\widetilde{H}_{rr}$ is not $(x,r)$, then we still have that $P$ and $Q$ do not share a pair of antiparallel edges, and so $u,v$ are twinless strongly connected in $\widetilde{H}_{rr}$. Otherwise, since $P$ and $Q$ must end in an $\mathit{oo}$ vertex, for every occurrence of $d(r)$ in them they must pass from the path $(r,d(r)),(d(r),x),(x,r)$. Then we can simply discard every occurrence of this triangle. This shows that $u,v$ are twinless strongly connected in $\widetilde{H}_{rr}\setminus{e}$. Now let $e=(r,x)$. Since $u\stackrel[]{H_{\mathit{rr}}}\leftrightarrow_{2et}v$, we have that $u,v$ are twinless strongly connected in $H_{rr}\setminus{(r,d(r))}$. This means that there is a path $P$ in $H_{rr}\setminus{(r,d(r))}$ from $u$ to $v$, and a path $Q$ in $H_{rr}\setminus{(r,d(r))}$ from $v$ to $u$, such that there is no edge $(z,w)\in P$ with $(w,z)\in Q$. Since $(r,d(r))$ is the only incoming edge to $d(r)$ in $H_{rr}$, this means that $P,Q$ do not pass from $d(r)$. Also, $P,Q$ do not use the edge $(r,x)$. This means that $u,v$ are twinless strongly connected in $\widetilde{H}_{rr}\setminus{(r,x)}$. Thus we have $u\stackrel[]{\widetilde{H}_{\mathit{rr}}}\leftrightarrow_{2et}v$.

Conversely, suppose that $u,v$ are not $2$-edge twinless strongly connected in $H_{rr}$. Then there is an edge $e\in H_{rr}$ and a pair of antiparallel edges $(z,w),(w,z)\in H_{rr}$, such that for every path $P$ in $H_{rr}\setminus{e}$ from $u$ to $v$ and every path $Q$ in $H_{rr}\setminus{e}$ from $v$ to $u$, we have $(z,w)\in P$ and $(w,z)\in Q$. Suppose first that $e$ is not one of $(r,d(r)),(d(r),x)$. Now let us assume, for the sake of contradiction, that there is a path $P$ in $\widetilde{H}_{rr}\setminus{e}$ from $u$ to $v$ that does not contain $(z,w)$. Then we can construct a path $P'$ in $H_{rr}$ by replacing every occurrence of $(r,x)$ in $P$ with the segment $(r,d(r)),(d(r),x)$. Observe that $(z,w)$ is neither $(r,d(r))$ nor $(d(r),x)$, because $(w,z)$ is also an edge of $H_{rr}$ (and there cannot be an edge $(d(r),r)$, because $(d(r),x)$ is the only outcoming edge of $d(r)$; and also there cannot be an edge $(x,d(r))$, because $(r,d(r))$ is the only incoming edge to $d(r)$ in $H_{rr}$, and $x\neq r$ because $x$ is auxiliary in $H_r$ but $r$ is ordinary in $H_r$). Thus $P'$ is a path in $H_{rr}\setminus{e}$ from $u$ to $v$ that does not use $(z,w)$ - a contradiction. Similarly, we can show that there is no path $Q$ in $\widetilde{H}_{rr}\setminus{e}$ from $v$ to $u$ that does not use $(w,z)$. This means that $u,v$ are not twinless strongly connected in $\widetilde{H}_{rr}\setminus{e}$. Now suppose that $e$ is either $(r,d(r))$ or $(d(r),x)$. Then we can see that $u,v$ are not twinless strongly connected in $\widetilde{H}_{rr}\setminus{(r,x)}$, because every path in $\widetilde{H}_{rr}\setminus{(r,x)}$ from $u$ to $v$, or from $v$ to $u$, is also a path in $H_{rr}\setminus{e}$.
\end{proof}

\begin{lemma}
\label{lemma:tilde_hrr_scc}
Let $e$ be an edge of $\widetilde{H}_{rr}$. Then $e$ is a strong bridge of $\widetilde{H}_{rr}$ if and only if it is a strong bridge of $H_{rr}$. The strongly connected components of $\widetilde{H}_{rr}\setminus{e}$ are given essentially by Lemma~\ref{lemma:Hrr} (where we replace ``$H_{rr}$'' with ``$\widetilde{H}_{rr}$'').
\end{lemma}
\begin{proof}
First observe that $(r,x)$ is not a strong bridge of $\widetilde{H}_{rr}$, because $x$ is an ordinary vertex of $H_{rr}$, and so, by removing any edge from $H_{rr}$, $r$ can still reach $x$ in $H_{rr}$. In particular, by removing $(r,d(r))$ or $(d(r),x)$ from $H_{rr}$, $r$ can still reach $x$. From this, we can conclude that $(r,x)$ is not a strong bridge of $\widetilde{H}_{rr}$.

Now let $e$ be an edge of $\widetilde{H}_{rr}$, but not $(r,x)$. Then $e$ is also an edge of $H_{rr}$. Let $u,v$ be two vertices of $\widetilde{H}_{rr}$. We will show that $u$ can reach $v$ in $\widetilde{H}_{rr}\setminus{e}$ if and only if $u$ can reach $v$ in $H_{rr}\setminus{e}$. Let $P$ be a path in $\widetilde{H}_{rr}\setminus{e}$ from $u$ to $v$. We can construct a path in $H_{rr}\setminus{e}$ by replacing every occurrence of $(r,x)$ in $P$ with the segment $(r,d(r)),(d(r),x)$. This shows that $u$ can reach $v$ in $H_{rr}\setminus{e}$. Conversely, let $P$ be a path in $H_{rr}\setminus{e}$ from $u$ to $v$. If $P$ passes from $d(r)$, then it must necessarily use the segment $(r,d(r)),(d(r),x))$ (because $(r,d(r))$ is the only incoming edge to $d(r)$ in $H_{rr}$, and $(d(r),x)$ is the only king of outgoing edge from $d(r)$ in $H_{rr}$). Then we can construct a path in $\widetilde{H}_{rr}\setminus{r}$ by replacing every occurrence of the segment $(r,d(r)),(d(r),x))$ with $(r,x)$. This shows that $u$ can reach $v$ in $\widetilde{H}_{rr}\setminus{e}$. Thus we have that the strongly connected components of $\widetilde{H}_{rr}\setminus{e}$ coincide with those of $H_{rr}\setminus{e}$ (by excluding $d(r)$ from the strongly connected component of $H_{rr}\setminus{e}$ that contains $d(r)$).
\end{proof}

Before we move on to describe the structure of the strongly connected components of the remaining graphs, we will need the following two lemmas.

\begin{lemma}
\label{lemma:dr-reachability}
Let $G_s$ be a strongly connected digraph and let $r$ be a marked vertex of $G_s$. Let $(x,y)$ be an edge of $H_r$ such that either both $x$ and $y$ are ordinary in $H_r$, or $y$ is auxiliary in $H_r$ and $x=d(y)$ in $G_s$. Let $u$ be a vertex of $H_r$. If $u$ reaches $s$ in $G_s\setminus{(x,y)}$, then $u$ reaches $d(r)$ in $H_r\setminus{(x,y)}$.
\end{lemma}
\begin{proof}
 Let $P$ be a path from $u$ to $s$ in $G_s\setminus{(x,y)}$. Since $r$ is a marked vertex of $G_s$, by Lemma~\ref{lemma:internal_paths} we have that there is a path $Q$ from $s$ to $d(r)$ that avoids every vertex of $D(r)$. In particular, $Q$ does not use the edge $(x,y)$. Let $P'=P+Q$ be the concatenation of $P$ and $Q$ in $G_s$. This is a path from $u$ to $d(r)$ in $G_s\setminus{(x,y)}$. Now consider the corresponding path $P_H$ of $P'$ in $H_r$. By Lemma~\ref{lemma:cor_paths}, this is a path from $u$ to $d(r)$ in $H_r$ that avoids $(x,y)$.
\end{proof}


\begin{lemma}
\label{lemma:dr-reachability-cor_edge}
Let $G_s$ be a strongly connected digraph and let $r$ be a marked vertex of $G_s$. Let $x$ be a marked vertex of $G_s$ such that $d(x)\in T(r)$. Let also $(x_0,y)$ be an edge of $G_s$ such that $x_0$ is a descendant of $x$ in $D(G_s)$, $y$ is not a descendant of $x$ in $D(G_s)$, and $y\in T(r)$. Let $u$ be a vertex of $H_r$. If $u$ reaches $s$ in $G_s\setminus{(x_0,y)}$, then $u$ reaches $d(r)$ in $H_r\setminus{(x,y)}$.
\end{lemma}
\begin{proof}
Let $u$ be a vertex of $H_r$ and let $P$ be a path from $u$ to $s$ in $G_s\setminus{(x_0,y)}$. Since $r$ is a marked vertex of $G_s$, by Lemma~\ref{lemma:internal_paths} we have that there is a path $Q$ from $s$ to $d(r)$ that avoids every vertex of $D(r)$. In particular, $Q$ does not use the edge $(x_0,y)$. Let $P'=P+Q$ be the concatenation of $P$ and $Q$ in $G_s$. This is a path from $u$ to $d(r)$ in $G_s\setminus{(x,y)}$. Now consider the corresponding path $P_H$ of $P'$ in $H_r$. $P_H$ has the same endpoints as $P'$. Thus, if $P_H$ avoids the edge $(x,y)$ of $H_r$, then we are done. Otherwise, this means that $P'$ uses an edge $(x_1,y)$ of $G_s$, distinct from $(x_0,y)$ (although we may have $x_1=x_0$, since we allow multiple edges), such that $x_1$ is a descendant of $x$ in $D(G_s)$. But then we must have that $(x,y)$ is a double edge of $H_r$, and therefore $H_r\setminus{(x,y)}$ is strongly connected (since $H_r$ is strongly connected, by Corollary~\ref{corollary:sc_of_H}). Thus $u$ can reach $d(r)$ in $H_r\setminus{(x,y)}$.
\end{proof}

\begin{lemma}
\label{lemma:Hsr}
Let $G_s$ be a strongly connected digraph, and let $r$ be a marked vertex of $H_s^R$. Let $(x,y)$ be a strong bridge of $H_{sr}$ where none of $x$ and $y$ is $d(r)$ in $H_s^R$. Then the strongly connected components of $H_{sr}\setminus{(x,y)}$ are given by one of the following:
\begin{enumerate}[label={(\roman*)}]
\item{$\{x\}$ and $H_{sr}\setminus\{x\}$, where $x$ is auxiliary in $H_s$ and $y=d(x)$ in $G_s$ (see case $(1.3)$ below)}
\item{$\{x\}$ and $H_{sr}\setminus\{x\}$, where $x$ is auxiliary in $H_{sr}$ and $(x,y)$ is the only outgoing edge of $x$ in $H_{sr}$ (see case $(3.3)$ below)}
\item{$\{y\}$ and $H_{sr}\setminus\{y\}$, where $y$ is auxiliary in $H_{sr}$ and $x=d(y)$ in $H_s^R$ (see cases $(2.1)$ and $(2.2)$ below)}
\item{$\{x\}$, $\{y\}$ and $H_{sr}\setminus\{x,y\}$, where $x$ is auxiliary in $H_s$, $y$ is auxiliary in $H_{sr}$, $x=d(y)$ in $H_s^R$ and $y=d(x)$ in $G_s$ (see case $(2.3)$ below)}
\end{enumerate}
\end{lemma}
\begin{proof}
Let $(x,y)$ be an edge of $H_{sr}$, where none of $x$ and $y$ is $d(r)$ in $H_s^R$. We distinguish four different cases, depending on whether $x$ and $y$ are ordinary or auxiliary vertices of $H_{sr}$.

$(1)$ Let both $x$ and $y$ be ordinary vertices of $H_{sr}$. This implies that $(y,x)$ is an edge in $H_s$. Here we distinguish four cases, depending on whether $x$ and $y$ are ordinary or auxiliary in $H_s$. $(1.1)$ Let $x$ be $\mathit{oo}$ and $y$ be $\mathit{oo}$. Then, by Lemma~\ref{lemma:r-reachability} we have that $s$ can reach all vertices of $H_s$ in $H_s\setminus{(y,x)}$. This implies that $s$ can be reached by all vertices of $H_s^R$ in $H_s^R\setminus{(x,y)}$. Then Lemma~\ref{lemma:dr-reachability} implies that all vertices of $H_{sr}$ can reach $d(r)$ in $H_{sr}\setminus{(x,y)}$. Conversely, as an implication of Lemma~\ref{lemma:r-reachability}, we have that $d(r)$ can reach every vertex of $H_{sr}$ in $H_{sr}\setminus{(x,y)}$. We conclude that $H_{sr}\setminus{(x,y)}$ is strongly connected. $(1.2)$ Let $x$ be $\mathit{oo}$ and $y$ be $\mathit{oa}$. Then we can argue as in $(1.1)$ in order to conclude that $H_{sr}\setminus{(x,y)}$ is strongly connected. $(1.3)$ Let $x$ be $\mathit{oa}$ and $y$ be $\mathit{oo}$. Since $H_s$ contains no critical vertex, we have that $x$ is a marked vertex of $G_s$ and $y=d(x)$ in $G_s$. Then, by Lemma~\ref{lemma:bridge-r-reachability} we have that $s$ can reach all vertices of $H_s$ in $H_s\setminus{(y,x)}$, except $x$, which is unreachable from $s$ in $H_s\setminus{(y,x)}$. This implies that $s$ is reachable from all vertices of $H_s^R$ in $H_s^R\setminus{(x,y)}$, except from $x$, which cannot reach $s$ in $H_s^R\setminus{(x,y)}$. Then Lemma~\ref{lemma:dr-reachability} implies that all vertices of $H_{sr}$ can reach $d(r)$ in $H_{sr}\setminus{(x,y)}$, except possibly $x$. Now, as an implication of Lemma~\ref{lemma:r-reachability}, we have that $d(r)$ can reach all vertices of $H_{sr}$ in $H_{sr}\setminus{(x,y)}$. We conclude that either $H_{sr}\setminus{(x,y)}$ is strongly connected, or its strongly connected components  are $\{x\}$ and $H_{sr}\setminus\{x\}$. $(1.4)$ Let $x$ be $\mathit{oa}$ and $y$ be $\mathit{oa}$. But this implies that both $x$ and $y$ are auxiliary vertices of $H_s$, which is impossible to occur.

$(2)$ Let $x$ be an ordinary vertex of $H_{sr}$ and let $y$ be an auxiliary vertex of $H_{sr}$. Since $y\neq d(r)$ in $H_s^R$, this means that $y$ is a marked vertex of $H_s^R$ and $x=d(y)$ in $H_s^R$. Thus $(y,x)$ is an edge of $H_s$. Now we distinguish four different cases, depending on whether $x$ and $y$ are ordinary or auxiliary in $H_s$. $(2.1)$ Let $x$ be $\mathit{oo}$ and $y$ be $\mathit{ao}$. Then we can argue as in $(1.1)$ in order to show that $d(r)$ is reachable from every vertex of $H_{sr}$ in $H_{sr}\setminus{(x,y)}$. Now, as an implication of Lemma~\ref{lemma:bridge-r-reachability}, we have that $d(r)$ reaches every vertex of $H_{sr}$ in $H_{sr}\setminus{(x,y)}$, except $y$, which is unreachable from $d(r)$ in $H_{sr}\setminus{(x,y)}$. Thus we conclude that the strongly connected components of $H_{sr}\setminus{(x,y)}$ are $\{y\}$ and $H_{sr}\setminus\{y\}$. $(2.2)$ Let $x$ be $\mathit{oo}$ and $y$ be $\mathit{aa}$. Then we can argue as in $(2.1)$ in order to conclude that the strongly connected components of $H_{sr}\setminus{(x,y)}$ are $\{y\}$ and $H_{sr}\setminus\{y\}$. $(2.3)$. Let $x$ be $\mathit{oa}$ and $y$ be $\mathit{ao}$. Then, by Lemma~\ref{lemma:bridge-r-reachability} we have that $s$ can reach all vertices of $H_s$ in $H_s\setminus{(y,x)}$, except $x$, which is unreachable from $s$ in $H_s\setminus{(y,x)}$. This implies that $s$ can be reached by all vertices of $H_s^R$ in $H_s^R\setminus{(x,y)}$, except from $x$, which cannot reach $s$ in $H_s^R\setminus{(x,y)}$. By Lemma~\ref{lemma:dr-reachability}, this implies that $d(r)$ is reachable from all vertices of $H_{sr}$ in $H_{sr}\setminus{(x,y)}$, except from $x$, which cannot reach $d(r)$ in $H_{sr}\setminus{(x,y)}$. Now, as an implication of Lemma~\ref{lemma:bridge-r-reachability} we have that $d(r)$ can reach all vertices of $H_{sr}$ in $H_{sr}\setminus{(x,y)}$, except $y$, which is unreachable from $d(r)$ in $H_{sr}\setminus{(x,y)}$. We conclude that the strongly connected components of $H_{sr}\setminus{(x,y)}$ are $\{x\}$, $\{y\}$ and $H_{sr}\setminus\{x,y\}$.
$(2.4)$ Let $x$ be $\mathit{oa}$ and $y$ be $\mathit{aa}$. But this implies that both $x$ and $y$ are auxiliary vertices of $H_s$, which is impossible to occur.

$(3)$ Let $x$ be an auxiliary vertex of $H_{sr}$ and let $y$ be an ordinary vertex of $H_{sr}$. Since $x\neq d(r)$ in $H_s^R$, this means that $x$ is a marked vertex of $H_s^R$ with $d(x)\in T(r)$ in $H_s^R$. Now we distinguish four different cases, depending on whether $x$ and $y$ are ordinary or auxiliary in $H_s$. In any case, as an implication of Lemma~\ref{lemma:r-reachability} we have that $d(r)$ can reach every vertex of $H_{sr}$ in $H_{sr}\setminus{(x,y)}$.
$(3.1)$ Let $x$ be $\mathit{ao}$ and $y$ be $\mathit{oo}$. Let $(x_0,y)$ be an edge of $H_s^R$ that corresponds to $(x,y)$ in $H_{sr}$. Then $(y,x_0)$ is an edge of $H_s$. If $x_0$ is an ordinary vertex of $H_s$, then by Lemma~\ref{lemma:r-reachability} we have that $s$ reaches all vertices of $H_s$ in $H_s\setminus{(y,x_0)}$. This implies that every vertex of $H_s^R$ can reach $s$ in $H_s^R\setminus{(x_0,y)}$.
%
Then Lemma~\ref{lemma:dr-reachability-cor_edge} implies that every vertex of $H_{sr}$ can reach $d(r)$ in $H_{sr}\setminus{(x,y)}$, and so we have that $H_{sr}\setminus{(x,y)}$ is strongly connected.
Similarly, if $x_0$ is an auxiliary vertex of $H_s$, then, by Lemma~\ref{lemma:bridge-r-reachability}, we have that $s$ reaches all vertices of $H_s$ in $H_s\setminus{(y,x_0)}$, except $x_0$, which is unreachable from $s$ in $H_s\setminus{(y,x_0)}$. This implies that every vertex of $H_s^R$ can reach $s$ in $H_s^R\setminus{(x_0,y)}$, except $x_0$, which cannot reach $s$ in $H_s^R\setminus{(x_0,y)}$. Since $x$ is an ordinary vertex of $H_s$, we have that $x\neq x_0$, and so $x$ can reach $s$ in $H_s^R\setminus{(x_0,y)}$.
Then Lemma~\ref{lemma:dr-reachability-cor_edge} implies that every vertex of $H_{sr}$ can reach $d(r)$ in $H_{sr}\setminus{(x,y)}$, and so we have that $H_{sr}\setminus{(x,y)}$ is strongly connected.
$(3.2)$ Let $x$ be $\mathit{ao}$ and $y$ be $\mathit{oa}$. Let $\{(x_1,y),\dots,(x_k,y)\}$ be the set of all the incoming edges to $y$ in $H_s^R$. Then $\{(y,x_1),\dots,(y,x_k)\}$ is a collection of edges of $H_s$ that stem from $y$. Since $H_s$ contains no critical vertex, we have that $y$ is a marked vertex of $G_s$ with $d(y)\in T(s)$. By Lemma~\ref{lemma:r-reachability2} we have that $s$ can reach all vertices of $H_s$ in $H_s\setminus\{(y,x_1),\dots,(y,x_k)\}$. This implies that $s$ can be reached from all vertices of $H_s^R$ in $H_s^R\setminus\{(x_1,y),\dots,(x_k,y)\}$. Furthermore, since $y$ is dominated by $d(r)$ in $H_s^R$, we have that there is a path from $s$ to $d(r)$ in $H_s^R$ that avoids $y$; therefore, this path also avoids all edges $\{(x_1,y),\dots,(x_k,y)\}$. Thus we infer that $d(r)$ is reachable from all vertices of $H_s^R$ in $H_s^R\setminus\{(x_1,y),\dots,(x_k,y)\}$. In particular, let $P$ be a path from $x$ to $d(r)$ in $H_s^R\setminus\{(x_1,y),\dots,(x_k,y)\}$. Then, by Lemma~\ref{lemma:cor_paths} the corresponding path $P_H$ in $H_{sr}$ has the same endpoints as $P$ and avoids the edge $(x,y)$. Thus, even by removing $(x,y)$ from $H_{sr}$, all vertices from $H_{sr}$ can reach $d(r)$. We conclude that $H_{sr}\setminus{(x,y)}$ is strongly connected.
$(3.3)$ Let $x$ be $\mathit{aa}$ and $y$ be $\mathit{oo}$. Let $(x_0,y)$ be an edge of $H_s^R$ that corresponds to $(x,y)$ in $H_{sr}$. Then $(y,x_0)$ is an edge of $H_s$. Now, if $x_0$ is ordinary in $H_s$ or it is auxiliary in $H_s$ but $x_0\neq x$, then we can argue as in $(3.1)$ in order to establish that $H_{sr}\setminus{(x,y)}$ is strongly connected. Otherwise, suppose that $x_0=x$. Then, since $x$ is auxiliary in $H_s$ and $y$ is ordinary in $H_s$ (and $H_s$ contains no critical vertex), we have that $y=d(x)$ in $G_s$. Now, by Lemma~\ref{lemma:bridge-r-reachability} we have that $s$ reaches all vertices of $H_s$ in $H_s\setminus{(y,x)}$, except $x$, which is unreachable from $s$ in $H_s\setminus{(y,x)}$. This implies that every vertex of $H_s^R$ can reach $s$ in $H_s^R\setminus{(x,y)}$, except $x$, which cannot reach $s$ in $H_s^R\setminus{(x,y)}$. Now Lemma~\ref{lemma:dr-reachability-cor_edge} implies that every vertex of $H_{sr}$ can reach $d(r)$ in $H_{sr}\setminus{(x,y)}$, except possibly $x$. Thus, we have that either $H_{sr}\setminus{(x,y)}$ is strongly connected, or its strongly connected components are $\{x\}$ and $H_{sr}\setminus\{x\}$.
$(3.4)$ Let $x$ be $\mathit{aa}$ and $y$ be $\mathit{oa}$. Then we can use the same argument that we used in $(3.2)$ in order to conclude that $H_{sr}\setminus{(x,y)}$ is strongly connected.

$(4)$ Since $y\neq d(r)$ in $H_s^R$, it is impossible that both $x$ and $y$ are auxiliary in $H_{sr}$.
\end{proof}

\begin{lemma}
\label{lemma:Hrr'}
Let $G_s$ be a strongly connected digraph, let $r$ be a marked vertex of $G_s$, and let $r'$ be a marked vertex of $H_r^R$. Let $(x,y)$ be a strong bridge of $H_{rr'}$ such that neither $x$ nor $y$ is $d(r')$ in $H_r^R$. Furthermore, if $r'=d(r)$ in $G_s$, then we also demand that neither $x$ nor $y$ is $r'$.  Then the strongly connected components of $H_{rr'}\setminus{(x,y)}$ are given by one of the following:
\begin{enumerate}[label={(\roman*)}]
\item{$\{x\}$ and $H_{rr'}\setminus\{x\}$, where $x$ is auxiliary in $H_r$ and $y=d(x)$ in $G_s$ (see case $(1.3)$ below)}
\item{$\{x\}$ and $H_{rr'}\setminus\{x\}$, where $x$ is auxiliary in $H_{rr'}$ and $(x,y)$ is the only outgoing edge of $x$ in $H_{rr'}$ (see case $(3.3)$ below)}
\item{$\{y\}$ and $H_{rr'}\setminus\{y\}$, where $y$ is auxiliary in $H_{rr'}$ and $x=d(y)$ in $H_r^R$ (see cases $(2.1)$ and $(2.2)$ below)}
\item{$\{y\}$ and $H_{rr'}\setminus\{y\}$, where $x$ is auxiliary in $H_{r}$ and $y=d(x)$ in $G_s$ (see case $(2.3)$ below)}
\item{$\{x\}$, $\{y\}$ and $H_{rr'}\setminus\{x,y\}$, where $x$ is auxiliary in $H_r$, $y$ is auxiliary in $H_{rr'}$, $x=d(y)$ in $H_r^R$ and $y=d(x)$ in $G_s$ (see case $(2.3)$ below)}
\item{$\{x\}$, $\{y\}$, $\{d(r)\}$ and $\{r\}$, where $r'=d(r)$, $x$ is auxiliary in $H_r$, $y$ is auxiliary in $H_{rr'}$, $x=d(y)$ in $H_r^R$, $y=d(x)$ in $G_s$, and $(d(r),x)$ is the only kind of outgoing edge of $d(r)$ in $H_{rr'}$ (see case $(2.3)$ below)}
\end{enumerate}
\end{lemma}
\begin{proof}
Let $(x,y)$ be an edge of $H_{rr'}$ such that neither $x$ nor $y$ is $d(r')$ in $H_r^R$, and also, if $r'=d(r)$ in $G_s$, then neither $x$ nor $y$ is $r'$. Notice that, if $r'=d(r)$ in $G_s$, then $r'$ is auxiliary in $H_r$, and therefore $r'$ is an $\mathit{oa}$-vertex of $H_{rr'}$. Conversely, $d(r)$ appears as an ordinary vertex in $H_{rr'}$ only when $r'=d(r)$. Therefore, if $x$ or $y$ is ordinary in $H_{rr'}$, then we can immediately infer, due to our demand in the statement of the Lemma, that $x$ or $y$, respectively, is not $d(r)$. Now we distinguish four cases, depending on whether $x$ and $y$ are ordinary or auxiliary in $H_{rr'}$.

$(1)$ Let both $x$ and $y$ be ordinary vertices of $H_{rr'}$. This implies that $(x,y)$ is also an edge of $H_r^R$, and therefore $(y,x)$ is an edge of $H_r$. Now we distinguish four cases, depending on whether $x$ and $y$ are ordinary or auxiliary in $H_r$. Notice that, in any case, as an implication of Lemma~\ref{lemma:r-reachability}, we have that $d(r')$ reaches all vertices of $H_{rr'}$ in $H_{rr'}\setminus{(x,y)}$.
$(1.1)$ Let $x$ be $\mathit{oo}$ and $y$ be $\mathit{oo}$. Then, as an implication of Lemma~\ref{lemma:r-reachability}, we have that $d(r)$ reaches all vertices of $H_r$ in $H_r\setminus{(y,x)}$. This implies that all vertices of $H_r^R$ can reach $d(r)$ in $H_r^R\setminus{(x,y)}$. By Lemma~\ref{lemma:dr-reachability}, this implies that all vertices of $H_{rr'}$ can reach $d(r')$ in $H_{rr'}$ in $H_{rr'}\setminus{(x,y)}$. We conclude that $H_{rr'}\setminus{(x,y)}$ is strongly connected.
$(1.2)$ Let $x$ be $\mathit{oo}$ and $y$ be $\mathit{oa}$. Then we can argue as in $(1.1)$ in order to conclude that $H_{rr'}\setminus{(x,y)}$ is strongly connected.
$(1.3)$ Let $x$ be $\mathit{oa}$ and $y$ be $\mathit{oo}$. Since $x\neq d(r)$, we must have that $x$ is a marked vertex of $G_s$ and $y=d(x)$ in $G_s$. Then, by Lemma~\ref{lemma:bridge-r-reachability}, we have that $r$ can reach every vertex of $H_r$ in $H_r\setminus{(y,x)}$, except $x$, which is unreachable from $r$ in $H_r\setminus{(y,x)}$, and possibly $d(r)$. This implies that $r$ is reachable from every vertex of $H_r^R$ in $H_r^R\setminus{(x,y)}$, except from $x$, which cannot reach $r$ in $H_r^R\setminus{(x,y)}$, and possibly $d(r)$.
Now let us assume, for the sake of contradiction, that $d(r)$ cannot reach $r$ in $H_r^R\setminus{(x,y)}$, and also that $d(r)\in H_{rr'}$. Since $r$ is reachable from every vertex of $H_r^R\setminus\{x,d(r)\}$ in $H_r^R\setminus{(x,y)}$, we must have that $(d(r),x)$ is the only kind of outgoing edge of $d(r)$ in $H_r^R$. Now, since $x\in H_{rr'}$ and the immediate dominator of $d(r)$ in $H_r^R$ is $r$, we have that $x$ is dominated by $d(r)$ in $H_r^R$; and since $(d(r),x)$ is an edge of $H_r^R$, we must have that $d(r)$ is the immediate dominator of $x$ in $H_r^R$. Now, since $y\in H_{rr'}$, we must also have that $y$ is dominated by $d(r)$ in $H_r^R$. And then, since $(d(r),x)$ is the only kind of outgoing edge of $d(r)$ in $H_r^R$, and $(x,y)$ is the only outgoing edge of $x$ in $H_r^R$ (since $x$ is marked in $G_s$ and $y=d(x)$ in $G_s$), we must have that $(x,y)$ is a bridge of $H_r^R$, and so $y$ is a marked vertex of $H_r^R$ and $x$ is the immediate dominator of $y$ in $H_r^R$. Since $y$ is a marked vertex of $H_r^R$, but also $y$ is ordinary in $H_{rr'}$, we must have that $r'=y$. But then $x$ is the critical vertex of $r'$ in $H_{rr'}$ - which contradicts the fact that $x$ is ordinary in $H_{rr'}$.
Thus we have that either $d(r)$ can reach $r$ in $H_r^R\setminus{(x,y)}$, or that $d(r)\notin H_{rr'}$. In any case, since $r$ is reachable from every vertex of $H_r^R$ in $H_r^R\setminus{(x,y)}$, except from $x$, which cannot reach $r$ in $H_r^R\setminus{(x,y)}$, by Lemma~\ref{lemma:dr-reachability} we must have that $d(r')$ is reachable from every vertex of $H_{rr'}$ in $H_{rr'}\setminus{(x,y)}$, except possibly from $x$. We conclude that either $H_{rr'}\setminus{(x,y)}$ is strongly connected, or its strongly connected components are $\{x\}$ and $H_{rr'}\setminus\{x\}$.
$(1.4)$ Let $x$ be $\mathit{oa}$ and $y$ be $\mathit{oa}$. This means that both $x$ and $y$ are auxiliary in $H_r$, which implies that $y=d(r)$ in $G_s$. But since $y$ is ordinary in $H_{rr'}$, this case is not permitted to occur.

$(2)$ Let $x$ be an ordinary vertex of $H_{rr'}$ and let $y$ be an auxiliary vertex of $H_{rr'}$. Since $y\neq d(r')$ in $H_r^R$, this means that $y$ is a marked vertex of $H_r^R$ and $x=d(y)$ in $H_r^R$. Thus $(y,x)$ is an edge of $H_r$. Now we distinguish four different cases, depending on whether $x$ and $y$ are ordinary or auxiliary in $H_r$.
$(2.1)$ Let $x$ be $\mathit{oo}$ and $y$ be $\mathit{ao}$ in $H_{rr'}$. Then we can argue as in $(1.1)$ in order to get that all vertices of $H_{rr'}$ can reach $d(r')$ in $H_{rr'}$ in $H_{rr'}\setminus{(x,y)}$. Now, as an implication of Lemma~\ref{lemma:bridge-r-reachability}, we have that $d(r')$ reaches all vertices of $H_{rr'}$ in $H_{rr'}\setminus{(x,y)}$, except $y$, which is unreachable from $d(r')$ in $H_{rr'}\setminus{(x,y)}$. We conclude that the strongly connected components of $H_{rr'}\setminus{(x,y)}$ are $\{y\}$ and $H_{rr'}\setminus\{y\}$.
$(2.2)$ Let $x$ be $\mathit{oo}$ and $y$ be $\mathit{ao}$ in $H_{rr'}$. Then we can argue as in $(2.1)$ in order to conclude that the strongly connected components of $H_{rr'}\setminus{(x,y)}$ are $\{y\}$ and $H_{rr'}\setminus\{y\}$.
$(2.3)$ Let $x$ be $\mathit{oa}$ and $y$ be $\mathit{ao}$. First observe that, as an implication of Lemma~\ref{lemma:bridge-r-reachability}, we have that $d(r')$ reaches all vertices of $H_{rr'}$ in $H_{rr'}\setminus{(x,y)}$, except $y$, which is unreachable from $d(r')$ in $H_{rr'}\setminus{(x,y)}$. Now, we can argue as in $(1.3)$ in order to get that $r$ is reachable from every vertex of $H_r^R$ in $H_r^R\setminus{(x,y)}$, except from $x$, which cannot reach $r$ in $H_r^R\setminus{(x,y)}$, and possibly $d(r)$. Then, if we have that either $d(r)$ can also reach $r$ in $H_r^R\setminus{(x,y)}$, or $d(r)\notin H_{rr'}$, by Lemma~\ref{lemma:dr-reachability} we must have that $d(r')$ is reachable from every vertex of $H_{rr'}$ in $H_{rr'}\setminus{(x,y)}$, except possibly from $x$. Thus, we have that the strongly connected components of $H_{rr'}\setminus{(x,y)}$ are either $\{y\}$ and $H_{rr'}\setminus\{y\}$, or $\{x\}$, $\{y\}$ and $H_{rr'}\setminus\{x,y\}$.
Otherwise, let us assume that $d(r)$ cannot reach $r$ in $H_r^R\setminus{(x,y)}$, and also that $d(r)\in H_{rr'}$. Since $r$ is reachable from every vertex of $H_r^R\setminus\{x,d(r)\}$ in $H_r^R\setminus{(x,y)}$, we must have that $(d(r),x)$ is the only kind of outgoing edge of $d(r)$ in $H_r^R$. Now, since $x\in H_{rr'}$ and the immediate dominator of $d(r)$ in $H_r^R$ is $r$, we have that $x$ is dominated by $d(r)$ in $H_r^R$; and since $(d(r),x)$ is an edge of $H_r^R$, we must have that $d(r)$ is the immediate dominator of $x$ in $H_r^R$. Now, since $y\in H_{rr'}$, we must also have that $y$ is dominated by $d(r)$ in $H_r^R$. And then, since $(d(r),x)$ is the only kind of outgoing edge of $d(r)$ in $H_r^R$, and $(x,y)$ is the only outgoing edge of $x$ in $H_r^R$ (since $x$ is marked in $G_s$ and $y=d(x)$ in $G_s$), we must have that $(x,y)$ is a bridge of $H_r^R$, and so $y$ is a marked vertex of $H_r^R$ and $x$ is the immediate dominator of $y$ in $H_r^R$. Since $x$ is an ordinary vertex of $H_{rr'}$, and $y$ is marked vertex of $H_r^R$ such that $x=d(y)$ in $H_r^R$, we must have that either $x$ is marked in $H_r^R$ and $r'=x$, or that $r'=d(x)$ in $H_r^R$. In any case, recall that $(d(r),x)$ is the only kind of outgoing edge of $d(r)$ in $H_r^R$, and $(x,y)$ is the only outgoing edge of $x$ in $H_r^R$. Thus, if $r'=x$, then the vertex set of $H_{rr'}$ is $\{d(r),x,y\}$. Furthermore, the strongly connected components of $H_{rr'}\setminus{(x,y)}$ in this case are $\{x\}$, $\{y\}$ and $\{d(r)\}$ ($=H_{rr'}\setminus{(x,y)}$). Otherwise, if $r'=d(r)$, then the vertex set of $H_{rr'}$ is $\{r,d(r),x,y\}$.  Furthermore, the strongly connected components of $H_{rr'}\setminus{(x,y)}$ in this case are $\{x\}$, $\{y\}$,  $\{d(r)\}$ and $\{r\}$.
$(2.4)$ Let $x$ be $\mathit{oa}$ and $y$ be $\mathit{aa}$. This means that both $x$ and $y$ are auxiliary in $H_r$, which implies that $x=d(r)$ in $G_s$. But since $x$ is ordinary in $H_{rr'}$, this case is not permitted to occur.

$(3)$ Let $x$ by an auxiliary vertex of $H_{rr'}$ and let $y$ be an ordinary vertex of $H_{rr'}$. Since $x\neq d(r')$ in $H_r^R$, this means that $x$ is a marked vertex of $H_r^R$ with $d(x)\in T(r')$ in $H_r^R$. Now we distinguish four different cases, depending on whether $x$ and $y$ are ordinary or auxiliary in $H_r$. In any case, as an implication of Lemma~\ref{lemma:r-reachability}, we have that $d(r')$ can reach every vertex of $H_{rr'}$ in $H_{rr'}\setminus{(x,y)}$.
$(3.1)$ Let $x$ be $\mathit{ao}$ and $y$ be $\mathit{oo}$. Let $(x_0,y)$ be an edge of $H_r^R$ that corresponds to $(x,y)$ in $H_{rr'}$. Then $(y,x_0)$ is an edge of $H_r$. If $x_0$ is an ordinary vertex of $H_r$, then by Lemma~\ref{lemma:r-reachability} we have that $r$ reaches all vertices of $H_r$ in $H_r\setminus{(y,x_0)}$. This implies that every vertex of $H_r^R$ can reach $r$ in $H_r^R\setminus{(x_0,y)}$. Then Lemma~\ref{lemma:dr-reachability-cor_edge} implies that every vertex of $H_{rr'}$ can reach $d(r')$ in $H_{rr'}\setminus{(x,y)}$, and so we have that $H_{rr}\setminus{(x,y)}$ is strongly connected. Similarly, if $x_0$ is an auxiliary vertex of $H_r$, then, by Lemma~\ref{lemma:bridge-r-reachability}, we have that $r$ reaches all vertices of $H_r$ in $H_r\setminus{(y,x_0)}$, except $x_0$, which is unreachable from $r$ in $H_r\setminus{(y,x_0)}$, and possibly $d(r)$. This implies that every vertex of $H_r^R$ can reach $r$ in $H_r^R\setminus{(x_0,y)}$, except $x_0$, which cannot reach $r$ in $H_r^R\setminus{(x_0,y)}$, and possibly $d(r)$. Since $x$ is an ordinary vertex of $H_r$, we have that $x\neq x_0$, and so $x$ can reach $r$ in $H_r^R\setminus{(x_0,y)}$. Now Lemma~\ref{lemma:dr-reachability-cor_edge} implies that $x$ can reach $d(r')$ in $H_{rr'}\setminus{(x,y)}$. Thus, every vertex of $H_{rr'}$ can reach $d(r')$ in $H_{rr'}\setminus{(x,y)}$, and so we have that $H_{rr'}\setminus{(x,y)}$ is strongly connected.
$(3.2)$ Let $x$ be $\mathit{ao}$ and $y$ be $\mathit{oa}$. Let $\{(x_1,y),\dots,(x_k,y)\}$ be the set of all the incoming edges to $y$ in $H_r^R$. Then $\{(y,x_1),\dots,(y,x_k)\}$ is a collection of edges of $H_r$ that stem from $y$. Since $y$ is auxiliary in $H_r$ and ordinary in $H_{rr'}$, we have that $y\neq d(r)$ in $G_s$, and therefore $y$ is a marked vertex of $G_s$ such that $d(y)\in T(r)$. Thus, by Lemma~\ref{lemma:r-reachability2}, we have that $r$ can reach all vertices of $H_r$ in $H_r\setminus\{(y,x_1),\dots,(y,x_k)\}$. This implies that $r$ can be reached from all vertices of $H_r^R$ in $H_r^R\setminus\{(x_1,y),\dots,(x_k,y)\}$. Furthermore, since $y$ is dominated by $d(r')$ in $H_r^R$, we have that there is a path from $r$ to $d(r')$ in $H_r^R$ that avoids $y$; therefore, this path also avoids all edges $\{(x_1,y),\dots,(x_k,y)\}$. Thus we infer that $d(r')$ is reachable from all vertices of $H_r^R$ in $H_r^R\setminus\{(x_1,y),\dots,(x_k,y)\}$. In particular, let $P$ be a path from $x$ to $d(r')$ in $H_r^R\setminus\{(x_1,y),\dots,(x_k,y)\}$. Then, by Lemma~\ref{lemma:cor_paths} the corresponding path $P_H$ in $H_{rr'}$ has the same endpoints as $P$ and avoids the edge $(x,y)$. Thus, even by removing $(x,y)$ from $H_{rr'}$, all vertices from $H_{rr'}$ can reach $d(r')$. We conclude that $H_{rr'}\setminus{(x,y)}$ is strongly connected.
$(3.3)$ Let $x$ be $\mathit{aa}$ and $y$ be $\mathit{oo}$. Let $(x_0,y)$ be an edge of $H_r^R$ that corresponds to $(x,y)$ in $H_{rr'}$. Then $(y,x_0)$ is an edge of $H_r$. Now, if $x_0$ is ordinary in $H_r$ or it is auxiliary in $H_r$ but $x_0\neq x$, then we can argue as in $(3.1)$ in order to establish that $H_{rr'}\setminus{(x,y)}$ is strongly connected.
Otherwise, suppose that $x_0=x$. First, let us assume that $d(r)\notin H_{rr'}$. Now, since $x$ is auxiliary in $H_r$ and $y$ is ordinary in $H_r$ (and $x\neq d(r)$), we have that $y=d(x)$ in $G_s$. Now, by Lemma~\ref{lemma:bridge-r-reachability} we have that $r$ reaches all vertices of $H_r$ in $H_r\setminus{(y,x)}$, except $x$, which is unreachable from $r$ in $H_r\setminus{(y,x)}$, and possibly $d(r)$. This implies that every vertex of $H_r^R$ can reach $r$ in $H_r^R\setminus{(x,y)}$, except $x$, which cannot reach $r$ in $H_r^R\setminus{(x,y)}$, and possibly $d(r)$. Now Lemma~\ref{lemma:dr-reachability-cor_edge} implies that every vertex of $H_{rr'}$ can reach $d(r')$ in $H_{rr'}\setminus{(x,y)}$, except possibly $x$ and $d(r)$. Since we have assumed that $d(r)\notin H_{rr'}$, we conclude that either $H_{rr'}\setminus{(x,y)}$ is strongly connected, or its strongly connected components are $\{x\}$ and $H_{rr'}\setminus\{x\}$.
Now let us assume that $d(r)\in H_{rr'}$. There are two cases to consider: either $r'=d(r)$, or $d(r)$ is the immediate dominator of $r'$ in $H_r^R$, and so it is contained in $H_{rr'}$ as the critical vertex. In either case, we have that $x\neq d(r)$. Thus we can argue as previously, in order to get that every vertex of $H_{rr'}$ can reach $d(r')$ in $H_{rr'}\setminus{(x,y)}$, except $x$, which cannot reach $d(r')$ in $H_{rr'}$, and possibly $d(r)$. Now, if $d(r)$ is the critical vertex of $H_{rr'}$, then we have that $d(r)=d(r')$, and so we conclude that the strongly connected components of $H_{rr}\setminus{(x,y)}$ are $\{x\}$ and $H_{rr}\setminus\{x\}$. Otherwise, let $r'=d(r)$. Then, since $H_{rr'}$ is strongly connected, we have that every vertex of $H_{rr'}$ reaches $d(r')$ in $H_{rr'}$. And since $y=d(r)$ is forbidden (since $r'=d(r)$), we have that $(x,y)$ is not an incoming edge to $d(r)$. Thus, every vertex of $H_{rr'}$ can reach $d(r')$ in $H_{rr'}\setminus{(x,y)}$, except possibly $x$. We conclude that the strongly connected components of $H_{rr'}\setminus{(x,y)}$ are $\{x\}$ and $H_{rr'}\setminus\{x\}$.
$(3.4)$ Let $x$ be $\mathit{aa}$ and $y$ be $\mathit{oa}$. Then we can use the same argument that we used in $(3.2)$ in order to conclude that $H_{rr'}\setminus{(x,y)}$ is strongly connected.

$(4)$ Since $y\neq d(r')$ in $H_r^R$, it is impossible that both $x$ and $y$ are auxiliary in $H_{rr'}$.
\end{proof}

\begin{figure*}[t!]
\begin{center}
\centerline{\includegraphics[trim={0 0 0 7.5cm}, clip=true, width=1\textwidth]{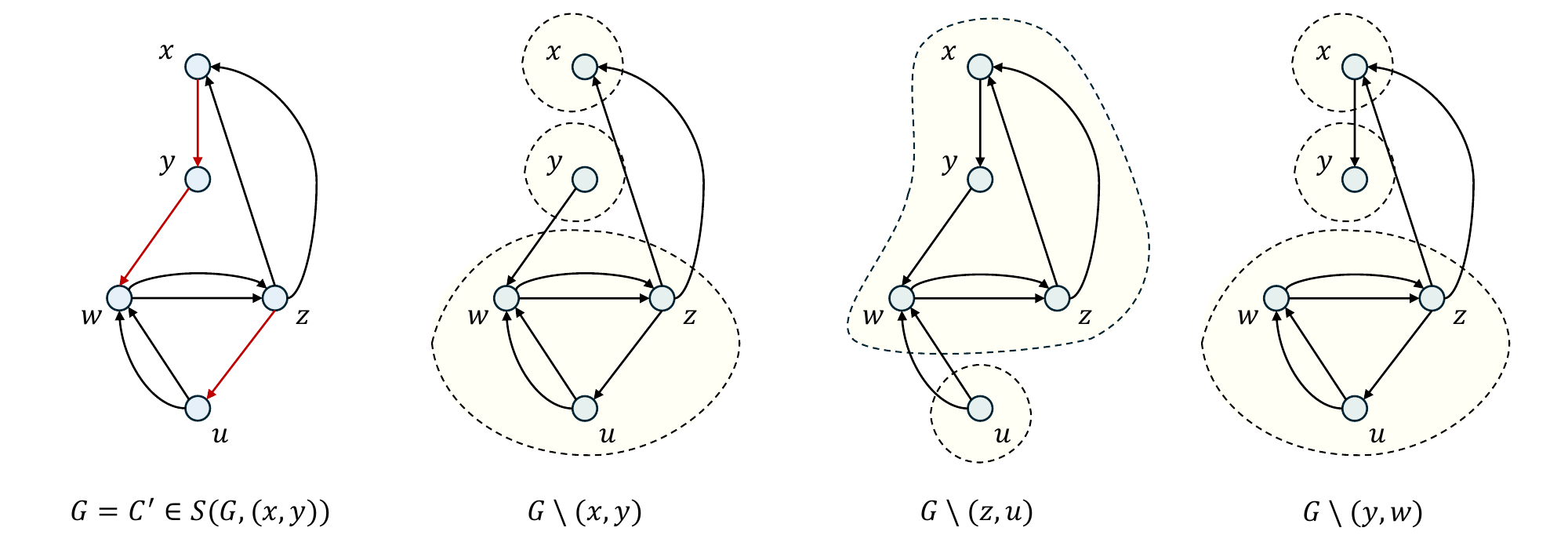}}
\vspace{-.3cm}
\caption{A strongly connected graph $G$ with its strong bridges shown in red color. $C=\{w,u,z\}$ is a SCC of $G \setminus (x,y)$ and the corresponding graph $C' \in S(G,(x,y))$ is identical to $G$. Each SCC of $G \setminus e$ that does not contain $x$ or $y$, for any strong bridge $e$ of $G$, lies entirely within $C$. \label{figure:SopSCCs}}
\end{center}
\vspace{-.5cm}
\end{figure*}

The following Lemma shows that the $S$-operation maintains the strongly connected components of a digraph upon removal of strong bridges. (See also Figure~\ref{figure:SopSCCs}.)

\begin{lemma}
\label{lemma:S-components}
Let $G$ be a strongly connected digraph and let $(x,y)$ be a strong bridge of $G$. Let $C$ be a strongly connected component of $G\setminus{(x,y)}$, and let $C'$ be the corresponding graph in $S(G,(x,y))$. Let $e$ be an edge of $C'$.
\begin{enumerate}[label={(\arabic*)}]
\item{Suppose first that both endpoints of $e$ lie in $C$. Then $e$ is a strong bridge of $C'$ if and only if $e$ is a strong bridge of $G$. Furthermore, let $C_1,\dots,C_k$ be the strongly connected components of $G\setminus{e}$. Then at most one of $C_1,\dots,C_k$ contains both $x$ and $y$. All the other strongly connected components of $G\setminus{e}$ either lie entirely within $C$, or they do not intersect with $C'$ at all. Now let $i\in\{1,\dots,k\}$. Then, if $C_i$ contains both $x$ and $y$, but no vertex of $C$, then $\{x\}$ and $\{y\}$ are two distinct strongly connected components of $C'\setminus{e}$. Otherwise, $C_i\cap C'$ (if it is non-empty) is a strongly connected component of $C'\setminus{e}$.}

\item{Now suppose that at least one endpoint $p$ of $e$ lies outside of $C$. Then $e$ is a strong bridge of $C'$ if and only if it is the only incoming or outgoing edge of $p$ in $C'$. Furthermore, the strongly connected components of $C'\setminus{e}$ are $\{x\}$ (if $x\notin C$), $\{y\}$ (if $y\notin C$), and $C$.}
\end{enumerate}
\end{lemma}
\begin{proof}
$(1)$ Let $e$ be an edge of $C'$ such that both endpoints of $e$ lie in $C$. Let $u,v$ be two vertices of $C$. Then, as a consequence of Lemma~\ref{lemma:compressed_path}, we have that $u$ reaches $v$ in $C'\setminus{e}$ if and only if $u$ reaches $v$ in $G\setminus{e}$ $(*)$. This implies that $e$ is a strong bridge of $C'$ if and only if $e$ is a strong bridge of $G$. Now let $C_0$ be a strongly connected component of $G\setminus{e}$.
First, let us assume that $C_0$ contains both $x$ and $y$, but no vertex of $C$. This means that neither $x$ nor $y$ is a vertex of $C$. Then, by the definition of $C'$, we have that $(y,x)$ is not an edge of $C'$. However, since both endpoints of $e$ lie in $C$, we have that $e\neq (x,y)$, and so $(x,y)\in C'\setminus{e}$. Now let's assume, for the sake of contradiction, that $y$ can reach $x$ in $C'\setminus{e}$. Then there must be vertices $z,w\in C$ such that $(y,z)\in C'$, $(w,x)\in C'$, and $z,w$ remain strongly connected in $C'\setminus{e}$. The existence of the edge $(y,z)\in C'$ implies that there is a path $P_1$ from $y$ to $z$ in $G$, that uses no vertices of $C$ except $z$. Similarly, the existence of the edge $(w,x)\in C'$ implies that there is a path $P_2$ from $w$ to $x$ in $G$, that uses no vertices of $C$ except $w$. Now, since $z$ can reach $w$ in $C'\setminus{e}$, Lemma~\ref{lemma:compressed_path} implies that there is a path $P$ from $z$ to $w$ in $G\setminus{e}$. Thus the concatenation $P_1+P+P_2$ is a path from $y$ to $x$ in $G\setminus{e}$ that uses at least one vertex from $C$. But then the existence of $(x,y)$ in $G\setminus{e}$ implies that $\{x,y,z,w\}\subseteq C_0$ - a contradiction. Thus we have that $\{x\}$ and $\{y\}$ are two distinct strongly connected components of $C'\setminus{e}$.

Now let us assume that $C_0$ contains both $x$ and $y$, and also a vertex $z$ of $C$ (of course, it is possible that $z=x$ or $z=y$). Then $(*)$ implies that all vertices of $C_0\cap C$ remain strongly connected in $C'\setminus{e}$. Furthermore, there is no vertex in $C\setminus C_0$ that is strongly connected with vertices of $C_0\cap C$ in $C'\setminus{e}$. Now, since $z$ is strongly connected with $x$ in $G\setminus{e}$, there is a path $P_1$ from $z$ to $x$ in $G\setminus{e}$. Similarly, there is a path $P_2$ from $y$ to $z$ in $G\setminus{e}$. Now consider the concatenation $P=P_1+(x,y)+P_2$, and let $\tilde{P}$ be the compressed path of $P$ in $C'$. Then, by Lemma~\ref{lemma:compressed_path}, $\tilde{P}$ is a path from $z$ to $z$ in $C'$, that uses the edge $(x,y)$ and avoids the edge $e$. Thus, $z$ is strongly connected with both $x$ and $y$ in $C'\setminus{e}$. This means that $C_0\cap C'$ is a strongly connected component of $C'\setminus{e}$.

Finally, let us assume that $C_0$ contains a vertex $z$ of $C$, but not both $x$ and $y$. We will show that $C_0$ lies entirely within $C$. So let us assume, for the sake of contradiction, that there is a vertex $w\in C_0\setminus{C}$. Then, since $z$ and $w$ are strongly connected in $G\setminus{e}$, there is a path $P_1$ from $z$ to $w$ in $G\setminus{e}$. Similarly, there is a path $P_2$ from $w$ to $z$ in $G\setminus{e}$. Now, since $z\in C$ but $w\notin C$, at least one of $P_1$ and $P_2$ must use the edge $(x,y)$. But then the existence of the concatenation $P=P_1+P_2$ implies that $z$ is strongly connected with both $x$ and $y$ in $G\setminus{e}$ - a contradiction. Thus we have that $C_0\subseteq C$. Now we will show that $C_0$ is a strongly connected component of $C'\setminus{e}$. By $(*)$ we have that all vertices of $C_0$ are strongly connected in $C'\setminus{e}$. Furthermore, there is no vertex in $C\setminus C_0$ that is strongly connected with vertices of $C_0$ in $C'\setminus{e}$. Thus, $C_0$ is contained in a strongly connected component $C_0'$ of $C'\setminus{e}$ such that $C_0\subseteq C_0' \subseteq C_0\cup\{x,y\}$. Now let us assume that both $x$ and $y$ are contained in $C_0'$. Then, since $z\in C_0'$, there is a path $P'$ in $C'\setminus{e}$ that starts from $z$, ends in $z$, and uses the edge $(x,y)$. By Lemma~\ref{lemma:compressed_path}, there is a path $P$ in $G$ such that $\tilde{P}=P'$. This implies that $P$ starts from $z$, ends in $z$, uses the edge $(x,y)$ and avoids $e$. But this means that $z$ is strongly connected with both $x$ and $y$ in $G'\setminus{e}$ - a contradiction. Thus $C_0'$ cannot contain both $x$ and $y$. Now, if $C_0$ contains either $x$ or $y$, then $C_0'$ also contains $x$ or $y$, respectively, and we are done (i.e., we have $C_0'=C_0$, because $C_0'$ cannot contain both $x$ and $y$). Thus it is left to assume that $C_0$ contains neither $x$ nor $y$. Now there are three cases to consider: either $x\in C$ and $y\notin C$, or $x\notin C$ and $y\in C$, or $C\cap\{x,y\}=\emptyset$. Let us assume first that $x\in C$ and $y\notin C$. Now suppose, for the sake of contradiction, that $x\in C_0'$. This means that there is a path $P'$ from $z$ to $z$ in $C'\setminus{e}$ that contains $x$. Then, by Lemma~\ref{lemma:compressed_path}, we have that there is a path $P$ in $G$ such that $\tilde{P}=P'$. But this means that $z$ is strongly connected with $x$ in $G\setminus{e}$ - contradicting our assumption that $x\notin C_0$. Thus $x\notin C_0'$. Now suppose, for the sake of contradiction, that $y\in C_0'$. But since $y\notin C$, in order for $z\in C_0'$ to reach $y$ in $C'$, we must necessarily pass from the edge $(x,y)$. Then we can see that $x$ and $y$ are both strongly connected with $z$ in $C'\setminus{e}$, contradicting the fact that $C_0'$ cannot contain both $x$ and $y$. Thus $y\notin C_0'$, and so we have that $C_0'=C_0$. Similarly, we can see that if $x\notin C$ and $y\in C$, then $C_0'$ contains neither $x$ nor $y$, and so we have $C_0'=C_0$. Finally, let us assume that neither $x$ nor $y$ is in $C$. Now, if either $x$ or $y$ is in $C_0'$, then $y$ or $x$, respectively, must also be in $C_0'$, since $(x,y)$ is the unique outgoing (resp. incoming) edge of $x$ (resp. $y$) in $C'$. But this contradicts the fact that $C_0'$ cannot contain both $x$ and $y$. Thus we conclude that neither $x$ nor $y$ is in $C_0'$, and so $C_0'=C_0$.

$(2)$ Now let $e$ be an edge of $C'$ such that at least one endpoint $p$ of $e$ lies outside of $C$. Then we have that either $p=x$ or $p=y$. Let us assume that $p=x$ (the case $p=y$ is treated similarly). Now suppose that $e=(x,z)$, for some $z\in C'$. Then, by the definition of $C'$, we have that $z=y$. Also, $(x,y)$ is the only outgoing edge of $x$ in $C'$. Thus, the strongly connected components of $C'\setminus{(x,y)}$ are $\{x\}$, $\{y\}$ (if $y\notin C$), and $C$. Now suppose that $e=(z,x)$, for some $z\in C'$. Then, by the definition of $C'$, we have that $z\in C$. Thus, if there is another incoming edge $(w,x)$ of $x$ in $C'$, then $C'\setminus{(z,x)}$ is not a strong bridge of $C'$, since $z$ can still reach $w$ in $C'\setminus{(z,x)}$ using edges of $C$. Thus $(z,x)$ is a strong bridge of $C'$ if and only if it is the only incoming edge of $x$ in $C'$. In this case, the strongly connected components of $C'\setminus{(z,x)}$ are $\{x\}$, $\{y\}$ (if $y\notin C$), and $C$.
\end{proof}

\begin{corollary}
\label{corollary:Hsr}
Let $G_s$ be a strongly connected digraph and let $r$ be a marked vertex of $H_s^R$. Let $C$ be a strongly connected component of $H_{sr}\setminus{(d(r),r)}$, and let $H$ be the corresponding graph in $S(H_{sr},(d(r),r))$, where $d(r)$ is the immediate dominator of $r$ in $H_s^R$. Let $e$ be a strong bridge of $H$. Then the strongly connected components of $H\setminus{e}$ are given by one of the following:
\begin{enumerate}[label={(\roman*)}]
\item{$\{x\}$ and $H\setminus\{x\}$, where $x$ is auxiliary in $H_s$ and $y=d(x)$ in $G_s$}
\item{$\{x\}$ and $H\setminus\{x\}$, where $x$ is auxiliary in $H_{sr}$ and $(x,y)$ is the only outgoing edge of $x$ in $H_{sr}$}
\item{$\{y\}$ and $H\setminus\{y\}$, where $y$ is auxiliary in $H_{sr}$ and $x=d(y)$ in $H_s^R$}
\item{$\{x\}$, $\{y\}$ and $H\setminus\{x,y\}$, where $x$ is auxiliary in $H_s$, $y$ is auxiliary in $H_{sr}$, $x=d(y)$ in $H_s^R$ and $y=d(x)$ in $G_s$}
\item{$\{d(r)\}$, $\{r\}$ (if $r\notin C$), and $C$}
\end{enumerate}
\end{corollary}
\begin{proof}
Since $(d(r),r)$ is the only outgoing edge of $d(r)$ in $H_{sr}$, observe that one of the strongly connected components of $H_{sr}\setminus{(d(r),r)}$ is $\{d(r)\}$. It is trivial to verify the Corollary for this case, and so we may assume that $C\neq\{d(r)\}$. Now let us assume first that both endpoints of $e$ lie in $C$. Then case $(1)$ of Lemma~\ref{lemma:S-components} applies, and thus $e$ is subject to one of cases $(i)$ to $(iv)$ of Lemma~\ref{lemma:Hsr}. Thus we get cases $(i)$ to $(iv)$ in the statement of the Corollary. Now, if one of the endpoints of $e$ lies outside of $C$, then case $(2)$ of Lemma~\ref{lemma:S-components} applies, and thus we get case $(v)$ in the statement of the Corollary.
\end{proof}

\begin{corollary}
\label{corollary:Hrr'}
Let $G_s$ be a strongly connected digraph, let $r$ be a marked vertex of $G_s$, and let $r'$ be a marked vertex of $H_r^R$. Let $C$ be a strongly connected component of $H_{rr'}\setminus{(d(r'),r')}$, and let $H$ be the corresponding graph in $S(H_{rr'},(d(r'),r'))$, where $d(r')$ is the immediate dominator of $r'$ in $H_r^R$. Let $e$ be a strong bridge of $H$. Then the strongly connected components of $H\setminus{e}$ are given by one of the following:
\begin{enumerate}[label={(\roman*)}]
\item{$\{x\}$ and $H\setminus\{x\}$, where $x$ is auxiliary in $H_r$ and $y=d(x)$ in $G_s$}
\item{$\{x\}$ and $H\setminus\{x\}$, where $x$ is auxiliary in $H_{rr'}$ and $(x,y)$ is the only outgoing edge of $x$ in $H_{rr'}$}
\item{$\{y\}$ and $H\setminus\{y\}$, where $y$ is auxiliary in $H_{rr'}$ and $x=d(y)$ in $H_r^R$}
\item{$\{x\}$, $\{y\}$ and $H\setminus\{x,y\}$, where $x$ is auxiliary in $H_r$, $y$ is auxiliary in $H_{rr'}$, $x=d(y)$ in $H_r^R$ and $y=d(x)$ in $G_s$}
\item{$\{x\}$, $\{y\}$, $\{d(r)\}$ and $\{r\}$, where $r'=d(r)$, $x$ is auxiliary in $H_r$, $y$ is auxiliary in $H_{rr'}$, $x=d(y)$ in $H_r^R$, $y=d(x)$ in $G_s$, and $(d(r),x)$ is the only kind of outgoing edge of $d(r)$ in $H_{rr'}$}
\item{$\{d(r')\}$, $\{r'\}$ (if $r'\notin C$), and $C$}
\end{enumerate}
\end{corollary}
\begin{proof}
Arguing as in the proof of Corollary~\ref{corollary:Hsr}, we can see that this Corollary is a consequence of Lemma~\ref{lemma:S-components} and Lemma~\ref{lemma:Hrr'}.
\end{proof}

\ignore{
\begin{proposition}
\label{proposition:Hscc}
Let $H$ be one of the auxiliary graphs that appear in Corollary~\ref{corollary:2et_char_main}, and let $e=(x,y)$ be a strong bridge of $H$. Then the strongly connected components of $H\setminus{e}$ are given by one of the following\footnote{With some simplifications, that are expanded on in Appendix~\ref{sec:appendixA}.}:
\begin{enumerate}[label={(\roman*)}]
\item{$\{x\}$ and $H\setminus\{x\}$, where $x$ is an auxiliary vertex}
\item{$\{y\}$ and $H\setminus\{y\}$, where $y$ is an auxiliary vertex}
\item{$\{x\}$, $\{y\}$, and $H\setminus\{x,y\}$, where $x,y$ are both auxiliary vertices}
\end{enumerate}
\end{proposition}
\begin{proof}
See Lemma~\ref{lemma:Hss}, Lemma~\ref{lemma:tilde_hrr_scc}, Corollary~\ref{corollary:Hsr}, and Corollary~\ref{corollary:Hrr'}, in Appendix~\ref{sec:appendixA}.
\end{proof}
}

\section{Computing $2$-edge twinless strongly connected components}
\label{sec:linear}

We assume that $G$ is a twinless strongly connected digraph since otherwise we can compute the twinless strongly connected components in linear time and process each one separately.
We let $E_t$ denote the set of twinless strong bridges of $G$, and let $E_s$ denote the set of strong bridges of $G$. (Note that $E_s \subseteq E_t$.)
Algorithm~\ref{algorithm:twinless-simple} is a simple $O(mn)$-time algorithm for computing the
$2$-edge twinless strongly connected components of $G$. (It is essentially the same as in \cite{JABERI2021102969}.)
\vspace{.5cm}

\begin{algorithm}[H]
\caption{\textsf{Partition vertices of $G$ with respect to its twinless strong bridges.}}
\label{algorithm:twinless-simple}
\LinesNumbered
\DontPrintSemicolon
initialize a partition of the vertices $\mathcal{P} = \{ V(G) \}$\;
compute the set $E_t$ of the twinless strong bridges of $G$\;
\ForEach{$e \in E_t$}{
compute the twinless strongly connected components $C_1,\ldots,C_k$ of $G\setminus e$\;
let $\mathcal{P}=\{S_1,\ldots,S_l\}$ be the current partition of the vertices in $V(G)$\;
refine the partition by computing the intersections $C_i \cap S_j$ for all $i=1,\ldots,k$ and $j=1,\ldots,l$
}
\end{algorithm}
\vspace{.5cm}

Our goal is to provide a faster algorithm by processing separately the edges in $E_t \setminus E_s$ and the edges in $E_s$.
That is, we first partition the vertices according to the twinless strong bridges of $G$ that are not strong bridges, and then we refine this partition by considering the effect of strong bridges. We call the first partition the one that is \emph{``due to the twinless strong bridges that are not strong bridges"}, and the second partition the one that is \emph{``due to the strong bridges"}.

\begin{figure*}[t!]
\begin{center}
\centerline{\includegraphics[trim={0 0 0 16.5cm}, clip=true, width=1.3\textwidth]{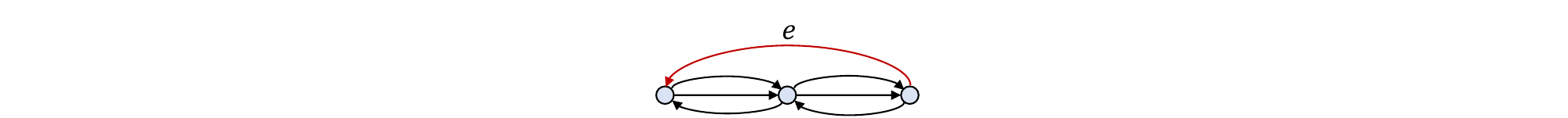}}
\vspace{-.3cm}
\caption{Example of an edge $e$ (colored red) that is a twinless strong bridge but not a strong bridge.
Note that the deletion of $e$ leaves the digraph strongly connected by not twinless strongly connected.
\label{figure:twinless-bridge}}
\end{center}
\vspace{-,5cm}
\end{figure*}

\subsection{Computing the partition of the 2eTSCCs due to the twinless strong bridges that are not strong bridges}
\label{sec:partition-twinlessSB}

Let $e$ be an edge in $E_t\setminus{E_s}$. (See Figure~\ref{figure:twinless-bridge}.) Then the TSCCs of $G\setminus{e}$ are given by the $2$-edge-connected components of $G^u\setminus\{e^u\}$, where $e^u$ is the undirected counterpart of $e$ \cite{Raghavan2006}. Thus, we can simply remove the bridges of $G^u\setminus\{e^u\}$, in order to get the partition into the TSCCs that is due to $e$. To compute the partition that is due to all edges in $E_t\setminus{E_s}$ at once, we may use the cactus graph $Q$ which is given by contracting the $3$-edge-connected components of $G^u$ into single nodes \cite{3cuts:Dinitz}. $Q$ comes together with a function $\phi:V(G^u)\rightarrow V(Q)$ (the quotient map) that maps every vertex of $G^u$ to the node of $Q$ that contains it, and induces a natural correspondence between edges of $G^u$ and edges of $Q$. The cactus graph of the $3$-edge-connected components provides a clear representation of the $2$-edge cuts of an undirected graph; by definition, it has the property that every edge of it belongs to exactly one cycle.  Thus, Algorithm~\ref{algorithm:twinless-Et} shows how we can compute the partition of 2eTSCCs that is due to the edges in $E_t\setminus{E_s}$.

\vspace{.5cm}
\begin{algorithm}[H]
\caption{\textsf{Compute the partition of 2eTSCCs of $G$ that is due to the twinless strong bridges that are not strong bridges.}}
\label{algorithm:twinless-Et}
\LinesNumbered
\DontPrintSemicolon
compute the cactus $Q$ of the $3$-edge-connected components of $G^u$, and let $\phi:V(G^u)\rightarrow V(Q)$ be the quotient map\;
\ForEach{edge $e$ of $Q$}{
  \lIf{$e$ corresponds to a single edge of $G$ that has no twin and is not a strong bridge}{
    remove from $Q$ the edges of the cycle that contains $e$
  }
}
let $Q'$ be the graph that remains after all the removals in the previous step\;
let $C_1,\dots,C_k$ be the connected components of $Q'$\;
\textbf{return} $\phi^{-1}(C_1),\dots,\phi^{-1}(C_k)$\;
\end{algorithm}
\vspace{.5cm}

\begin{proposition}
\label{proposition:alg_Et_correctness}
Algorithm~\ref{algorithm:twinless-Et} is correct and runs in linear time.
\end{proposition}
\begin{proof}
The correctness of Algorithm~\ref{algorithm:twinless-Et} is easily established due to the structure of the cactus of the $3$-edge-connected components. Since the strong bridges of a directed graph can be computed in linear time \cite{Italiano2012}, and the $3$-edge-connected components of an undirected graph can also be computed in linear time (see e.g., \cite{TSIN:3EC,TwinlessSAP}), we have that Algorithm~\ref{algorithm:twinless-Et} runs in linear time.
\end{proof}

\begin{figure*}[t!]
\begin{center}
\centerline{\includegraphics[trim={0 0 0 11cm}, clip=true, width=1\textwidth]{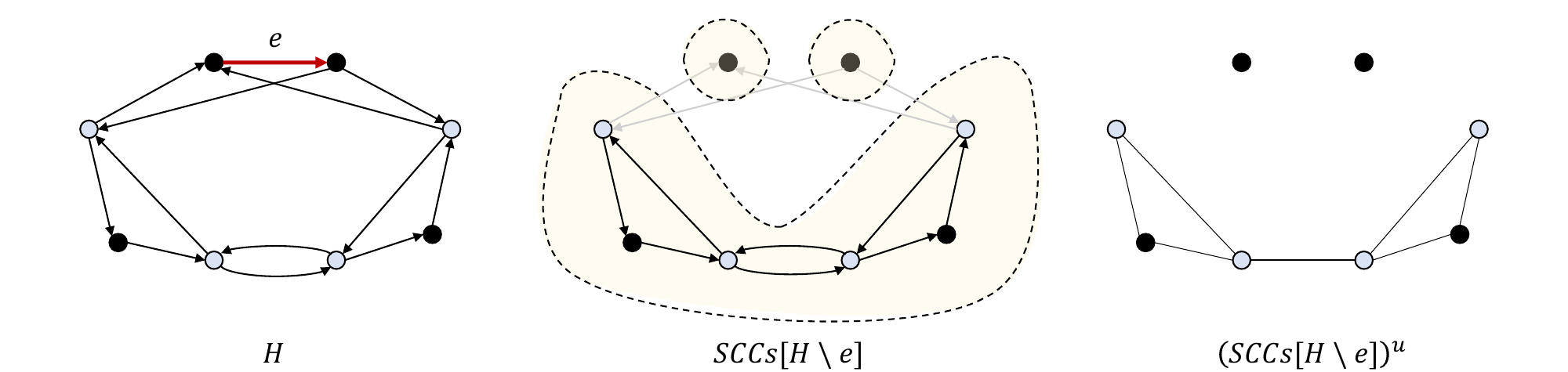}}
\caption{An auxiliary graph $H=H_{uu}$, corresponding to the digraph of Figure~\ref{figure:example0}, with auxiliary vertices colored black. Edge $e$ (shown in red) is a strong bridge of $H$.
The underlying undirected graph of the SCCs of $H \setminus{e}$.\label{figure:Xe}}
\end{center}
\vspace{-.5cm}
\end{figure*}

\subsection{Computing the partition of the 2eTSCCs due to the strong bridges}
\label{sec:partition-SB}

Let $G_s$ be a strongly connected digraph, and let $H$ be either $(1)$ $H_{ss}$, or $(2)$ $\widetilde{H}_{rr}$, or $(3)$ $S(H_{sr},(d(r),r))$, or $(4)$ $S(H_{rr'},(d(r'),r'))$ (for some marked vertices $r$ and $r'$ in $G_s$ and $H_r^R$, respectively). Depending on whether $H$ satisfies $(1)$, $(2)$, $(3)$, or $(4)$, we have a clear picture of the structure of the strongly connected components of $H$ upon removal of a strong bridge, thanks to Lemma~\ref{lemma:Hss}, Lemma~\ref{lemma:tilde_hrr_scc}, Corollary~\ref{corollary:Hsr}, or Corollary~\ref{corollary:Hrr'}, respectively. We can apply this information in order to compute the partition of the $2$-edge twinless strongly connected components of $H$ due to the strong bridges as follows.

Let $e=(x,y)$ be a strong bridge of $H$. Observe that in every case that appears in Lemma~\ref{lemma:Hss}, Lemma~\ref{lemma:tilde_hrr_scc}, Corollary~\ref{corollary:Hsr}, or Corollary~\ref{corollary:Hrr'}, the strongly connected components of $H\setminus{e}$ are given by one of the following: $(i)$, $\{x\}$ and $H\setminus\{x\}$, $(ii)$, $\{y\}$ and $H\setminus\{y\}$, $(iii)$, $\{x\}$, $\{y\}$ and $H\setminus\{x,y\}$, or $(iv)$ $\{x\}$, $\{y\}$, $\{d(r)\}$ and $\{r\}$.
In every case in which $\{x\}$ or $\{y\}$ appears as a separate strongly connected component, we have that $x$, $y$, or both, respectively, are not $\mathit{oo}$ vertices, and so they are not in the same $2$-edge twinless strongly connected component with any $\mathit{oo}$ vertex of $H$.
(We may ignore case $(iv)$, because this is due to a singular case in which $H$ consists of four vertices, where only one of them is $\mathit{oo}$.)
Now, depending on whether we are in case $(i)$, $(ii)$, or $(iii)$, we define $X_e=\{x\}$, $X_e=\{y\}$, or $X_e=\{x,y\}$, respectively. In other words, $X_e$ is the set of the non-$\mathit{oo}$ vertices that get (strongly) disconnected from the rest of the graph upon removal of $e$. Then we have that $X_e$ satisfies the following:

\begin{enumerate}[label={(\arabic*)}]
\item{$H[V\setminus{X_e}]$ is a strongly connected component of $H\setminus{e}$}
\item{$X_e$ contains no $\mathit{oo}$ vertex of $H$}
\end{enumerate}

Now we can apply the following procedure to compute the partition of $2$-edge twinless strongly connected components of (the $\mathit{oo}$ vertices of) $H$ due to the strong bridges. Initially, we let $\mathcal{P}$ be the trivial partition of $V$ (i.e., $\mathcal{P}=\{V\}$). Now, for every strong bridge $e$ of $H$, we compute the TSCCs of $H\setminus{X_e}$, and we refine $\mathcal{P}$ according to those TSCCs. By \cite{Raghavan2006}, the computation of the TSCCs of $H\setminus{X_e}$ is equivalent to determining the $2$-edge-connected components of $H^u\setminus{X_e}$. Observe that this procedure does not run in linear time in total, since it has to be performed for every strong bridge $e$ of $H$.

\begin{figure*}[t!]
\begin{center}
\centerline{\includegraphics[trim={0 0 0 11.5cm}, clip=true, width=\textwidth]{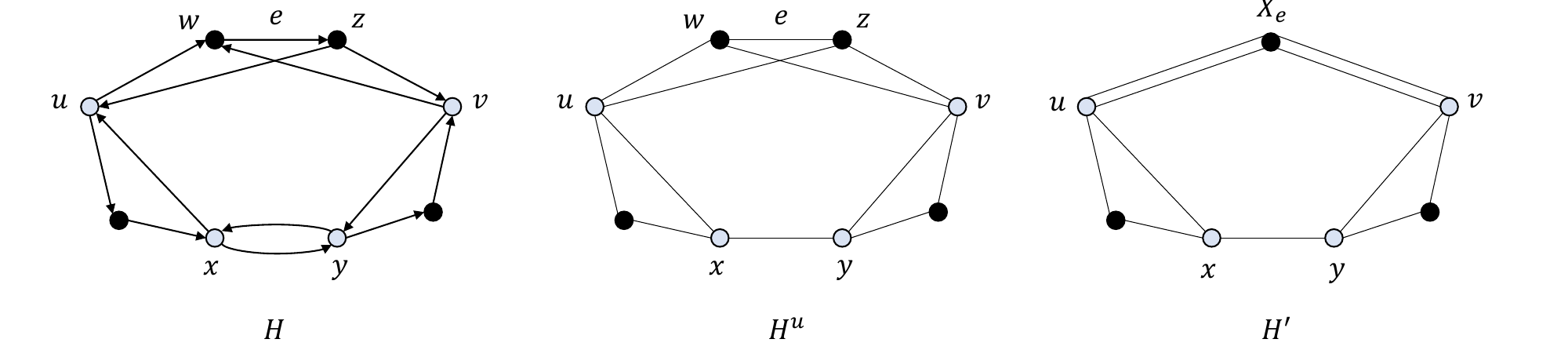}}
\caption{An auxiliary graph $H=H_{uu}$, corresponding to the digraph of Figure~\ref{figure:example0}, with auxiliary vertices colored black.
The underlying undirected graph $H^u$ of $H$, and the corresponding graph $H'$ resulting from $H^u$ after shrinking the vertex sets $X_e$ for all strong bridges $e$; here $X_e = \{w,z\}$.
Note that we have two parallel edges $\{X_e,u\}$ and $\{X_e,v\}$. \label{figure:example3}}
\end{center}
\vspace{-.5cm}
\end{figure*}

Our goal is to perform the above procedure for all strong bridges $e$ of $H$ at once. We can do this by first taking $H^u$, and then shrinking every $X_e$ in $H^u$ into a single marked vertex, for every strong bridge $e$ of $H$. Let $H'$ be the resulting graph. Then we simply compute the marked vertex-edge blocks of $H'$.
(See Figure~\ref{figure:example3} for an example of the process of contracting all $X_e$ into single marked vertices.)
This is shown in Algorithm~\ref{algorithm:linear_main}.
We note that given an auxiliary graph $H$ as above, we can compute all sets $X_e$ in linear time, by first computing all strong bridges of $H$ \cite{Italiano2012}, and then checking which case of Lemma~\ref{lemma:Hss}, Lemma~\ref{lemma:tilde_hrr_scc}, Corollary~\ref{corollary:Hsr}, or Corollary~\ref{corollary:Hrr'} applies for each strong bridge (which can be easily checked in $O(1)$ time).
\vspace{.5cm}

\begin{algorithm}[H]
\caption{\textsf{A linear-time algorithm for computing the partition of $2$-edge twinless strongly connected components of an auxiliary graph $H$ due to the strong bridges}}
\label{algorithm:linear_main}
\LinesNumbered
\DontPrintSemicolon
\SetKwInOut{Input}{input}
\SetKwInOut{Output}{output}
\Input{An auxiliary graph $H$ equipped with the following information: for every strong bridge $e$ of $H$, the set $X_e$ defined as above}
\Output{The partition of $2$-edge twinless strongly connected components of the ordinary vertices of $H$ due to the strong bridges}
\Begin{
compute the underlying undirected graph $H^u$\;
\ForEach{strong bridge $e$ of $H$}{
  contract $X_e$ into a single vertex in $H^u$, and mark it\;
}
let $H'$ be the graph with the marked contracted vertices derived from $H^u$\;
compute the partition $\mathcal{B}_{ve}$ of the marked vertex-edge blocks of $H'$\;
\label{line:alg_line}
let $\mathcal{O}$ be the partition of $V$ consisting of the set of the ordinary vertices of $H$ and the set of the auxiliary vertices of $H$\;
\textbf{return} $\mathcal{B}_{ve}$ refined by $\mathcal{O}$\;}
\end{algorithm}
\vspace{.5cm}

The following two lemmas describe conditions under which the simultaneous contraction of several subsets of the vertex set of an undirected graph maintains the marked vertex-edge blocks.

\begin{lemma}
\label{lemma:same_ve_block}
Let $G=(V,E)$ be a connected undirected graph and let $X_1,\dots,X_k$ be a collection of disjoint subsets of $V$, such that $G\setminus{X_i}$ is connected for every $i\in\{1,\dots,k\}$. Let also $u$ and $v$ be two vertices of $G$ that do not belong to any of $X_1,\dots,X_k$. Suppose that $u$ and $v$ belong to the same $2$-edge-connected component of $G\setminus{X_i}$, for every $i\in\{1,\dots,k\}$. Now let $G'$ be the graph that is derived from $G$ by contracting every $X_i$, for $i\in\{1,\dots,k\}$, into a single marked vertex. Then $u$ and $v$ belong to the same marked vertex-edge block of $G'$.
\end{lemma}
\begin{proof}
Let $z$ be a marked vertex of $G'$ that corresponds to a set $X\in\{X_1,\dots,X_k\}$. Then we have that $u$ and $v$ belong to the same $2$-edge-connected component of $G\setminus{X}$. Thus there is a cycle $C$ in $G\setminus{X}$ that contains $u$ and $v$, but no vertex of $X$. Then $C$ corresponds to a cycle $C'$ of $G'$ that contains both $u$ and $v$ and avoids $z$. Thus $u$ and $v$ are also $2$-edge-connected in $G'\setminus{z}$. Due to the generality of $z$, we have that $u$ and $v$ belong to the same marked vertex-edge block of $G'$.
\end{proof}

\begin{lemma}
\label{lemma:different_ve_blocks}
Let $G=(V,E)$ be a connected undirected graph and let $X_1,\dots,X_k$ be a collection of disjoint subsets of $V$, such that $G\setminus{X_i}$ is connected for every $i\in\{1,\dots,k\}$. Let also $u$ and $v$ be two vertices of $G$ that do not belong to any of $X_1,\dots,X_k$. Suppose that there is a $j\in\{1,\dots,k\}$ such that $u$ and $v$ belong to different $2$-edge-connected components of $G\setminus{X_j}$. Now let $G'$ be the graph that is derived from $G$ by contracting every $X_i$, for $i\in\{1,\dots,k\}$, into a single marked vertex. Then the following is a sufficient condition to establish that $u$ and $v$ belong to different marked vertex-edge blocks of $G'$: Let $T$ be the tree of the $2$-edge-connected components of $G\setminus{X_j}$; then, for every $i\in\{1,\dots,k\}\setminus\{j\}$, we have that either $(1)$ $X_i$ lies entirely within a node of $T$, or $(2)$ a part of $X_i$ constitutes a leaf of $T$, and the remaining part of $X_i$ lies within the adjacent node of that leaf in $T$.
\end{lemma}
\begin{proof}
Let $z$ be the marked vertex of $G'$ that corresponds to $X_j$. Let $T'$ be the tree of the $2$-edge-connected components of $G'\setminus{z}$. We will show that $u$ and $v$ belong to different nodes of $T'$. First, we have that $u$ and $v$ belong to different nodes of $T$. Now, $T'$ is derived from $T$ in the same way that $G'$ is derived from $G$, i.e., by contracting the nodes of $T$ that contain all the different parts of $X_i$ into a single node, for every $i\in\{1,\dots,k\}\setminus\{j\}$. Now take a $X_i\in\{X_1,\dots,X_k\}\setminus\{X_j\}$. If $(1)$ holds for $X_i$, then the contraction of $X_i$ in $G'$ induces no change on $T$. If $(2)$ holds for $X_i$, then the contraction of $X_i$ in $G'$ induces the following change on $T$: a leaf of $T$, that contains neither $u$ nor $v$, is contracted with its adjacent node in $T$. Thus we can see that $u$ and $v$ belong to different nodes of $T'$.
\end{proof}

\begin{proposition}
\label{proposition:alg_correctness_app}
Let $G_s$ be a strongly connected digraph, and let $H$ be a digraph that satisfies one of the following:
\begin{enumerate}[label={(\arabic*)}]
\item{$H=H_{ss}$}
\item{$H=\widetilde{H}_{rr}$, where $r$ is a marked vertex of $G_s$}
\item{$H\in S(H_{sr},(d(r),r))$, where $r$ is a marked vertex of $G_s$ and $(d(r),r)$ is the critical edge of $H_{sr}$}
\item{$H\in S(H_{rr'},(d(r'),r'))$, where $r$ is a marked vertex of $G_s$, $r'$ is a marked vertex of $H_r^R$, and $(d(r'),r')$ is the critical edge of $H_{rr'}$}
\end{enumerate}
Then Algorithm~\ref{algorithm:linear_main} correctly computes the partition of $2$-edge twinless strongly connected components of the $\mathit{oo}$ vertices of $H$ due to the strong bridges.
\end{proposition}
\begin{proof}
We will demonstrate how to handle case $(1)$, by relying on Lemma~\ref{lemma:Hss}. Cases $(2)$, $(3)$ and $(4)$ can be handled using similar arguments, by relying on Lemma~\ref{lemma:tilde_hrr_scc}, Corollary~\ref{corollary:Hsr}, and Corollary~\ref{corollary:Hrr'}, respectively. In any case, we have to establish the following. Let $\mathcal{E}$ be the collection of the strong bridges of $H$. Then the collection $\{X_e\mid e\in\mathcal{E}\}$ consists of sets that are either singletons or $2$-element sets. First we show that the $2$-element sets in $\{X_e\mid e\in\mathcal{E}\}$ are pairwise disjoint. Then we show that, despite the simultaneous contractions of the sets in $\{X_e\mid e\in\mathcal{E}\}$, the marked vertex-edge blocks of the underlying graph of $H$ are maintained. This is achieved by showing that the collection $\{X_e\mid e\in\mathcal{E}\}$ satisfies the preconditions in Lemmas \ref{lemma:same_ve_block} and \ref{lemma:different_ve_blocks}.

Now, in case $(1)$, by Lemma~\ref{lemma:Hss} we have that the sets $X_e$, for all strong bridges $e$ of $H_{ss}$, are given by cases $(i)$ to $(iv)$ in the statement of this Lemma. To be precise, in cases $(i)$ and $(ii)$ we set $X_e\leftarrow \{x\}$; in case $(iii)$ we set $X_e\leftarrow \{y\}$; and in case $(iv)$ we set $X_e\leftarrow\{x,y\}$. Thus, the last case is the only one in which $X_e$ is a $2$-element set.

Now let $e,e'$ be two strong bridges of $H_{ss}$ such that $X_e$ and $X_{e'}$ are two distinct $2$-element sets. We will show that $X_e\cap X_{e'}=\emptyset$.
So let $X_e=\{x,y\}$ and $X_{e'}=\{z,w\}$. Since these sets can occur only in case $(iv)$ of Lemma~\ref{lemma:Hss}, we may assume, without loss of generality, that $e=(x,y)$ and $e'=(z,w)$. Then we have that $y$ and $w$ are auxiliary, but not critical, vertices of $H_{ss}$. Since, then, $y$ and $w$ have only one incoming edge in $H_{ss}$, we must have that either $y=w$ and $x=z$, or $y\neq w$. The first case means that $e=e'$, and so it is impossible to occur. Thus we have $y\neq w$. By Lemma~\ref{lemma:Hss}, we have that $y=d(x)$ in $G_s$ and $w=d(z)$ in $G_s$. Thus, $y\neq w$ implies that $x\neq z$. Notice that none of the cases $x=w$ and $y=z$ can occur, because there is no edge between auxiliary but not critical vertices. We conclude that $\{x,y\}\cap\{z,w\}=\emptyset$.

Now let $H'$ be the graph that is derived from $H^u$ after we have contracted every $X_e$ into a single marked vertex. We will show that two $\mathit{oo}$-vertices of $H_{ss}$ are twinless strongly connected in $H_{ss}\setminus{e}$ for any strong bridge $e$, if and only if they belong to the same marked vertex-edge block of $H'$. So let $u,v$ be two $\mathit{oo}$-vertices of $H_{ss}$ that are twinless strongly connected in $H_{ss}\setminus{e}$ for any strong bridge $e$. Then, by \cite{Raghavan2006} we have that, for every strong bridge $e$ of $H_{ss}$, $u$ and $v$ belong to the same $2$-edge-connected component of $H_{ss}^u\setminus{X_e}$. Then Lemma~\ref{lemma:same_ve_block} implies that $u$ and $v$ belong to the same marked vertex-edge block of $H'$. Conversely, let $u,v$ be two $\mathit{oo}$-vertices of $H_{ss}$ such that there is a strong bridge $e$ for which $u$ and $v$ are not twinless strongly connected in $H_{ss}\setminus{e}$. By \cite{Raghavan2006}, this means that $u$ and $v$ belong to different $2$-edge-connected components of $H_{ss}^u\setminus{X_e}$. Now let $e'$ be a strong bridge of $H_{ss}$ such that $X_{e'}\neq X_{e}$. We will show that $X_{e'}$ satisfies one of conditions $(1)$ and $(2)$ of Lemma~\ref{lemma:different_ve_blocks}. Now, if $X_{e'}$ is a singleton set, then there is nothing to show, since there is no contraction induced by $X_{e'}$ in this case. So let $X_{e'}=\{x,y\}$. This can occur only in case $(iv)$, and so we may assume, without loss of generality, that $e'=(x,y)$. Now, if $x$ and $y$ belong to the same twinless strongly connected component of $H_{ss}\setminus{e}$, then condition $(1)$ of Lemma~\ref{lemma:different_ve_blocks} is satisfied. Otherwise, since $x$ and $y$ are connected with the edge $(x,y)$, we must have that $(y,x)$ exists. But since $(x,y)$ is the only incoming edge of $y$ in $H_{ss}$ (since $x=d(y)$ in $H_s^R$), we must have that $\{y\}$ constitutes a singleton twinless strongly connected component of $H_{ss}\setminus{e}$, and so condition $(2)$ of Lemma~\ref{lemma:different_ve_blocks} is satisfied. Thus, Lemma~\ref{lemma:different_ve_blocks} ensures that, by contracting every $\{X_e\mid e\in\mathcal{E}\}$ into a single marked vertex in $H^u$, $u$ and $v$ belong to different marked vertex-edge blocks of $H'$.
\end{proof}

The final 2eTSCCs of (the subset of the ordinary vertices of) an auxiliary graph are given by the mutual refinement of the partitions computed by Algorithms \ref{algorithm:twinless-Et} and \ref{algorithm:linear_main}. (The mutual refinement of two partitions can be computed in linear time using bucket sort.)
Hence, by Corollary~\ref{corollary:2et_char_main} and Propositions \ref{proposition:alg_Et_correctness} and \ref{proposition:alg_correctness_app}, we have that the
2eTSCCs of a strongly connected digraph can be computed in linear time.

It remains to establish that Algorithm~\ref{algorithm:linear_main} runs in linear time. For this, we provide a linear-time procedure for Step~\ref{line:alg_line}. Observe that the marked vertices of $H'$ have the property that their removal from $H$ leaves the graph strongly connected, and thus they are not articulation points of the underlying graph $H^u$. This allows us to reduce the computation of the marked vertex-edge blocks of $H'$ to the computation of marked vertex-edge blocks in biconnected graphs. Specifically, we first partition $H'$ into its biconnected components, which can be done in linear time \cite{dfs:t}. Then we process each biconnected component separately, and we compute the marked vertex-edge blocks that are contained in it. Finally, we ``glue'' the marked vertex-edge blocks of all biconnected components, guided by their common vertices that are articulation points of the graph. In the next section, we provide a linear-time algorithm for computing the marked vertex-edge blocks of a biconnected graph.

\section{Computing marked vertex-edge blocks}
\label{sec:spqr}
Let $G$ be a biconnected undirected graph.
An SPQR tree $\mathcal{T}$ for $G$ represents the triconnected components of $G$~\cite{SPQR:triconnected,SPQR:planarity}.
Each node $\alpha \in \mathcal{T}$ is associated with an undirected graph $G_\alpha$. Each vertex of $G_\alpha$ corresponds to a vertex of the original graph $G$.
An edge of $G_\alpha$ is either a \emph{virtual edge} that corresponds to a separation pair of $G$, or a \emph{real edge} that corresponds to an edge of the original graph $G$.
The node $\alpha$, and the graph $G_\alpha$ associated with it, has one of the following types:
\begin{itemize}
\item If $\alpha$ is an $S$-node, then $G_\alpha$ is a cycle graph with three or more vertices and edges.
\item If $\alpha$ is a $P$-node, then  $G_\alpha$ is a multigraph with two vertices and at least $3$ parallel edges.
\item If $\alpha$ is a $Q$-node, then $G_\alpha$ is a single real edge.
\item If $\alpha$ is an $R$-node, then $G_\alpha$ is a simple triconnected graph.
\end{itemize}
Each edge $\{\alpha,\beta\}$ between two nodes of the SPQR tree is associated with two virtual edges, where one is an edge in $G_\alpha$ and the other is an edge in $G_\beta$.
If $\{u,v\}$ is a separation pair in $G$, then one of the following cases applies:
\begin{itemize}
\item[(a)] $u$ and $v$ are the endpoints of a virtual edge in the graph $G_\alpha$ associated with an $R$-node $\alpha$ of $\mathcal{T}$.
\item[(b)] $u$ and $v$ are vertices in the graph $G_\alpha$ associated with a $P$-node $\alpha$ of $\mathcal{T}$.
\item[(c)] $u$ and $v$ are vertices in the graph $G_\alpha$ associated with an $S$-node $\alpha$ of $\mathcal{T}$, such that either $u$ and $v$ are not adjacent, or the edge $\{u,v\}$ is virtual.
\end{itemize}

In case (c), if $\{u,v\}$ is a virtual edge, then $u$ and $v$ also belong to a $P$-node or an $R$-node. If $u$ and $v$ are not adjacent then $G \setminus \{u,v\}$ consists of two components that are represented by two paths of the cycle graph $G_\alpha$ associated with the $S$-node $\alpha$ and with the SPQR tree nodes attached to those two paths.
Gutwenger and P. Mutzel~\cite{SPQR:GM} showed that an SPQR tree can be constructed in linear time, by extending the triconnected components algorithm of Hopcroft and Tarjan~\cite{3-connectivity:ht}.

Let $e=\{x,y\}$ be an edge of $G$ such that $\{v,e\}$ is a vertex-edge cut-pair of $G$.
Then, $\mathcal{T}$ must contain an $S$-node $\alpha$ such that $v$, $x$ and $y$ are vertices of $G_\alpha$ and $\{x,y\}$ is not a virtual edge.
The above observation implies that we can use $\mathcal{T}$ to identify all vertex-edge cut-pairs of $G$ as follows.
A vertex-edge cut-pair $(v,e)$ is such that $v\in V(G_{\alpha})$ and $e$ is a real edge of $G_{\alpha}$ that is not adjacent to $v$, where $\alpha$ is an $S$-node~\cite{TwinlessSAP,2halfconnectivity}.
Now we define the \emph{split operation} of $v$ as follows. Let $e_1$ and $e_2$ be the edges incident to $v$ in $G_{\alpha}$.
We split $v$ into two vertices $v_1$ and $v_2$, where $v_1$ is incident only to $e_1$ and $v_2$ is incident only to $e_2$. (In effect, this makes $S$ a path with endpoints $v_1$ and $v_2$.)
To find the connected components of $G \setminus \{v,e\}$, we execute a split operation on $v$ and delete $e$ from the resulting path. Note that $e \not= e_1,e_2$, and $e$ does not have a copy in any other node of the SPQR tree since it is a real edge. Then, the connected components of $G \setminus \{v,e\}$ are represented by the resulting subtrees of $\mathcal{T}$.

Here, we need to partition the ordinary vertices of $G$ according to the vertex-edge cut-pairs $(v,e)$, where $v$ is a marked auxiliary vertex. To do this efficiently, we can process all vertices simultaneously as follows. First, we note that we only need to consider the marked vertices that are in $S$-nodes that contain at least one real edge.
Let $\alpha$ be such an $S$-node. We perform the split operation on each marked (auxiliary) vertex $v$, and then delete all the real edges of $\alpha$.
This breaks $\mathcal{T}$ into subtrees, and the desired partition of the ordinary vertices is formed by the ordinary vertices of each subtree. See Algorithm~\ref{algorithm:spqr-partition}.
\vspace{.5cm}

\begin{algorithm}[H]
\caption{\textsf{Partition the ordinary vertices of a biconnected graph $G$ according to a set of marked vertex-edge cuts.}}
\label{algorithm:spqr-partition}
\LinesNumbered
\DontPrintSemicolon
\SetKwInOut{Input}{input}
\SetKwInOut{Output}{output}
\Input{An SPQR tree $\mathcal{T}$ of $G$, a set of marked vertices $v \in V(G_{\alpha})$ and real edges $e \in E(G_{\alpha})$, for all $S$-nodes $\alpha$ of $\mathcal{T}$.}
\Output{A partition of the ordinary vertices of $G$, so that two ordinary vertices in the same set of the partition remain in the same connected component of $G \setminus (v,e)$, for any vertex-edge cut $(v,e)$ such that $v$ is marked.}
\Begin{
\ForEach{$S$-node $\alpha$ of $\mathcal{T}$ that contains a marked vertex and a real edge}{
  perform a split operation on each marked vertex of $\alpha$\;
  delete all real edges of $G_\alpha$
}
break $\mathcal{T}$ into connected subtrees $\mathcal{T}_1,\ldots,\mathcal{T}_{\lambda}$ after the above operations\;
\ForEach{subtree $\mathcal{T}_i$}{
put all the ordinary vertices of $\mathcal{T}_i$ into the same set of the partition\;
}
}
\end{algorithm}
\vspace{.5cm}

\begin{theorem}
\label{theorem:marked-vertex-edge}
The marked vertex-edge blocks of an undirected graph can be computed in linear time.
\end{theorem}
\begin{proof}
First, we establish the correctness of Algorithm~\ref{algorithm:spqr-partition}.
To see why this algorithm works, consider two ordinary vertices $u$ and $w$ of $G$. First observe that if $u$ and $w$ are in different connected components of $G \setminus \{v,e\}$ for some marked vertex-edge cut $(v,e)$, then $u$ and $w$ will end up in different subtrees of $\mathcal{T}$.
Now suppose that $u$ and $w$ remain in the same connected component of $G \setminus \{v,e\}$, for any vertex-edge cut $(v,e)$ with $v$ marked.
Let $\beta$ and $\gamma$ be the nodes of $\mathcal{T}$ such that $u \in V(G_\beta)$ and $w \in V(G_\gamma)$.
Consider any $S$-node $\alpha$ that lies on the path of $\mathcal{T}$ between $\beta$ and $\gamma$ and contains at least one marked vertex and at least one real edge.
(If no such $S$-node exists, then clearly $u$ and $w$ cannot be separated by any marked vertex-edge cut.)
Let $e_u$ and $e_v$ be the virtual edges of $E(G_\alpha)$ that correspond to the paths from $\beta$ to $\alpha$ and from $\gamma$ to $\alpha$, respectively.
Also, let $P_1$ and $P_2$ be the two paths that connect $e_u$ and $e_v$ in $G_{\alpha}$.
Without loss of generality, we can assume that $P_1$ contains a marked vertex $v$. Then, $P_2$ cannot contain any real edge $e$, since otherwise
$(v,e)$ would be a marked vertex-edge cut separating $u$ and $w$. Hence, all real edges are on $P_1$.
But then, for the same reason, $P_2$ cannot contain a marked vertex. Hence, all marked vertices are also on $P_1$.
This implies that $\beta$ and $\gamma$ remain in the same subtree of $\mathcal{T}$ after the execution of Algorithm~\ref{algorithm:spqr-partition}.
The same arguments work if $u$ or $w$ (or both) are vertices of $V(G_\alpha)$.
Also, it is easy to see that Algorithm~\ref{algorithm:spqr-partition} runs in linear time. 
\end{proof}

Now, since Algorithm~\ref{algorithm:spqr-partition} computes the marked vertex-edge blocks in linear time, we have that Algorithm~\ref{algorithm:linear_main} also runs in linear time.
Hence, we obtain the following result:
\begin{theorem}
\label{theorem:twinless2ecc}
The $2$-edge twinless strongly connected components of a directed graph can be computed in linear time.
\end{theorem}

Finally, by the reduction of Section~\ref{sec:preliminaries}, we have:

\begin{theorem}
\label{theorem:mixed2ecc}
The edge-resilient strongly orientable blocks of a mixed graph can be computed in linear time.
\end{theorem}

\section{Concluding remarks}
\label{sec:concluding}

In this paper, we studied the notion of edge-resilient strongly orientable blocks of a mixed graph $G$. Each such block $C$ has the property that for any (directed or undirected) edge $e$ of $G$, there is an orientation $R$ of $G\setminus{e}$ that maintains the strong connectivity of the vertices in $C$.
We note that if we change in the definition of edge-resilient strongly orientable blocks the assumption that the edges that we allow to fail are both directed and undirected, and we demand that they are only directed, or only undirected, then we can easily modify Algorithm~\ref{algorithm:blocks2} so that we can compute those blocks in linear time (using again the reduction to computing $2$-edge twinless strongly connected components).

\begin{figure*}[t!]
\begin{center}
\centerline{\includegraphics[trim={0 0 0 17.7cm}, clip=true, width=1.3\textwidth]{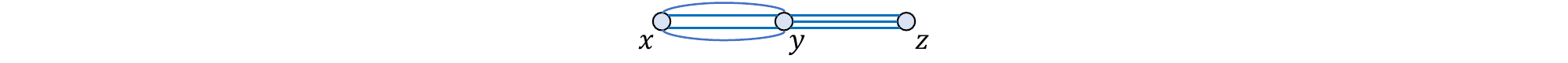}}
\caption{In this example $\{x,y\}$ is a $2$-edge strongly orientable block. No orientation can make $z$ and $y$ (or $z$ and $x$) $2$-edge strongly connected. $\{x,y,z\}$ is an edge-resilient strongly orientable block.\label{figure:block2}}
\end{center}
\vspace{-.5cm}
\end{figure*}

We may also introduce the similar concept of the \emph{$2$-edge strongly orientable blocks} of a mixed graph. These are the maximal sets of vertices $C_1,\dots,C_k$ with the property that, for every $i\in\{1,\dots,k\}$, there is an orientation $R$ of $G$ such that all vertices of $C_i$ are $2$-edge strongly connected in $R$.
There are some relations between the $2$-edge strongly orientable blocks and the edge-resilient strongly orientable blocks.
First, we can easily see that every $2$-edge strongly orientable block lies within an edge-resilient strongly orientable block.
Moreover, both these concepts coincide with the $2$-edge strongly connected components in directed graphs. However, Figure~\ref{figure:block2} shows that these concepts do not coincide in general.
Despite these connections, the efficient computation of the $2$-edge strongly orientable blocks seems to be an even more challenging problem than computing the edge-resilient strongly orientable blocks, since the former generalizes the notion of $2$-edge strong orientations of a mixed graph.

%

Finally, we note that our techniques may be useful for solving other connectivity problems in mixed graphs, such as related connectivity augmentation problems~\cite{MixedGraphAugmentation2,Gabow:Poset:TALG,MixedGraphAugmentation}.

\newcommand{\etalchar}[1]{$^{#1}$}

\end{document}